\documentstyle[12pt]{article}
\newcommand{\ltwid}{\raise.3ex\hbox{$<$\kern-.75em\lower1ex\hbox{$\sim$}}}
\newcommand{\ds}{\displaystyle}
\renewcommand{\theequation}{\arabic{section}.\arabic{equation}}
\author{{\sc Vadim~M.~Loktev$^{(1)}$, Sergei~G.~Sharapov$^{(1,2)}$}\\[1.5ex]
\it $^{(1)}$Bogolyubov Institute for Theoretical Physics\\
\it of the National Academy of Sciences of Ukraine \\
\it 14-b Metrologichna Str., 252143 Kyiv--143, Ukraine \\
\it E-mail: vloktev@gluk.apc.org \\[0.5ex]
\it $^{(2)}$Department of Physics, University of Pretoria, \\
\it 0002 Pretoria, South Africa\\
\it E-mail: sharapov@scientia.up.ac.za}

\title{SUPERCONDUCTING CONDENSATE FORMATION IN METALLIC SYSTEMS
WITH ARBITRARY CARRIER DENSITY}
\date{July 18, 1997}

\begin{document}
\setcounter{page}{1}
\maketitle
\begin{abstract}
This article gives a contemporary and to some extent pedagogical
review of the current theoretical understanding of the formation of
the superconducting state in metallic systems with a variable density
of carriers. We make an attempt to describe the crossover
from the Bose-Einstein condensation type (small densities) to
the Bardeen-Cooper-Schrieffer one (large densities).
The functional methods are used throughout the treatment.
The most of the results are considered in a review form for the first
time. Some of them (in particular, the possible opening of a pseudogap)
are used to explain the experimental data avaliable for
high-temperatures superconductors.
\end{abstract}
\newpage
\tableofcontents
\newpage
\section{Introduction}
\subsection{The general formulation of the problem}
     It is well known that there are two fundamental limiting descriptions
which allow us to understand the equilibrium low temperature
"superphenomena", such as superconductivity and superfluidity,
in 3D systems:

\noindent
{\bf i)} Bardeen-Cooper-Schrieffer (BCS) theory \cite{Schrieffer}, in which
the normal state is a degenerate Fermi liquid that undergoes a
cooperative pairing (Cooper) instability at the temperature $T_{P}$.
Here two processes, the formation of Cooper pairs and their condensation
(macroscopic occupation of a single quantum state), occur
simultaneously at the transition temperature, $T_{c}^{3D} = T_{P}$.

\noindent
{\bf ii)} Bose-Einstein condensation (BEC) of bosons (see e.g.
\cite{Kvasnikov}) in another single quantum state at the temperature
$T_{B}$. In fact, these bosons (for instance, $^{4}$He) are tightly
bound (composite) particles made up of an even number of fermions.
The bosonic particles are formed at some high temperature $T_{P}$ of
their binding (dissociation). However in contrast to the BCS case,
these "pre-formed" bosons  condense only at $T_{B} \ll T_{P}$. The
above-mentioned single quantum states as well as an intermediate case,
which we will discuss below in details, are usually described by an order
parameter which is {\em homogeneous} in the 3D case
assuming that there are no external fields.

	Most, if not all, of the real 3D systems clearly fall into
either category i) or ii).  For example, $^{3}$He and
essentially all metallic superconductors that we now understand are
Fermi superfluids described by the first category, whereas $^{4}$He
is a Bose superfluid described by the second one.

	Nevertheless, the general problem of the crossover (or
interpolation) from the BCS scenario of superconductivity with
cooperative Cooper pairing to the formation of composite (separate)
bosons and their BEC has remained of great interest for a long time.
It is connected primarily with a deeper understanding of the phenomenon
of superconductivity even for 3D systems which are clearly closer to
the BCS limit.

	The significance of the problem increases when one tries to
reduce the dimensionality of space from 3D to 2D.  In such a case the
problem acquires not only fundamental significance, but also practical
importance. The latter, as we shall discuss below, is mainly related
to the discovery of the high-temperature superconductors (HTSC)
\cite{Bednorz}. It is not the only reason for the widespread interest
in the problem of the crossover. Even from the theoretical point of view
the 2D crossover is far more mysterious than the 3D one. At first glance
the BCS limit with weak attraction and high fermion density is completely
different from the Bose limit with strong coupling and low boson density.
Indeed, the simplest BCS theory predicts a finite value of
$T_{c}^{2D} = T_{P}^{2D}$ while the BEC is forbidden in 2D systems
for massive bosons \cite{Kvasnikov} for which $T_{B}^{2D} \equiv 0$.
This is the reason one might ask whether it is possible to obtain a
2D crossover.

            It is worthwhile to note here that by virtue of the
Coleman-Mermin-Wagner-Hohenberg theorem \cite{Coleman} a
"super-behaviour"
\footnote{"Super-behaviour" refers to 2D ferro- or
antiferromagnetism in the case of the Mermin-Wagner theorem and to
superfluid behaviour in Hohenberg's theorem, while the Coleman theorem
is the field-theoretical generalization of the previous theorems
(see e.g. \cite{Miransky}).}
with {\em homogeneous} order parameter (the latter
related to breaking of some continuous symmetry) is not possible
for pure 2D systems. This is due to the fluctuations of the order
parameter (in particular of its phase) destroying the long-range
order \cite{Rice} (see also \cite{Emery}). Therefore, the temperature
$T_{c}^{2D}$ of establishing long-range order has to be zero. The
statement that $T_{B}^{2D} = 0$ is in agreement with this general theorem.
In fact the BCS result $T_{P}^{2D}$ does not contradict these theorems
if one remembers that as a rule the BCS theory presupposes the
{\em mean field} (MF) approach. Thus it does not take into account any
fluctuations and the result only means that
$T_{c}^{MF(2D)} = T_{P}^{2D} \neq 0$, while the temperature $T_{c}^{2D}$
of actually establishing the long-range order is zero in this case
as well. Such a conclusion, however, results in a general question:
how can one study 2D models of superconductors with finite $T_{c}$
values?

             There are at least two ways to answer this question.

\noindent {\em First}, 2D models are only a mathematical
idealization and any real system must be at least quasi-two
dimensional (quasi-2D). This immediately gives $T_{c} \neq 0$,
although $T_{c} < T_{c}^{MF}$.

\noindent {\em Second}, there is another interesting possibility
for 2D systems to pass into the superfluid state at a temperature
which is usually denoted by $T_{BKT}$. This corresponds to the so
called Berezinskii-Kosterlitz-Thouless (BKT) phase transition.
The BKT superfluid state has an {\em inhomogeneous} order parameter
and is therefore not forbidden by the above-mentioned 2D theorems.
The formation of this state is quite different from the ordinary
superconducting transition and results in a new way of condensation
(see Chapter~5).

	  The possibility for real systems to undergo the BKT
transition crucially depends on their spatial anisotropy. Thus, one
should study both 2D and quasi-2D cases.

             Another important aspect of
the crossover problem is related to the following unusual property of
the HTSC compounds. Almost all the physical characteristics of HTSC's
($T_{c}$ included) crucially depend on the itinerant carrier density.
This means that the crossover in superconducting systems cannot be
studied without taking into account the changing delocalized fermion
density. We believe that many of the HTSC   "anomalies" are caused by
the unusual (two-stage) formation of the superconducting state in 2D and
(very probably) quasi-2D systems with an arbitrary carrier density
\cite{JETPLetters}.

	  Hopefully, this short introduction has convinced the
potential readers that the superconducting condensate formation at the
different densities of interacting fermions (in particular, in the
BCS and Bose limiting cases) deserves a separate review.

\subsection{History}

The idea that the composite bosons (or, as they called, local pairs)
exist and define the superconducting properties of metals is in fact
more then 10 years older than the BCS theory.  As early as 1946,
a sensational communication appeared saying that the
chemist-experimentator Ogg had observed superconductivity in the
solution of Na in NH$_{3}$ at 77$K$ \cite{Ogg}. It is very interesting
that the researcher made an attempt to interpret his own result in
terms of the BEC of {\em paired electrons}. Unfortunately, the
discovery was not confirmed and both it  and his theoretical
concept, were soon completely forgotten (for the details see
e.g. \cite{Dmitrenko}).

	  It is appropriate to note that the history of
superconductivity has many similar examples. Probably, this explains
why Bednorz and M\"uller named their first paper "Possible high-$T_c$
superconductivity..." \cite{Bednorz}. And even today there appear
many unconfirmed communications about room-temperature superconducting
transitions.

	  A new step, the development of the local pair concept,
was taken in 1954 by Schafroth who in fact re-discovered the idea of
the electronic quasi-molecules \cite{Schafroth}. This idea was
further developed in the Schafroth, Blatt and Butler theory of
quasi-chemical equilibrium \cite{Blatt}, where superconductivity was
considered versus the BEC. Such a scenario, unfortunately, could not
compete with the BCS one due to some mathematical difficulties which
did not allow the authors to obtain the famous BCS results.
Then the triumph of the BCS theory replaced the far more obvious
concept of the local pairs and their BEC by the Cooper ones and their
instability which takes place in the necessary presence of a Fermi
surface (more precisely a finite density of states at the Fermi level).
In contrast for the local (i.e. separated) pairs the formation is
not in principle connected to this density of electronic states. Also
unlike the local pairs, the Cooper ones are highly overlapping
in real space.  Therefore, the Cooper pairing should be understood as a
{\em momentum} space pairing.

	 Later experiment reminded us about the local pairs. It was
Frederikse et al. \cite{Frederikse} who found
that the superconducting compound SrTiO$_{3}$ ($T_{c} \sim 0.3$K) has
a relatively low density of carriers and, moreover, this density is
controled by Zr doping. The first deep discussion of the possibility
of the BCS-Bose crossover and pairing above the
superconductivity transition temperature for a low density of
carriers  was carried out by Eagles \cite{Eagles.1969}
(see also his et al. relatively recent article
\cite{Eagles.1989}) in the context of SrTiO$_{3}$.

	  The further history of the investigation of the BCS-Bose
crossover has been often cited in the current literature. So, we
note only that the features of 3D crossover at $T = 0$ were considered
in \cite{Leggett}, and its extension to finite temperatures was first
given by Nozieres and Schmitt-Rink \cite{Nozieres}.

	   The 2D crossover in superfluid $^{3}$He was studied in
\cite{Miyake}, where the very natural and convenient physical
parameter $\varepsilon_{b}$, the bound pair state energy, was first
used.

	   The discovery of HTSC significantly stimulated the interest
in the problem of crossover and related phenomena. There are many
articles dealing with its study and we can only mention
some of them at this point
\cite{Randeria.2D,Uemura,Drechsler,Akh,GGL,Haussmann,Randeria.3D,Pistolesi,Quick,Carter,Mic,GLSh.Cond}
(see also the recent excellent review \cite{Randeria.book} devoted
to other aspects of the BCS-Bose crossover, e.g. lattice models,
time-dependent Ginzburg-Landau theory, numerical study of normal
state crossover etc., which we because of this shall not discuss).
We note only that some historical debates are associated with
the parallel (and generally speaking independent) development of the
macro-  and microphysics of superfluidity and superconductivity since
their discoveries were touched upon by Ginzburg in his new extremely
comprehensive survey \cite{Ginzburg}.

\subsection{Relevance to the high-temperature superconductors}
As we have already mentioned the crossover problem appears to be
relevant to the general problem of the understanding of the HTSC.
Indeed, these superconducting compounds have some peculiarities
which place them much closer to the Bose  or at least to the
crossover region than majority of low temperature superconductors.
We shall discuss the peculiarities in detail here.

	 Certainly, it is necessary to point out right away that the
crossover phenomena do not (and cannnot) address the problem of the
mechanism for HTSC. The problem of HTSC itself is very difficult and
controversial. Nevertheless the treatment of the crossover may shed
light on some of the features of HTSC.

	 Almost all HTSC reveal following common properties:
\begin{itemize}
\item a relatively low and easily varied density of itinerant
      (appeared owing to the doping) carriers;

\item  a 2D, or more exactly quasi-2D, character of the conductivity
      and the magnetism;

\item the block structure of the crystal lattices, i.e. from $1$ to $6$
superconducting CuO$_2$ layers per unit cell.
\end{itemize}

	 Let us consider these items. The ground state of the
copper-oxide based materials forms due to the  strong antiferromagnetic
spin fluctuations in the proximity of the metal-insulator transition
(see e.g. review \cite{Loktev.review}). It seems plausible that
further doping of these materials results in the appearance of weakly
interacting (i.e. non-strongly correlated)
itinerant carriers (holes). However the exact nature of the
ground state including the strong electron-electron correlations is in
fact not yet established. The density $n_{f}$ of these holes is not as
large as in ordinary metals, so the mean distance between them proves
to be comparable with a pair size $\xi_{0}$ or a coherence length
\footnote{It is
well to bear in mind that the coherence length strictly speaking is
distinguished from the pair size, especially at low carrier density
(this question will be treated in Section~3.4).}.  This situation is
significantly different from the conventional BCS theory where the
parameter $\xi_{0}$ greatly exceeds the mean distance between carriers
which is $\sim n_{f}^{1/2}$.  Experimentally the  dimensionless value of
$k_{F} \xi_{0}$ ($k_{F}$ is the Fermi momentum) which describes the
ratio of the pair size and the distance between carriers is about
$5$ -- $20$ for HTSC while for the low-temperature superconductors it is
about $10^{3}$ -- $10^{4}$ \cite{Randeria.2D,Loktev.review}.

	  It follows from the above that the new materials are likely
to be in an intermediate regime between the Cooper pairs and the
composite bosons, at least when the doping is not large and the value
of $T_{c}$ is far from the highest possible (optimal) one. Although
there is some evidence \cite{Quick} that HTSC at the optimal doping
are closer to the BCS limit than to the Bose one.

	  Besides, because the coherence length is less than a lattice
spacing in the direction perpendicular to the CuO$_2$ planes (c-direction),
the superconductivity in the copper oxides takes place mainly in  the
isolated CuO$_2$ layers (or their blocks). This is the reason why pure 2D
models of HTSC are commonly accepted. Undoubtedly these cuprate layers are
connected even if they are situated in different unit cells. The mechanisms
for this connection (by coherent or incoherent electronic transport) are not
yet known exactly. There is no question, however, that one has to take into
account the possibility of different (for instance, direct or indirect)
interlayer hoppings to develop a full theory of HTSC. Therefore, strictly
speaking, we need to consider quasi-2D models of these superconductors.
Moreover, as we have already mentioned, it is  necessary to extend the
system into a third-dimension in order to have a finite value for
$T_{c}$. It can be seen from the anisotropy of the conductivity
that the influence of the third-dimension varies strongly from one
family of cuprates to the other. For example, the anisotropy reaches
$10^{5}$ for the Bi- and Tl-based cuprates, while its value for
YBaCu$_{3}$O$_{6+\delta}$ compound is close to $10^{2}$. It seems
plausible that in the Bi(Tl)-compounds where the transport in the
c-direction is incoherent that the scenario with BKT transition is
more probable.  In the Y-ones this transport is rather coherent and
the superconducting state, at least at high doping, has a 3D character,
with a homogeneous condensate which appears in the ordinary way.

	  In general, it is now evident  that, depending on the crystal
anisotropies (and/or the relationship between the intensity of c-transport
and the doping value) the quasi-2D systems
can undergo two or even three phase transitions. The first
one is from the normal phase to another normal (with a pseudogap) phase
with a finite density of incoherent pairs. The second one is from the
latter phase to the BKT phase with an algebraic order. The third and final
one is from the BKT phase to the superconducting phase with long-range
order. These possibilities will be discussed below.

	  The role of interplanar effects inevitably increases when
the superconducting planes are situated in the same unit cell so that
the carriers in the adjacent layers are strongly coupled and can form
interplanar pairs. This possibility together with the concentration
effects brings about the simultaneous existence of the Cooper and local
pairs in the system \cite{GLSh.FNT.1995}.

\subsection{Outline}
The remainder of this review is organized as follows.
In Chapter~2 we first describe the model for studying
the BCS-Bose crossover, especially in 3D systems. Then in
Section~2.2 we introduce in detail the functional integral formalism which
will be used throughout.  The rest of this Chapter is devoted to applying
the methods developed to the 3D crossover.

        The zero temperature 2D crossover is discussed in Chapter~3.
Considerable attention will be given to the effective potential (see also
Appendix A), which allows one to obtain the main equations. The
kinetic terms of the effective action are obtained in Section~3.4. This gives
one the possibility to investigate the dependencies of the coherence length
and the penetration depth on the carrier density.

        In Chapter~4 the quasi-2D crossover is studied.

        Finally, the finite temperature 2D crossover is addressed in
Chapter~5. We describe how it proves necessary to modify the accepted
methods to consider the BKT phase formation (refer also to Appendices B
and C) and its dependence on carrier density correctly. An attempt to
explain some observable normal state anomalies HTSC is also made.

\section*{Acknowledgments}
We would like to thank Profs. V.P.~Gusynin and V.A.~Miransky,
Drs. E.V.~Gorbar, I.A.~Shovkovy and V.M.~Turkowskii for their fruitful
collaboration and numberless discussions which helped to make clear
some deep questions of the low dimensional phase transitions. We also
indebted to Profs. V.G.~Bar'yakhtar, P.I.~Fomin and I.V.~Stasyuk
whose stimulating support
at the different stages of the work allowed us to complete the
review.  Particularly we  thank Prof. R.M.~Carter and Dr. I.A.~Shovkovy for
many thoughtful comments on the earlier version of this manuscript.  At last,
one of us (S.G.Sh) is grateful to the members of the Department of Physics of
the University of Pretoria, especially Prof. R.M.~Carter and Dr.
N.J.~Davidson, for very useful points and hospitality.

\section{3D crossover: critical temperature}
\setcounter{equation}{0}
It is most convenient for studying the crossover to start with the 3D
case where there is no problem with establishing the long-range
order below $T_{c}^{3D}$. We shall mainly follow here the paper
\cite{Randeria.3D} (see also the review \cite{Randeria.book}) in
which the corresponding results were obtained by applying the more
appropriate in this case the functional integral formalism rather
than to the original paper \cite{Nozieres}.

\subsection{Model}
Let us introduce a continuum field model of fermions with an attractive
two-body interaction. Our goal is to consider how the critical
temperature $T_{c}^{3D}$ changes as a function of attraction and to
establish the main factors which determine it in the BCS and Bose
regimes.

	The simplest model is described by the Hamiltonian density
\begin{equation}
{\cal H} = - \psi^{\dagger}_{\sigma}(x) \left(
	   \frac{\nabla^{2}}{2 m} + \mu \right) \psi_{\sigma}(x)
	 - V \psi^{\dagger}_{\uparrow}(x) \psi^{\dagger}_{\downarrow}(x)
       \psi_{\downarrow}(x) \psi_{\uparrow}(x) ,
						       \label{Hamiltonian.3D}
\end{equation}
where $x \equiv \mbox{\bf r}, \tau$ ($\mbox{\bf r}$ is a 3D vector);
$\psi_{\sigma}(x)$ is the Fermi field; $m$ is the fermion effective
mass; $\sigma = \uparrow, \downarrow$ is the fermion spin;
$V > 0$ is the attraction constant. The chemical potential $\mu$
fixes the average density $n_{f}$ of the free (bare) carriers.
We choose units in which $\hbar = k_{B} = 1$ and the system occupies
the volume $v$.

	  Since we want to describe the Bose regime in terms of its
constituent fermions, we have to allow the magnitude of the attraction
$V$ to be arbitrary. In addition, we cannot use a simple BCS-like
cutoff $\omega_{D} \ll \epsilon_{F}$ (usually $\omega_{D}$ is the Debye
frequency) since we must allow the possibility for {\em all} the
fermions to be affected by the interaction, not just for a small
fraction ($\sim \omega_{D}/\epsilon_{F}$) in a shell around the Fermi
energy $\epsilon_{F}$.  We shall thus study a dilute Fermi gas in which
the range of the attractive interactions can be characterized by a shape
independent parameter, the scattering length $a_{s}$ in three dimensions.
We shall describe in Section~2.3 how the "renormalized" coupling
$a_{s}^{-1}$ replaces "bare" $V$.

\subsection{Formalism}
As was already noted, the functional integral approach along
with the Matsubara thermal technique is more appropriate for the
problem that is studied here and in the subsequent Chapters. Thus,
let us consider the formalism used.

First of all, introducing Nambu spinors for fermion fields
\cite{Nam}
(see also \cite{Schrieffer})
\begin{equation}
\Psi(x) = \left( \begin{array}{c}
\psi_{\uparrow}(x) \\  \psi_{\downarrow}^{\dagger}(x)
\end{array} \right), \qquad
\Psi^{\dagger}(x) = \left( \begin{array}{cc}
\psi_{\uparrow}^{\dagger}(x) \quad \psi_{\downarrow}(x)
\end{array} \right)
                            \label{Nambu.variables}
\end{equation}
one has to rewrite (\ref{Hamiltonian.3D}) in a more suitable form:
\begin{equation}
{\cal H} =  -\Psi^{\dagger}(x)
\left( \frac{\nabla^{2}}{2 m} + \mu \right) \tau_{3} \Psi(x)
	    - V \Psi^{\dagger}(x) \tau_{+} \Psi(x)
		\Psi^{\dagger}(x) \tau_{-} \Psi(x) ,
					\label{Hamiltonian.Nambu}
\end{equation}
where $\tau_{3}$, $\tau _{\pm} \equiv (\tau _{1} \pm i \tau _{2}) / 2$
are Pauli matrices.

	   Now the partition function is expressed through the
Hamiltonian (\ref{Hamiltonian.Nambu}) as:
\begin{equation}
Z(v, \mu, T) = \int {\cal D} \Psi {\cal D} \Psi^{\dagger}
\exp \left\{-\int_{0}^{\beta} d \tau \int d \mbox{\bf r}
[\Psi^{\dagger}(x) \partial_{\tau} \Psi(x) + {\cal H}(\mbox{\bf r})]
\right\},                    \label{partition.main}
\end{equation}
where $\beta \equiv 1/T$ and ${\cal D} \Psi {\cal D} \Psi^{\dagger}$
denotes the measure of the integration over the Grassman variables
$\Psi$ and $\Psi^{\dagger}$, satisfying the antiperiodic boundary
conditions: $\Psi(\tau, \mbox{\bf r}) = -\Psi(\tau+\beta, \mbox{\bf r})$
and $\Psi^{\dagger}(\tau, \mbox{\bf r}) =
-\Psi^{\dagger}(\tau+\beta, \mbox{\bf r})$.

      If it was possible to calculate the partition function
(\ref{partition.main}) one could, in principle, obtain all
thermodynamical functions from the thermodynamical potential
\begin{equation}
\Omega(v,\mu,T) = - T \ln Z(v, \mu, T).
\label{thermopotential}
\end{equation}

	    Using now the auxiliary Hubbard-Stratonovich complex scalar
field in the usual way one can represent (\ref{partition.main}) in
an equivalent form:
\begin{eqnarray}
&& Z(v, \mu, T) =
\int {\cal D} \Psi {\cal D} \Psi^{\dagger}
{\cal D} \Phi {\cal D} \Phi^{\ast}
\exp \left\{-\int_{0}^{\beta} d \tau \int d \mbox{\bf r}
\left[ \frac{|\Phi(x)|^{2}}{V} +  \right. \right.
\nonumber                                       \\
&& \quad \left. \left.
\Psi^{\dagger}(x) \left[ \partial_{\tau} \hat I -
\tau_{3} \left( \frac{\nabla^{2}}{2 m} + \mu \right) -
\tau_{-} \Phi(x) - \tau_{+} \Phi^{\ast}(x)
\right]\Psi(x) \right]
\right\}.                    \label{partition.Hubbard}
\end{eqnarray}
The main virtue of this representation is a nonperturbative
introduction of the composite fields
$\Phi(x) = V \Psi^{\dagger}(x) \tau_{+} \Psi(x) = $
$V \psi^{\dagger}_{\uparrow}(x) \psi^{\dagger}_{\downarrow}(x)$,
$\Phi^{\ast}(x) = $ \\
$V \Psi^{\dagger}(x) \tau_{-} \Psi(x) = $
$V\psi_{\downarrow}(x) \psi_{\uparrow}(x)$
and the possibility to develop a consistent approach.
Specifically, the expression (\ref{partition.Hubbard})
turns out to be rather convenient for studying such a
nonperturbative phenomenon as, for example, superconductivity.
In this case the complex Hubbard-Stratonovich field naturally
describes the order parameter arising due to the formation of
the Cooper or the local pairs. The average value of
$|\Phi|( \equiv \Delta)$ is proportional to the density of pairs,
on the one hand, and determines the gap in the one-particle
Fermi-spectrum, on the other.

	    The integration over fermion fields in
(\ref{partition.Hubbard}) can be done formally even though $\Phi$
and $\Phi^{\ast}$ depend on the spatial and temporal coordinates.
Thus, one obtains (formally exactly)
\begin{equation}
Z(v, \mu, T) =
\int {\cal D} \Phi {\cal D} \Phi^{\ast}
\exp [ - \beta \Omega(v, \mu, T, \Phi(x), \Phi^{\ast}(x))],
		    \label{partition.effective}
\end{equation}
where
\begin{equation}
\beta \Omega(v, \mu, T, \Phi(x), \Phi^{\ast}(x)) =
\frac{1}{V} \int_{0}^{\beta} d \tau \int d\mbox{\bf r} |\Phi(x)|^{2}
- \mbox{TrLn}G^{-1} + \mbox{TrLn}G_{0}^{-1}
		    \label{Effective.Action}
\end{equation}
is the one-loop effective action.
This action includes in itself a series of terms containing
derivatives with respect to $\Phi(x)$ and $\Phi^{\ast}(x)$. In
the lowest orders it corresponds to the Ginzburg-Landau effective
action.

           The operation Tr in (\ref{Effective.Action}) is taken with
respect to the space $\mbox{\bf r}$, the imaginary time $\tau$ and
the Nambu indices. The action (\ref{Effective.Action}) is expressed
through the fermion Green function which obeys the equation:
\begin{equation}
\left[-\hat I \partial_{\tau} +
\tau_{3} \left(\frac{\nabla^{2}}{2 m} + \mu \right)
+ \tau_{-} \Phi(\tau, \mbox{\bf r})
+ \tau_{+} \Phi^{\ast}(\tau, \mbox{\bf r}) \right]
G(\tau, \mbox{\bf r}) = \delta(\tau) \delta(\mbox{\bf r})
			 \label{Green.fermion}
\end{equation}
with boundary condition
\begin{equation}
G(\tau + \beta, \mbox{\bf r}) = - G(\tau, \mbox{\bf r}).
                                     \label{boundary}
\end{equation}
The free Green function
\begin{equation}
G_{0}(\tau, \mbox{\bf r}) =
G(\tau, \mbox{\bf r})|_{\Phi, \Phi^{\ast}, \mu = 0}
                          \label{Green.free}
\end{equation}
in (\ref{Effective.Action}) is needed to provide the regularization
in the calculation of \\
$\Omega(v, \mu, T, \Phi, \Phi^{\ast})$.
The representation (\ref{partition.effective}),
(\ref{Effective.Action}) is exact, although to perform the
calculation in practice it is necessary to restrict ourselves to some
approximation. Below we shall use the assumption, generally accepted
in the 3D case (and in some quasi-2D systems), that the
approximation including only the quadratic (Gauss)
fluctuations of the fields $\Phi(x)$ and $\Phi^{\ast}(x)$ (about
their equilibrium values which will now also be denoted as $\Phi$)
describes the system quite well
\footnote{We note that this Gaussian
approximation does not work in the 2D case at $T \neq 0$ and, thus,
we must modify it to analyze this case (see Chapter~5).}.

	 The thermodynamical potential $\Omega$ as well as the
partition sum $Z$ now depend on $\Phi$ and $\Phi^{\ast}$ which
play the role of the order parameter. The order parameter appears
due to the fact that one uses the effective potential
$\Omega(v, \mu, T, \Phi, \Phi^{\ast})$ instead of the exact potential
$\Omega(v, \mu, T)$.
Thus, it becomes necessary to write down the
additional equations which determine the values of $\Phi$ and
$\Phi^{\ast}$.

Moreover, in
some cases it is sufficient to use the mean field approximation
as is done in the original BCS theory. Here, however, we
include the fluctuations also. Therefore, one obtains
\begin{eqnarray}
&& \!  \! \! \! \! \! \! \! \! \! \! \!
Z(v, \mu, T, \Phi, \Phi^{\ast})  =
\exp[-\beta \Omega_{pot}(v, \mu, T, \Phi, \Phi^{\ast})]
\times \nonumber                          \\ && \! \! \! \! \! \! \!
\! \! \!  \int {\cal D} (\Delta \Phi) {\cal D} (\Delta \Phi^{\ast})
\exp \left\{-\int_{0}^{\beta} d \tau_{1} \int_{0}^{\beta} d \tau_{2}
\int d \mbox{\bf r$_{1}$} \int d \mbox{\bf r$_{2}$} \times \right.
\nonumber                          \\
&& \! \! \! \! \! \! \! \! \!
\left( \Delta \Phi^{\ast}(\tau_{1}, \mbox{\bf r$_{1}$})
\left. \frac{\beta \delta^{2} \Omega (\Phi, \Phi^{\ast})}
{\delta \Phi^{\ast}(\tau_{1}, \mbox{\bf r$_{1}$})
 \delta \Phi(\tau_{2}, \mbox{\bf r$_{2}$})}
\right|_{\Phi, \Phi^{\ast} = const}
\Delta \Phi(\tau_{2}, \mbox{\bf r$_{2}$}) \right. +
\nonumber                          \\
&& \! \! \! \! \! \! \!
\Delta \Phi(\tau_{1}, \mbox{\bf r$_{1}$})
\left.
\frac{1}{2}
\frac{\beta \delta^{2} \Omega (\Phi, \Phi^{\ast})}
{\delta \Phi(\tau_{1}, \mbox{\bf r$_{1}$})
 \delta \Phi(\tau_{2}, \mbox{\bf r$_{2}$})}
\right|_{\Phi, \Phi^{\ast} = const}
\Delta \Phi(\tau_{2}, \mbox{\bf r$_{2}$}) +
\nonumber                          \\
&& \! \! \! \! \!
\Delta \Phi^{\ast}(\tau_{1}, \mbox{\bf r$_{1}$})
\left. \left. \left.
\frac{1}{2}
\frac{\beta \delta^{2} \Omega (\Phi, \Phi^{\ast})}
{\delta \Phi^{\ast}(\tau_{1}, \mbox{\bf r$_{1}$})
 \delta \Phi^{\ast}(\tau_{2}, \mbox{\bf r$_{2}$})}
\right|_{\Phi, \Phi^{\ast} = const}
\Delta \Phi^{\ast}(\tau_{2}, \mbox{\bf r$_{2}$}) \right)\right\},
		    \label{partition.work}
\end{eqnarray}
where
\begin{eqnarray}
&& \! \! \! \!
\Omega_{pot}(v, \mu, T, \Phi, \Phi^{\ast}) =
\nonumber                     \\
&& \left. \left(
\frac{1}{V}  \int d\mbox{\bf r} |\Phi(x)|^{2}
- T \mbox{TrLn}G^{-1} + T \mbox{TrLn}G_{0}^{-1}
\right) \right|_{\Phi = \Phi^{\ast} = const}
		    \label{Omega.Potential}
\end{eqnarray}
is the mean field thermodynamical potential;
$\Delta \Phi(x) = \Phi(x) - \Phi$ and
$\Delta \Phi^{\ast}(x) = \Phi^{\ast}(x) - \Phi^{\ast}$ are the
fluctuation deviations from the equilibrium value  of the order parameter.

	 In present review we restrict the consideration of
the fluctuations to the critical line ($\Phi = \Phi^{\ast} = 0$)
only. So, the partition function (\ref{partition.work}) after
integration over $\Delta \Phi$ and $\Delta \Phi^{\ast}$
acquires the form
\begin{equation}
Z(v, \mu, T) = \exp [-\beta \Omega_{pot}(v, \mu, T, 0,0) -
\mbox{TrLn} \Gamma^{-1}],      \label{partition.final}
\end{equation}
where now
\begin{eqnarray}
\Gamma^{-1} (\tau, \mbox{\bf r}) \equiv
\left. \frac{\beta \delta^{2} \Omega (\Phi, \Phi^{\ast})}
{\delta \Phi^{\ast}(\tau, \mbox{\bf r})
 \delta \Phi(0, 0)}
\right|_{\Phi, \Phi^{\ast} = 0} =
\nonumber                        \\
\frac{1}{V} \delta(\tau) \delta(\mbox{\bf r}) +
\mbox{tr} [G(\tau, \mbox{\bf r}) \tau_{-}
G(-\tau, -\mbox{\bf r}) \tau_{+}] |_{\Phi, \Phi^{\ast} = 0}
		    \label{Gamma.def}
\end{eqnarray}
is the inverse Green function of the order parameter fluctuations.

       To avoid possible misunderstanding, we shall
write down the formulae for the  Fourier transformations which are
used throughout the paper; they  connect the coordinate
and momentum representations in the usual manner:
\begin{equation}
F(i \omega_{n}, \mbox{\bf k}) = \int_{0}^{\beta} d \tau \int
d\mbox{\bf r} F(\tau, \mbox{\bf r}) \exp(i \omega_{n} \tau - i
\mbox{\bf k} \mbox{\bf r}), \label{Fourier.momentum}
\end{equation}
\begin{equation}
F(\tau, \mbox{\bf r}) =
T \sum_{n = -\infty}^{+\infty}
\int \frac{d {\mbox{\bf k}}}{(2 \pi)^{d}} F(i \omega_{n}, \mbox{\bf k})
\exp(-i \omega_{n} \tau + i \mbox{\bf k} \mbox{\bf r}),
\label{Fourier.coordinate}
\end{equation}
where $\omega_{n} = \pi T (2n + 1)$ are the fermion (odd) Matsubara
frequencies and $d$ is the dimensionality of the space (remember
that in this Chapter $d = 3$). In the case of bosons the frequencies
should be replaced by even ones: $\Omega_{n} = 2 \pi n T$.

	     For example, the Green function (\ref{Green.fermion})
has, in the momentum representation, the following form:
\begin{equation}
G(i \omega_{n}, \mbox{\bf k}) = -
\frac{i \omega_{n} {\hat I} + \tau_{3} \xi(\mbox{\bf k})
-\tau_{-} \Phi - \tau_{+} \Phi^{\ast}}
{\omega_{n}^{2} + \xi^{2}(\mbox{\bf k})  + |\Phi|^{2}},
		       \label{Green.fermion.momentum}
\end{equation}
where $\xi(\mbox{\bf k}) = \varepsilon(\mbox{\bf k}) - \mu$ with
$\varepsilon(\mbox{\bf k}) = \mbox{\bf k}^{2}/2m$ and
$\Phi$ and $\Phi^{\ast}$ are already taken to be constants.

\subsection{The mean field analysis}
Substituting (\ref{Green.fermion.momentum}) into
(\ref{Omega.Potential}), one arrives at (see Appendix~A)
\begin{eqnarray}
&&\Omega_{pot}(v, \mu, T, \Phi, \Phi^{\ast})  =
\nonumber                        \\
&& \qquad
v \left\{\frac{|\Phi|^{2}}{V} \right.  -
2 T  \int \frac{d {\mbox{\bf k}}}{(2 \pi)^{d}}
\left[
\ln \cosh \frac{\sqrt{\xi^{2}(\mbox{\bf k}) + |\Phi|^{2}}}{2T}
- \xi(\mbox{\bf k}) \right] +
\nonumber                        \\
&& \qquad \qquad
\left. 2 T  \int \frac{d {\mbox{\bf k}}}{(2 \pi)^{d}}
\left[
\ln \cosh \frac{\varepsilon(\mbox{\bf k})}{2T}
- \varepsilon(\mbox{\bf k}) \right] \right\}.
\label{Omega.Potential.expr}
\end{eqnarray}
where $d=3$.

	    The stationary condition
\begin{equation}
\left.
\frac{\partial \Omega_{pot}(v, \mu, T_{c}^{MF}, \Phi, \Phi^{\ast})}
{\partial \Phi} \right|_{\Phi = \Phi^{\ast} = 0} = 0
				      \label{gap}
\end{equation}
results in the standard gap equation
\begin{equation}
\frac{1}{V} = \int \frac{d {\mbox{\bf k}}}{(2 \pi)^{3}}
\frac{1}{2 \xi(\mbox{\bf k})}
\tanh \frac{\xi(\mbox{\bf k})}{2T_{c}^{MF}}.
				      \label{gap.3D}
\end{equation}
Here we use the notation $T_{c}^{MF}$ rather than $T_{c}^{3D}$
because $\Omega_{pot}$ describes the system in the mean field
approximation only. In such a case $T_{c}^{MF}$ does not
depend significantly on the dimensionality of the space.
Besides, as it will be seen, the value of $T_{c}^{MF}$ may
significantly exceed $T_{c}^{3D}$ (or the temperature of real
condensation) in the strong coupling regime.

	  Before proceeding further, we need to describe how we
regulate the ultraviolet divergence in the gap equation
(\ref{gap.3D}). The idea is to replace the bare $V$ by the low energy
limit of the two body $T$-matrix (in the absence of a medium). In
the 3D case we use the formula \cite{Randeria.book}
\begin{equation}
\frac{m}{4 \pi a_{s}} =
-\frac{1}{V} + \int \frac{d {\mbox{\bf k}}}{(2 \pi)^{3}}
\frac{1}{2 \varepsilon(\mbox{\bf k})}
				      \label{scattering}
\end{equation}
which defines the above-mentioned $s$-wave scattering length $a_{s}$.
As a function of the bare interaction, $a_{s}^{-1}$ increases
monotonically from $- \infty$ for a very weak attraction to
$+ \infty$  for strong attractive interaction. Above the two-body bound state
threshold in vacuum ($a_{s}^{-1} = 0$), $a_{s}$
in fact is the "size" of this bound state with energy $E_{b} = - 1/m
a_{s}^{2}$. The dimensionless coupling constant in the dilute gas
model is then $1/k_{F} a_{s}$, which ranges from $- \infty$ in the
weak coupling (BCS) limit to $+ \infty$ in the strong coupling (Bose)
one.

     Using (\ref{scattering}) and (\ref{gap.3D}) one obtains directly the
equation for the introduced transition temperature $T_{c}^{MF}$ in terms
of the renormalized coupling $a_{s}$:
\begin{equation}
- \frac{m}{4 \pi a_{s}} =
\int \frac{d {\mbox{\bf k}}}{(2 \pi)^{3}}
\left[ \frac{1}{2 \xi(\mbox{\bf k})}
\tanh \frac{\xi(\mbox{\bf k})}{2T_{c}^{MF}} -
\frac{1}{2 \varepsilon(\mbox{\bf k})} \right].
                     \label{gap.scattering}
\end{equation}
There are two unknown quantities in this equation $T_{c}^{MF}$ and
$\mu$, and thus we need another equation,
\begin{equation}
\left.
- \frac{1}{v}
\frac{\partial \Omega_{pot}(v, \mu, T_{c}^{MF}, \Phi, \Phi^{\ast})}
{\partial \mu} \right|_{\Phi = \Phi^{\ast} = 0} = n_{f},
				      \label{number.mean}
\end{equation}
which, as was already said, fixes the chemical potential for a
given density.

                Note that unlike the BCS analysis in which one
usually assumes that $\mu = \epsilon_{F}$ (the non-interacting Fermi
energy), in the crossover problem $\mu$ will turn out
to be a strongly dependent function of the coupling as one goes into
the Bose regime where {\em all} the particles are affected by the
attractive interaction.  It directly follows from
(\ref{number.mean}) and (\ref{Omega.Potential.expr}) that the number
equation is given by
\begin{equation} n_{F}(\mu, T_{c}^{MF}) = \int
\frac{d {\mbox{\bf k}}}{(2 \pi)^{3}} \left[1 - \tanh
\frac{\xi(\mbox{\bf k})}{2T_{c}^{MF}} \right] = n_{f}.
\label{number.mean.expr}
\end{equation}

	   In the weak coupling limit,
$1/ k_{F} a_{s} \to - \infty$, we find from the system
(\ref{gap.scattering}) and (\ref{number.mean.expr})  the well-known
BCS result
\begin{equation}
\mu = \epsilon_{F}, \qquad
T_{c}^{MF} = \frac{8 \gamma}{\pi e^{2}} \epsilon_{F}
\exp \left(-\frac{\pi}{2k_{F} |a_{s}|} \right),
		    \label{temperature.mean}
\end{equation}
where $\gamma \simeq 1.781$.

	    The equations can also be solved analytically in the
strong coupling limit, where, however, one observes that the roles of the
gap and number equations are reversed: the gap equation
(\ref{gap.scattering}) determines $\mu$, while the number equation
determines $T_{c}^{MF}$.  In this limit $1/ k_{F} a_{s} \to + \infty$
and one finds tightly bound (separate) pairs with the energy $|E_{b}|
= 1/m a_{s}^{2}$ (and $|E_{b}| \gg \epsilon_{F}$).  The
non-degenerate Fermi system has here $\mu \approx - |E_{b}|/2$ and
its $T_{c}^{MF} \simeq |E_{b}|/2 \ln(|E_{b}|/\epsilon_{F})^{3/2}$.
But really such a system can be hardly recognized as a Fermi system
because all the fermions are bound in the pairs, so
(at least at rather small temperatures) they in fact form Bose
(local pair) system.

	    This unbounded growth  of the "transition temperature"
is an artifact of the approximation and there is, in fact, no sharp
phase transition at $T_{c}^{MF}$ (outside of weak coupling where
$T_{c}^{MF} = T_{c}^{3D}$ ignoring the small effects of thermal
fluctuations). The point is that the mean field approximation becomes
progressively worse with increasing coupling: the effective potential
(\ref{Omega.Potential.expr}) at $\Phi = \Phi^{\ast} = 0$ can only
describe a normal state consisting of essentially non-interacting
fermions.

            While this is adequate for weak coupling, in the strong
coupling limit unbound fermions exist in the normal state only at
very high temperatures. In this limit, where the system is completely
non-degenerate, a simple "chemical equilibrium" analysis
(boson $\rightleftharpoons$ 2 fermions) yields a dissociation (pairing)
temperature
$T_{dissoc} = |E_{b}|/ \ln(|E_{b}|/\epsilon_{F})^{3/2} (\approx T_{P})$.
We thus get convinced that for strong coupling $T_{c}^{MF}$ is related
to the pair dissociation scale rather than the
$T_{c}^{3D}(\ll T_{c}^{MF})$ at which the coherence is established.
As it will be seen below the temperature $T_{P}$ has an evident
physical meaning in the 2D case: in the region
$T_{BKT} < T < T_{P}$ (further $T_{\rho}$) the system does not
have a condensate (superconducting state) but its one-particle
spectrum acquires some features (in particular, the pseudogap) of
superconductivity.

\subsection{Beyond the mean field approximation}
As we have seen the mean field thermodynamical potential
(\ref{Omega.Potential}) (or its direct form
(\ref{Omega.Potential.expr})) cannot be used to access
the strong coupling limit. Therefore, to achieve this limit we
have to include the effects of the order parameter fluctuations.
It was already shown in Section~2.2 that treating the
fluctuations at the Gaussian level is done by the partition function
(\ref{partition.work}). It allows to incorporate the fluctuation into
both the gap (\ref{gap}) and number (\ref{number.mean}) equations
by adding the fluctuation correction to
$\Omega_{pot}(v, \mu, T, \Phi,\Phi^{\ast})$. As discussed in
\cite{Nozieres,Randeria.book} in the 3D case it is sufficient,
however, to take into account the fluctuation effects through the
number equation only. That is why we do not need the correction
$\partial (\Omega - \Omega_{pot})/ \partial \Phi |_{\Phi =
\Phi^{\ast}=0}=0$ to the gap equation.

                 Consequently, we can restrict
ourselves to calculation of the partition function at the critical
temperature given by (\ref{gap.3D}) that follows from
(\ref{partition.final}) only.  This permits us to get the fluctuation
correction to the number equation (\ref{number.mean}). Thus, one
finally obtains from (\ref{partition.final}) that
\begin{eqnarray}
\Omega (v, \mu, T, \Phi = \Phi^{\ast}=0)  =
\! \! \! \! &&
\Omega_{pot}(v, \mu, T, 0,0) +
\nonumber                      \\
&& \frac{v T}{(2 \pi)^{3}}
\sum_{n = - \infty}^{\infty} \int d \mbox{\bf K}
\ln \Gamma^{-1}(i \Omega_{n}, \mbox{\bf K}),
                          \label{thermopotential.final}
\end{eqnarray}
where
\begin{eqnarray}
\Gamma^{-1}(i \Omega_{n}, \mbox{\bf K}) = \frac{1}{V} & - &
\frac{1}{2} \int \frac{d \mbox{\bf k}}{(2 \pi)^{3}}
\frac{1}{\xi(\mbox{\bf k} + \mbox{\bf K}/2) + \xi(\mbox{\bf k} -
\mbox{\bf K}/2) - i \Omega_{n}} \times
\nonumber                          \\
&&\left[\tanh \frac{\xi(\mbox{\bf k} + \mbox{\bf K}/2)}{2T} +
\tanh \frac{\xi(\mbox{\bf k} - \mbox{\bf K}/2)}{2T}\right].
				    \label{Gamma.momentum}
\end{eqnarray}
Further, one has to remove the ultraviolet divergences from
(\ref{Gamma.momentum}) by applying the same regularization procedure
using the scattering length $a_{s}$ (see definition
(\ref{scattering})) as in case of the gap equation (\ref{gap.3D}).
With this substitution (\ref{Gamma.momentum}) transforms to the form:
\begin{eqnarray}
&&\Gamma^{-1}(i \Omega_{n}, \mbox{\bf K})  =
\frac{1}{2} \int \frac{d \mbox{\bf k}}{(2 \pi)^{3}}
\left[ \frac{1}{\varepsilon(\mbox{\bf k})} -
\frac{1}{\xi(\mbox{\bf k} + \mbox{\bf K}/2) + \xi(\mbox{\bf k} -
\mbox{\bf K}/2) - i \Omega_{n}} \times
\right.
\nonumber                      \\
&& \left. \qquad \qquad \qquad
\left(\tanh \frac{\xi(\mbox{\bf k} + \mbox{\bf K}/2)}{2T} +
\tanh \frac{\xi(\mbox{\bf k} - \mbox{\bf K}/2)}{2T}\right)
\right]
- \frac{m}{4 \pi a_{s}}.
				     \label{Gamma.renormalized}
\end{eqnarray}
According to \cite{Nozieres}, it is convenient to rewrite
$\Omega(v, \mu, T, \Phi = \Phi^{\ast}=0)$ in terms of a phase
shift defined by
$\Gamma(\omega \pm i0, \mbox{\bf K}) =|\Gamma(\omega, \mbox{\bf K})|
\exp (\pm i \delta(\omega, \mbox{\bf K}))$.
Then the new number equation is given by the equality
("conservation law")
\begin{equation}
n_{F} (\mu, T_{c}^{3D}) +  2 n_{B} (\mu, T_{c}^{3D}) = n_{f},
                        \label{number.fluctuation}
\end{equation}
with
\begin{equation}
n_{B} (\mu, T_{c}^{3D}) \equiv \int \frac{d \mbox{\bf K}}{(2 \pi)^{3}}
\int _{- \infty}^{\infty} \frac{d \omega}{2 \pi} n_{B} (\omega)
\frac{\partial \delta (\omega , \mbox{\bf K})}{\partial \mu},
                        \label{boson.definition}
\end{equation}
where $n_{B}(\omega) \equiv [\exp(\omega/T) - 1]^{-1}$ is the Bose
distribution function.

                     One sees from (\ref{number.fluctuation})
that the system of fermions is separated into two coexisting and
dynamically bounded subsystems: Fermi particles, or unbound fermions
with density $n_{F}(\mu, T_{c}^{3D})$, and local pairs, or bosons
with density $n_{B}(\mu, T_{c}^{3D})$. Such an interpretation is
possible and natural on the critical line only, when the finite
density of condensate $|\Phi|$ is not included nonperturbatively in
$n_{F}(\mu, T)$. Besides, as we shall demonstrate in Chapter 5, if the
order parameter has a nonzero value, one can study the number
equation without taking into account the bosonic contribution.

	    The temperature $T_{c}^{3D}$ at which homogeneous
long-range order is established is defined by the solution of
(\ref{gap.scattering}) together with (\ref{number.fluctuation}).
At weak coupling (large $n_{f}$), the results are essentially
unaffected by the inclusion of Gaussian fluctuations in the number
equation and $T_{c}^{3D}$ is the same as the mean field $T_{c}^{MF}$
obtained above (see (\ref{temperature.mean})).

       In fact the equations can be solved in the Bose limit too.
>From (\ref{gap.scattering}) we find $\mu(T_{c}^{3D}) = -|E_{b}|/2$,
which is one-half the energy required to break a pair. Further,
(\ref{boson.definition}) can be simplified because the inequality
$|\mu|/T \gg 1$ is satisfied. From this inequality it follows that the
isolated pole of $\Gamma (\omega, \mbox{\bf K})$ on the real axis is
situated far from the branch cut. The pole and the cut represent a
two-body bound state with the center-of-mass momentum $\mbox{\bf K}$
and the continuum of two-particle fermionic excitations,
respectively. The low energy physics for $T \ll T_{c}^{MF}$ is thus
dominated by this pole, and one can approximate the phase at the pole
by $\delta (\omega , \mbox{\bf K}) \simeq \pi \theta(\omega -
\mbox{\bf K}^{2}/4m + 2\mu + |E_{b}|)$, so that \begin{equation}
n_{B} (\mu, T_{c}^{3D}) = \frac{1}{4 \pi^{3}} \int d \mbox{\bf K}
n_{B} \left(\frac{\mbox{\bf K}^{2}}{4m} - 2\mu - |E_{b}| \right).
			 \label{number.boson}
\end{equation}
Finally, one gets
\begin{equation}
T_{c}^{3D} = \left(\frac{n_{f}}{2 \zeta(3/2)} \right)^{2/3}
\frac{\pi}{m} = 0.218 \epsilon_{F},
			 \label{temperature.bose.3D}
\end{equation}
which is simply the BEC result for bosons of mass $2m$ and
density $n_{f}/2$.

	    Summarizing the underlying physics in these
limiting cases it is necessary to stress the following \cite{Nozieres}:

\noindent
{\bf i)} The value of $T_{c}^{3D}$ in the weak coupling limit
results from thermal excitations of {\em individual particles}.

\noindent
{\bf ii)} In strong coupling case, $T_{c}^{3D}$ results from thermal
excitations of {\em collective modes}. These excitations are outside
the range of mean field theory. That is one of the reasons why we
took into account the fluctuations.

\noindent
The physics is thus quite different in the two limits considered:
the pair breaking in the first case and the motion of bound pairs in
the other.

     The results of the numerical solution of the equations are
presented in \cite{Randeria.3D} (see also \cite{Randeria.book}).
It may be worth commenting that the numerical result for
$T_{c}^{3D}$ is a non-monotonic function of $1/k_{F} a_{s}$
with a maximum value at intermediate coupling which is slightly larger
than the BEC value. The last is independent of the coupling constant.
The situation is completely  different for
the discrete (lattice) model of the 3D crossover \cite{Nozieres}
where $T_{c}^{3D}$ decreases as the coupling increases. So, certainly
there is an optimal value of the coupling for $T_{c}^{3D}$.
The latter will also take place for the quasi-2D model
which is in fact discrete in the third direction.  We shall discuss
such a model in Chapter~4.

\section{2D crossover: $T = 0$}
\setcounter{equation}{0}
      As we discussed in Introduction the 2D case has attracted much
attention partly because its possible relevance to the layered HTSC.

      To avoid the above-mentioned problem of how to determine
the $T_{c}$ value correctly in the strictly 2D system it is
worthwhile to start with the zero temperature case. The problem thus
becomes, due to the integration over frequency (which replaces the
summation over Matsubara frequencies), effectively a 3D one.
Therefore, the 2D theorems \cite{Coleman} are not
applicable to this case and one can speak about long-range order
in the 2D systems at $T = 0$.

       We shall follow here the paper \cite{GGL}
(see also its extended version in \cite{GGL.Preprint,Gorbar.PhD})
where the most general functional methods were used throughout.
However, it is more convenient to go to the limit $T = 0$
in the expressions from the previous Chapter rather than use the
zero temperature technique as it was done in \cite{GGL}.

\subsection{Model}
	The model Hamiltonian that we shall consider coincides
with (\ref{Hamiltonian.3D}) with one very important exception:
the dimensionality of the space here $d = 2$. This fact is crucial
in the choice of the parameter which one has to change to
trace the crossover. The fact is that 3D bound states in vacuum are
known to form only if the corresponding coupling constant $V$
exceeds some threshold. Thus, for the real cases one cannot achieve
the Bose regime even at very low carrier densities if the
attraction is not strong enough. This is the main reason why we
have studied the 3D crossover as the function of the "renormalized"
coupling $a_{s}^{-1}$ which corresponds to the "size" (radius) of the
bound state at $a_{s}^{-1} \geq 0$ only.

                 A thresholdless bound state formation in the 2D space
\cite{Landau} leads us to the important conclusion that one can reach
the Bose regime by decreasing the density $n_{f}$ of bare fermions at
{\em any} coupling. So, to study the 2D crossover it is convenient and
quite natural to regulate the density of carriers, or the Fermi energy
$\epsilon_{F}$, which is the same for 2D metals with the simplest
quadratic dispersion law:  $\epsilon_{F} = \pi n_{f}/m$. In doing so,
the coupling $V$ should be replaced by its renormalized value
$\varepsilon_{b}$ which is the energy of the bound state in vacuum.
We stress here once again that this parameter can be defined in the
2D case at any bare coupling $V$. The dimensionless coupling constant
in the 2D case -- the physical analog (see above) of $1/k_{F} a_{s}$
in 3D -- is thus given by the ratio $\epsilon_{F} / |\varepsilon_{b}|$,
which changes from $0$ (in the Bose limit $n_{f} \to 0$) to $\infty$
(in the BCS limit when $n_{f} \to \infty$).

\subsection{The Effective Action and Potential}
As we have already noted in the general case, (\ref{Effective.Action})
is impossible for $\Phi$ dependent on $x$. However, if one assumes
the gradients of $\Phi$ and $\Phi^{\ast}$ to be small the action
(\ref{Effective.Action}) can be naturally divided into kinetic and
potential parts
\begin{equation}
\Omega(v,\mu, \Phi(x),\Phi^{\ast}(x)) =
\Omega_{kin}(v, \mu, \Phi(x),\Phi^{\ast}(x)) +
\Omega_{pot}(v, \mu, \Phi,\Phi^{\ast}),
                    \label{kinetic+potential}
\end{equation}
where the effective potential has been defined by
(\ref{Omega.Potential}) and calculated in Appendix~A (its final
expression is given by (\ref{Omega.Potential.expr})). The terms
$\Omega_{kin}(\Phi(x),\Phi^{\ast}(x))$ with derivatives in
expansion (\ref{kinetic+potential}) contains the important physical
information, therefore we shall consider it in Section~3.4.

	  Let us return to the effective potential.
Going  to the limit $T \to 0$ one obtains from
(\ref{Omega.Potential.expr})
\begin{equation}
\Omega_{pot}(v, \mu, \Phi,\Phi^{\ast}) = v
\left[ \frac{|\Phi|^{2}}{V} - \int \frac{d \mbox{\bf k}}{(2 \pi)^{2}}
\left(\sqrt{\xi^{2}(\mbox{\bf k}) + |\Phi|^{2}} -
\xi (\mbox{\bf k}) \right) \right],
                        \label{Omega.Potential.T=0}
\end{equation}
where the terms which do not depend on $\Phi,\Phi^{\ast}$ and $\mu$
are omitted.

      It is interesting that, in virtue of the invariance of
      the partition function (\ref{partition.Hubbard})
with respect to the phase transformation of the group $U(1)$
\begin{eqnarray}
&& \Psi(x) \to e^{i \alpha \tau_{3}} \Psi(x), \qquad
\Psi^{\dagger}(x) \to \Psi^{\dagger}(x) e^{-i \alpha \tau_{3}};
\nonumber                           \\
&& \Phi(x) \to e^{-2 i \alpha} \Phi(x), \qquad
\Phi^{\ast}(x) \to  \Phi^{\ast}(x) e^{2 i \alpha},
		     \label{transformation}
\end{eqnarray}
with real $\alpha$ the potential $\Omega_{pot}(v, \mu, \Phi,\Phi^{\ast})$
(\ref{Omega.Potential}) (see also (\ref{Omega.Potential.expr}) and
(\ref{Omega.Potential.T=0})) can
only be dependent on the invariant product $\Phi^{\ast} \Phi$
\footnote{There is another transformation
(when the sign of the phase $\alpha$ is defined by the fermion spin
rather than the charge) under which the Hamiltonian (\ref{Hamiltonian.3D})
(or (\ref{Hamiltonian.Nambu})) is also invariant.
Such a transformation proves to be important for
fermion-fermion repulsion (i.e. $V<0$), or for the fermion-antifermion
(electron-hole) channel of pairing.  Apart from this difference
the physics for the case of a repulsive interaction is
identical to that under consideration. The complete set of gauge
transformations for the Hamiltonian under consideration were originally
given by Nambu \cite{Nam}.}.

        The analytic solution of the problem for the 2D case that we
consider here is easier than the 3D one.
Indeed, after performing the integration over $\mbox{\bf k}$ in
(\ref{Omega.Potential.T=0}) one directly obtains that
\begin{eqnarray}
&& \! \! \! \!
\Omega_{pot}(v, \mu, \Phi,\Phi^{\ast}) =  v
|\Phi|^{2} \left\{ \frac{1}{V} - \frac{m}{4 \pi} \left[ \ln \frac{W -
\mu + \sqrt{(W - \mu)^{2} + |\Phi|^{2}}} {\sqrt{\mu^{2} + |\Phi|^{2}}
- \mu} \right. \right.
\nonumber                 \\
&& + \left.
\left.  \frac{W - \mu}{W - \mu + \sqrt{(W - \mu)^{2} + |\Phi|^{2}}} +
\frac{\mu}{\sqrt{\mu^{2} + |\Phi|^{2}} - \mu}, \right] \right\},
\label{Omega.Potential.2D.expr} \end{eqnarray} where the value $W =
\mbox{\bf k}_{B}^{2}/2m$ is the conduction bandwidth and $\mbox{\bf
k}_{B}$ is the Brillouin boundary momentum.

\subsection{Main equations and analysis of solution}
     If the quantity $\Delta$
\footnote{This is the parameter that is responsible for the
appearance of a new (ordered, or with lowered symmetry) phase (see
Section~2.2.).}
is defined as the average value of $|\Phi|$,
then the equation for the extremum
\begin{equation}
\left. \frac{\partial \Omega_{pot}(v, \mu, \Phi, \Phi^{\ast})}
{\partial \Phi} \right|_{\Phi = \Phi^{\ast} = \Delta} = 0,
\end{equation}
yields, according to (\ref{Omega.Potential.2D.expr})
\begin{equation}
\Delta \left[ \frac{1}{V} - \frac{m}{4 \pi}
\ln \frac{W - \mu + \sqrt{(W - \mu)^{2} + \Delta^{2}}}
{\sqrt{\mu^{2} + \Delta^{2}} - \mu} \right] = 0,
		     \label{gap.2D.band}
\end{equation}
while the condition
\begin{equation}
-\frac{1}{v} \left. \frac{\partial
\Omega_{pot}(v, \mu, \Phi, \Phi^{\ast})}{\partial \mu}
\right|_{\Phi = \Phi^{\ast} = \Delta} = n_{f},
\end{equation}
which sets the density of the particles in the system,
takes the form
\begin{equation}
W - \sqrt{(W - \mu)^{2} + \Delta^{2}} + \sqrt{\mu^{2} + \Delta^{2}}
= 2 \epsilon_{F},        \label{number.2D.band}
\end{equation}
where we made use of the relation between $\epsilon_{F}$ and
$n_{f}$ in the 2D case.

	      Equations (\ref{gap.2D.band}) and
(\ref{number.2D.band}), which were obtained in the mean field
approximation which is quite sufficient at $T=0$, make up a set for
finding the quantities $\Delta$ and $\mu$ as functions of $W$ and
$\epsilon_{F}$ (or $n_{f}$). It differs from the similar set of
\cite{Randeria.2D} by the explicit dependence on $W$ that, in
principle, can be important for the case of narrow or multi-band
systems \cite{GLSh.FNT.1995}.

	       It should also be noted (see also the discussion
after the equation (\ref{number.mean})) that the necessity to utilize
the system of equations in order to find self-consistently $\Delta$
and $\mu$ has been known for a long time (see \cite{Schrieffer}).
However, the fermion density in real 3D metals is virtually
unchanged in practice so that, as a rule, the equation for $\mu$ is
trivialized to the equation $\mu = \epsilon_{F}$ and only the
value $\Delta$ is regarded as unknown. The importance of
the second equation for small particle densities was earlier pointed
out in the papers \cite{Eagles.1969,Leggett}.

		Equations (\ref{gap.2D.band}) and
(\ref{number.2D.band}) allow, along with the trivial solution
($\Delta=0$, $\mu= \epsilon_{F}$) the nontrivial one:
\begin{eqnarray}
&& \Delta^{2} =
\frac{\epsilon_{F}(W - \epsilon_{F})}{\sinh^{2}(2 \pi/ m V)};
\nonumber                         \\
&& \mu = \epsilon_{F} \coth \frac{2 \pi}{m V} - \frac{W}{2}
\left( \coth \frac{2 \pi}{m V} - 1 \right), \label{solution.2D.band}
\end{eqnarray}
which is valid for any physically reasonable values of the relevant
parameters. It is also very interesting that for small
$\epsilon_{F}$, there is a region where $\mu < 0$, and that the sign
changeover occurs at a definite point ${\bar \epsilon_{F}} = W/2 [1 -
\tanh(2 \pi/m V)]$.

        The expressions that are found in \cite{Miyake,Randeria.2D}
follow directly from (\ref{solution.2D.band}) if, treating $W$ as
large and the attraction $V$ as small, we introduce the 2D two-body
binding energy
\begin{equation}
\varepsilon_{b} = -2W \exp \left(-\frac{4 \pi}{m V} \right),
                \label{bound.energy}
\end{equation}
which does not include any many-particle effects.
The introduction of the expression (\ref{bound.energy}) enables one to
take the limit $W \gg \epsilon_{F}$ and thus justifies to a certain
degree the use of the parabolic dispersion law. Second the fitting
parameter $\varepsilon_{b}$ is more physically relevant. For example
it is well-defined even for potentials with repulsion. We stress that the
introduction of $\varepsilon_{b}$ instead of $V$ also allows us to
regulate the ultraviolet divergence, which is in fact present in
the gap equation (\ref{gap.2D.band}). So, this step is quite similar
to  the step from equation (\ref{gap.3D}) to (\ref{gap.scattering}).

	  It should be mentioned that, in a dilute gas model,
the existence of
a two-body bound state in vacuum is a {\em necessary} (and
sufficient) condition for a Cooper instability
\cite{Miyake,Randeria.2D,Randeria.book}. This statement becomes
nontrivial if one considers two-body potentials $V(r)$ with
short-range repulsion (e.g., hard-core plus attraction), so that one
has to cross a finite (but really very weak \cite{Landau}) threshold
in the attraction before a bound state forms in vacuum.

	   Thus, by making use of (\ref{bound.energy}) it is easy to
simplify (\ref{solution.2D.band})
\cite{Miyake,Randeria.2D,Randeria.book}:
\begin{equation}
\Delta = \sqrt{2 |\varepsilon_{b}| \epsilon_{F}}; \qquad
\mu = - \frac{|\varepsilon_{b}|}{2} + \epsilon_{F},
			   \label{solution.2D}
\end{equation}
with ${\bar \epsilon_{F}} = |\varepsilon_{b}|/2$.

	    To understand the physical significance of these
remarkably simple results we look at the two limits of this
solution. For very weak attraction (or high density) the two-particle
binding energy is extremely small, i.e. $|\varepsilon_{b}| \ll
\epsilon_{F}$, and it is seen that we recover the well-known BCS
results with strongly overlapping in $\mbox{\bf r}$-space Cooper pairs.
The chemical potential $\mu \simeq \epsilon_{F}$, and the gap function
$\Delta \ll \epsilon_{F}$.

         In the opposite limit of very strong attraction
(or a very low particle density) we have a deep two-body bound state
$|\varepsilon_{b}| \gg \epsilon_{F}$, and find that we are in a
regime in which there is BEC of composite bosons, or "diatomic
molecules". The chemical potential here
$\mu \simeq -|\varepsilon_{b}|/2$, which is one half the energy of
pair dissociation for tightly bound (local) pairs.

	      It should also be kept in mind that in the local pair
regime ($\mu < 0$) the gap $E_{gap}$ in the quasiparticle
excitation spectrum equals not $\Delta$ (as in the case $\mu > 0$)
but rather $\sqrt{\mu^{2} + \Delta^{2}}$ (see \cite{Leggett} and the
review \cite{Randeria.book}).

	   Leaving aside the analysis of $\Omega_{pot}$ for arbitrary
values of the parameters (see \cite{GGL.Preprint}), let us consider
the most interesting case $W \to \infty$, $V \to 0$ with finite
$\varepsilon_{b}$. Then finding from (\ref{bound.energy}) the
expression for $4 \pi/m V$ and substituting it into
(\ref{Omega.Potential.2D.expr}) we obtain
\begin{equation}
\Omega_{pot}(v, \mu, \Phi,\Phi^{\ast}) = v
\frac{m}{4\pi} |\Phi|^{2} \left( \ln \frac{\sqrt{\mu^{2} +
|\Phi|^{2}} - \mu}{|\varepsilon_{b}|} - \frac{\mu}{\sqrt{\mu^{2} +
|\Phi|^{2}} - \mu} - \frac{1}{2} \right),
\label{Omega.Potential.2D.eps}
\end{equation}
whence we arrive at the expressions near point $\Phi, \Phi^{\ast} \to
0$:  \begin{eqnarray} && \! \! \! \! \!  \Omega_{pot}(v, \mu,
\Phi,\Phi^{\ast})|_{\Phi, \Phi^{\ast} \to 0} \approx \nonumber
\\ && \left\{ \begin{array}{rl} v \frac{\ds m}{\ds 4\pi} |\Phi|^{2}
\left( \ln \frac{\ds 2 |\mu|}{\ds |\varepsilon_{b}|} + \frac{\ds
|\Phi|^{2}}{\ds 8 \mu^{2}} \right), & \mu < 0 \\ v \frac{\ds m}{\ds
4\pi} \left[ |\Phi|^{2} \left( \ln \frac{\ds |\Phi|^{2}}{\ds 2 \mu
|\varepsilon_{b}|} - \frac{\ds 1}{\ds 2} \right) - 2 \mu^{2} \right],
& \mu > 0.  \end{array} \right.  \label{Omega.Potential.2D.small}
\end{eqnarray} Equations (\ref{gap.2D.band})
and (\ref{number.2D.band}) can be written as
\begin{equation}
\sqrt{\mu^{2} + \Delta^{2}} - \mu = |\varepsilon_{b}|; \qquad
\sqrt{\mu^{2} + \Delta^{2}} + \mu = 2\epsilon_{F},
\label{gap+number.2D}
\end{equation}
respectively; their solution, (\ref{solution.2D}), is quoted above.
As it is evident from
(\ref{Omega.Potential.2D.small}), the potential term $\Omega_{pot}$ in the
region $\mu < 0$ corresponds to particles with repulsion, which accounts for
their BEC. We note here that for the 3D case this repulsion was obtained in
\cite{Haussmann} using a diagrammatic technique.  In the region of high
$n_{f}$, where $\mu > 0$, the $\Omega_{pot}$ cannot be represented as
a series even for small $\Phi$, which reflects the specific
properties of the effective potential in 2D systems.

	    If, proceeding from (\ref{Omega.Potential.2D.expr}) and
(\ref{solution.2D.band}), we find the thermodynamic potential
difference of the trivial and non-trivial solutions, we can easily
obtain the result that it takes form
\begin{equation}
\Omega_{pot}(v, \mu, \Delta) - \Omega_{pot}(v, \mu, 0) =
- v \frac{m}{2 \pi} \left[
\epsilon_{F}^{2} \coth \frac{2 \pi}{m V} - \mu^{2} \theta(\mu)
\right],                  \label{energy}
\end{equation}
which demonstrates (for various $\mu \leq \epsilon_{F} \leq W/2$)
that at $T = 0$ the nontrivial solution is always more favorable
than the trivial one, and the point $\Phi =
\Phi^{\ast} = 0$ is unstable since here
$\partial^{2} \Omega_{pot}/ \partial \Phi \partial \Phi^{\ast} < 0$
for all allowed values of the parameters.

         As follows from (\ref{Omega.Potential.2D.eps}),
both  $\Omega_{pot}(\mu, \Delta)$ and $\Omega_{pot}(\epsilon_{F}, 0)$
are equal to $-v m \epsilon_{F}^{2}/2 \pi$, i.e. the potentials of the
superconducting and normal phases turn out to be equal in the limit
$W \to \infty$, $V \to 0$. Nevertheless, it can be shown, just as in
\cite{Randeria.2D}, that the superconducting phase has
the lowest internal energy. In addition, the relevant
difference is proportional to $v \rho(\epsilon_{F})\Delta^{2}$
($\rho(\epsilon_{F}) = m/2\pi)$ being the density of states, which is
energy independent in the 2D case); in other words,
we arrive at the standard result of the BCS theory \cite{Schrieffer},
which is clearly valid for all values of the ratio
$\epsilon_{F}/|\varepsilon_{b}|$.

\subsection{The gradient terms of the effective action}
	  Now we calculate the terms $\Omega_{kin}$ which contain the
derivatives in expansion (\ref{kinetic+potential}). As before, we
shall assume the inhomogeneities of $\Phi$ and $\Phi^{\ast}$ to be
small having restricted ourselves only to the terms with lowest
derivatives. For simplicity we shall also consider the stationary
case and calculate the terms with the second-order spatial
derivatives only, which makes it possible to determine the coherence
length $\xi$ and the penetration depth $\lambda_{H}$ of the magnetic
field in a 2D superconductor.

	    With these restrictions, taking into account the
invariance of $\Omega(v, \mu, \Phi, \Phi^{\ast})$ (see equation
(\ref{Effective.Action})) with respect to the phase transformations
(\ref{transformation}) one can
write a general form for the kinetic part of the action:
\begin{eqnarray}
&& \Omega_{kin}(v, \mu, T, \Phi(x),\Phi^{\ast}(x)) =
T \int_{0}^{\beta} d \tau \int d \mbox{\bf r}
T_{kin}(\Phi, \Phi^{\ast}, \nabla \Phi, \nabla \Phi^{\ast}) =
\nonumber                       \\
&& \qquad
T \int_{0}^{\beta} d \tau \int d \mbox{\bf r}
\left[ T_{1}(|\Phi|^{2}) |\nabla \Phi|^{2} +
\frac{1}{2} T_{2}(|\Phi|^{2}) (\nabla |\Phi|^{2})^{2} \right],
			\label{Omega.Kinetic}
\end{eqnarray}
where there are no items with a total derivative since the
boundary effects are regarded as unessential, and the
coefficients $T_{1,2}(|\Phi|^{2})$ are assumed to be unknown
quantity. It follows from (\ref{Omega.Kinetic}) that in the
second approximation in derivatives the variations in both
the direction (phase) of the field $\Phi$ and its absolute
value are taken into account.

	 To calculate the coefficients $T_{1,2}(|\Phi|^{2})$ we
shall follow the paper \cite{Gusynin.1992} according to which
one can determine the variation derivatives
\begin{eqnarray}
&& \left. \frac{\delta^{2}\Omega_{kin}(\Phi,\Phi^{\ast})}
{\delta \Phi^{\ast}(\mbox{\bf r}) \delta \Phi (0)}
\right|_{\Phi, \Phi^{\ast} = const} =
-T \int_{0}^{\beta} d \tau
[T_{1}(|\Phi|^{2}) + |\Phi|^{2} T_{2}(|\Phi|^{2})]
\nabla^{2} \delta(\mbox{\bf r});
\nonumber                      \\
&& \left. \frac{\delta^{2}\Omega_{kin}(\Phi,\Phi^{\ast})}
{\delta \Phi(\mbox{\bf r}) \delta \Phi (0)}
\right|_{\Phi, \Phi^{\ast} = const} =
-T \int_{0}^{\beta} d \tau
(\Phi^{\ast})^{2} T_{2}(|\Phi|^{2})
\nabla^{2} \delta(\mbox{\bf r}).
			\label{variational.derivatives}
\end{eqnarray}
Multiplying both equations (\ref{variational.derivatives})
by $\mbox{\bf r}^{2}$ and integrating over $d \mbox{\bf r}$
one can obtain
\begin{eqnarray}
&& \int d \mbox{\bf r} \mbox{\bf r}^{2}
\left. \frac{\delta^{2}\Omega_{kin}(\Phi,\Phi^{\ast})}
{\delta \Phi^{\ast}(\mbox{\bf r}) \delta \Phi (0)}
\right|_{\Phi, \Phi^{\ast} = const} =
-4T \int_{0}^{\beta} d \tau
[T_{1}(|\Phi|^{2}) + |\Phi|^{2} T_{2}(|\Phi|^{2})];
\nonumber                      \\
&&  \int d \mbox{\bf r} \mbox{\bf r}^{2}
\left. \frac{\delta^{2}\Omega_{kin}(\Phi,\Phi^{\ast})}
{\delta \Phi(\mbox{\bf r}) \delta \Phi (0)}
\right|_{\Phi, \Phi^{\ast} = const} =
-4 T \int_{0}^{\beta} d \tau
(\Phi^{\ast})^{2} T_{2}(|\Phi|^{2}),
			\label{coefficients.superposition}
\end{eqnarray}
which allows one to determine the coefficients required. Indeed,
let us define the correlation functions
\begin{equation}
K_{i j} (\mbox{\bf q}) \equiv
\int d \mbox{\bf r} \exp (-i \mbox{\bf q} \mbox{\bf r})
\left. \frac{\delta^{2}\Omega_{kin}(\Phi,\Phi^{\ast})}
{\delta \Phi_{i}(\mbox{\bf r}) \delta \Phi_{j} (0)}
\right|_{\Phi, \Phi^{\ast} = const},
\quad
\Phi_{1} \equiv \Phi^{\ast}, \Phi_{2} \equiv \Phi.
			 \label{correlation}
\end{equation}
It is then readily seen from (\ref{coefficients.superposition})
and (\ref{correlation}) that $T_{1,2}(|\Phi|^{2})$ are determined
from the derivatives of the correlators $K_{i j}(\mbox{\bf q})$
in components of the vector $\mbox{\bf q}$ at $\mbox{\bf q} = 0$:
\begin{eqnarray}
&& T_{1}(|\Phi|^{2}) = \frac{1}{4} \left. \left[
\frac{\partial^{2} K_{1 2}(\mbox{\bf q})}{\partial \mbox{\bf q}^{2}} -
\frac{\Phi}{\Phi^{\ast}}
\frac{\partial^{2} K_{2 2}(\mbox{\bf q})}{\partial \mbox{\bf q}^{2}}
\right] \right|_{\mbox{\bf q} = 0};
\nonumber                       \\
&& T_{2}(|\Phi|^{2}) = \frac{1}{4 (\Phi^{\ast})^{2}} \left.
\frac{\partial^{2} K_{2 2}(\mbox{\bf q})}{\partial \mbox{\bf q}^{2}}
\right|_{\mbox{\bf q} = 0},
			      \label{coefficients}
\end{eqnarray}
it being apparently sufficient to know $K_{1 2}(\mbox{\bf q})$ and
$K_{2 2}(\mbox{\bf q})$ only. Thus, the problem has been simply
reduced to calculating the correlators.

	 On the other hand, the effective action for the case of
the imaginary time-independent fields can be written
(see (\ref{Effective.Action})) in the following form:
\begin{eqnarray}
&& \! \! \! \! \! \! \! \! \! \! \! \! \! \! \! \! \! \! \! \! \! \! \! \!
\Omega(v, \mu, T, \Phi(\mbox{\bf r}), \Phi^{\ast}(\mbox{\bf r})) =
T \int_{0}^{\beta} d \tau \int d \mbox{\bf r} \left\{
\frac{1}{V} |\Phi(\mbox{\bf r})|^{2} - \right.
\nonumber                       \\
&& \! \! \! \! \! \! \! \! \! \! \! \! \! \! \! \! \! \! \! \! \!
\left.
T \sum_{n = - \infty}^{\infty} \mbox{tr}
\langle \mbox{\bf r} | \mbox{Ln}
\left[- i \omega_{n} {\hat I}  +
\tau_{3} \left(\frac{\nabla^{2}}{2 m} + \mu \right)
+ \tau_{-} \Phi(\mbox{\bf r})
+ \tau_{+} \Phi^{\ast}(\mbox{\bf r}) \right]
| \mbox{\bf r} \rangle \right\},
		    \label{Effective.Action.momentum}
\end{eqnarray}
where the operation tr refers to Pauli matrices, and
the normalization term with $G_{0}^{-1}$ is omitted.
Comparing (\ref{correlation}) and (\ref{Effective.Action.momentum})
and using the definition (\ref{Green.fermion}) after the
Fourier-transformation (\ref{Fourier.coordinate}) we obtain directly:
\begin{eqnarray}
&& K_{1 2}(\mbox{\bf q}) =
T \sum_{n = -\infty}^{+\infty}
\int \frac{d {\mbox{\bf k}}}{(2 \pi)^{2}} \mbox{tr}
[G(i \omega_{n}, \mbox{\bf k}) \tau_{+}
G(i \omega_{n}, \mbox{\bf k} + \mbox{\bf q}) \tau_{-}];
\nonumber                      \\
&& K_{2 2}(\mbox{\bf q}) =
T \sum_{n = -\infty}^{+\infty}
\int \frac{d {\mbox{\bf k}}}{(2 \pi)^{2}} \mbox{tr}
[G(i \omega_{n}, \mbox{\bf k}) \tau_{-}
G(i \omega_{n}, \mbox{\bf k} + \mbox{\bf q}) \tau_{-}],
			 \label{correlation.main}
\end{eqnarray}
where $G(i \omega_{n}, \mbox{\bf k})$ is the Green's function defined by
(\ref{Green.fermion.momentum}). Calculating the traces in
(\ref{correlation.main}) it is easy to arrive at the
final expression for the correlators:
\begin{eqnarray}
&& K_{1 2}(\mbox{\bf q}) =
T \sum_{n = -\infty}^{+\infty}
\int \frac{d {\mbox{\bf k}}}{(2 \pi)^{2}}
\frac{\left(i \omega_{n} + \frac{\ds \mbox{\bf k}^{2}}{\ds 2m}
- \mu \right)
\left(i \omega_{n} - \frac{\ds (\mbox{\bf k} + \mbox{\bf q})^{2}}{\ds 2m}
+ \mu \right)}
{[\omega_{n}^{2} + {\cal E}^{2}(\mbox{\bf k})]
[\omega_{n}^{2} + {\cal E}^{2}(\mbox{\bf k} + \mbox{\bf q})]};
\nonumber                      \\
&& K_{2 2}(\mbox{\bf q}) =
T \sum_{n = -\infty}^{+\infty}
\int \frac{d {\mbox{\bf k}}}{(2 \pi)^{2}}
\frac{(\Phi^{\ast})^{2}}
{[\omega_{n}^{2} + {\cal E}^{2}(\mbox{\bf k})]
[\omega_{n}^{2} + {\cal E}^{2}(\mbox{\bf k} + \mbox{\bf q})]},
			 \label{correlation.tr}
\end{eqnarray}
where we used the notation
${\cal E}^{2}(\mbox{\bf k}) \equiv \xi^{2}(\mbox{\bf k}) + |\Phi|^{2}$.

   Summing over the Matsubara frequencies and then going to the
limit $T = 0$ one can obtain the final formulae:
\begin{eqnarray}
&& K_{1 2}(\mbox{\bf q}) = -
\int \frac{d {\mbox{\bf k}}}{8 \pi^{2}}
\frac{1}{{\cal E}(\mbox{\bf k})+{\cal E}(\mbox{\bf k}+\mbox{\bf q})}
\left\{1 + \frac{
\left[\frac{\ds \mbox{\bf k}^{2}}{\ds 2m} - \mu \right]
\left[\frac{\ds(\mbox{\bf k}+\mbox{\bf q})^{2}}{\ds 2m}-\mu \right]}
{{\cal E}(\mbox{\bf k}){\cal E}(\mbox{\bf k}+\mbox{\bf q})} \right\};
\nonumber                      \\
&& K_{2 2}(\mbox{\bf q}) =
\int \frac{d {\mbox{\bf k}}}{8 \pi^{2}}
\frac{1}{{\cal E}(\mbox{\bf k})+{\cal E}(\mbox{\bf k}+\mbox{\bf q})}
\frac{(\Phi^{\ast})^{2}}
{{\cal E}(\mbox{\bf k}){\cal E}(\mbox{\bf k}+\mbox{\bf q})}.
			 \label{correlation.final}
\end{eqnarray}
A somewhat tedious but otherwise straightforward calculation
(see \cite{GGL.Preprint,Gorbar.PhD}) now yields for the desired
coefficients of (\ref{Omega.Kinetic}):
\begin{eqnarray}
T_{1}(|\Phi|^{2}) & = &
\frac{1}{16 \pi} \int_{-\mu}^{W-\mu} d u \left[
\frac{2 |\Phi|^{2} - u^{2}}{(u^{2} + |\Phi|^{2})^{5/2}} (u + \mu) +
\frac{u}{(u^{2} + |\Phi|^{2})^{3/2}} \right]
\nonumber                      \\
& = & \left. -\frac{1}{16 \pi}
\frac{|\Phi|^{4} - 2 u \mu |\Phi|^{2} - u^{3} \mu}
{|\Phi|^{2} (u^{2} + |\Phi|^{2})^{3/2}}
\right|_{u = - \mu}^{u = W - \mu};
			      \label{coefficient1.final}
\end{eqnarray}
\begin{eqnarray}
T_{2}(|\Phi|^{2}) & = &  \frac{1}{16 \pi}
\int_{-\mu}^{W-\mu} d u \left[
\frac{5 u^{2} (u + \mu)}{(u^{2} + |\Phi|^{2})^{7/2}} -
\frac{3 (2 u + \mu)}{2 (u^{2} + |\Phi|^{2})^{5/2}} \right] =
\nonumber                        \\
&& \left. \frac{1}{16 \pi} \frac{1}{6}
\frac{2 |\Phi|^{6} - 4 |\Phi|^{4} u ^{2} - 2 u^{5} \mu -
5 u^{3} \mu |\Phi|^{2} - 9 u \mu |\Phi|^{4}}
{|\Phi|^{4}(u^{2} + |\Phi|^{2})^{5/2}}
\right|_{u = - \mu}^{u = W - \mu}.
			      \label{coefficient2.final}
\end{eqnarray}
The expressions (\ref{coefficient1.final}) and (\ref{coefficient2.final})
together with (\ref{Omega.Potential.2D.expr}) complete the
calculation of all the terms of the effective action
(\ref{kinetic+potential}) for the 2D metal with attraction between
fermions.

\subsection{Correlation length and penetration depth {\em versus} doping}
        Knowing $T_{1}(|\Phi|^{2})$ and $T_{2}(|\Phi|^{2})$ one
can find the values for different observables. For practical
purposes it suffices to restrict ourselves to considering the
coefficients obtained at the point $|\Phi| = \Delta$ of the effective
potential minimum. Besides, instead of (\ref{coefficient2.final})
it is convenient to introduce the combination
${\tilde T}_{2}(|\Phi|^{2}) \equiv T_{1}(|\Phi|^{2}) +
2 \Delta^{2} T_{2}(|\Phi|^{2})$ which determines the change in the
$|\Phi|$ value only and arises as the coefficient at $(\nabla|\Phi|)^{2}$.

         The analysis of (\ref{coefficient1.final}) and
(\ref{coefficient2.final}) reveals that at $|\Phi| = \Delta$ both
these functions are positive for $\epsilon_{F} \leq W/2$
where they change their sign, which simply reflects the necessity to
go over the antiparticle (from electron to hole or vice versa) picture
for the region $\epsilon_{F} > W/2$. On the other hand, in
the region $\epsilon_{F} \leq W/2$ the positiveness of these functions
demonstrates the stability of the homogeneous ground state of the
model concerned.

	    One can succeed in simplifying (\ref{coefficient1.final})
and (\ref{coefficient2.final}) if takes into account that, as a rule,
$\epsilon_{F} \ll W$ (for example, in HTSC metal-oxides
$\epsilon_{F} \sim 0.1W$\cite{Uemura,Loktev.review}). Then, using
(\ref{bound.energy}) and (\ref{solution.2D}), one can rewrite
$T_{1}(\Delta^{2})$ and ${\tilde T}_{2}(\Delta^{2})$ in the very
simple form:
\begin{equation}
T_{1}(\Delta^{2}) = \frac{1}{16 \pi \hbar |\varepsilon_{b}|};
\qquad
{\tilde T}_{2}(\Delta^{2}) = \frac{1}{24 \pi \hbar^{2}}
\frac{(2 \epsilon_{F} - |\varepsilon_{b}|)^{2}}
{(2 \epsilon_{F} + |\varepsilon_{b}|)^{3}},
		   \label{coefficients.energy}
\end{equation}
where for completeness we have restored the Planck constant again.

	  The explicit form of $T_{1}(\Delta^{2})$ and also
$a \equiv v^{-1} \partial^{2} \Omega_{pot}(\Phi,\Phi^{\ast})/
\partial \Phi \partial \Phi^{\ast}|_{|\Phi|^{2} = \Delta^{2}}$
allows one to calculate the coherence length and the penetration depth.
Since (see (\ref{Omega.Potential.2D.expr}))
\begin{eqnarray}
a & = & \frac{m}{\pi \hbar^{2}}
\frac{\epsilon_{F} (W -\epsilon_{F})
\coth \frac{\ds 2 \pi \hbar^{2}}{\ds mV}}
{W^{2} \coth^{2} \frac{\ds 2 \pi \hbar^{2}}{\ds mV} -
(W -2 \epsilon_{F})^{2}}
\nonumber                                   \\
& \to &  \frac{m}{2 \pi \hbar^{2}}
\frac{\epsilon_{F}}{2 \epsilon_{F} + |\varepsilon_{b}|} \qquad
(W \to \infty, V \to 0),
					   \label{a}
\end{eqnarray}
then, according to general theory of fluctuation phenomena
\cite{Lifshitz}, one gets that
\begin{equation}
\xi = \hbar \left[\frac{T_{1}(\Delta^{2})}{a} \right]^{1/2} =
\hbar \left(\frac{2 \epsilon_{F} + |\varepsilon_{b}|}
{8 m \epsilon_{F} |\varepsilon_{b}|} \right)^{1/2}.
					    \label{coherence}
\end{equation}
This formula shows the dependence of $\xi$ on $\epsilon_{F}$
(or $n_{f}$). It is very interesting and useful to compare
(\ref{coherence}) with the definition of the pair size
\cite{Quick,Carter} (also incorrectly referred to as the
coherence length \cite{Randeria.2D,Pistolesi}), namely
\begin{equation}
\xi_{0}^{2} = \frac{ds \int d \mbox{\bf r} g(\mbox{\bf r}) \mbox{r}^{2}}
{\ds \int d \mbox{\bf r} g(\mbox{\bf r})}, \label{size.general}
\end{equation}
where
\begin{equation}
g(\mbox{\bf r}) = \frac{1}{n_{f}^{2}} \langle
\Psi_{BCS} | \psi^{\dagger}_{\uparrow} (\mbox{\bf r})
\psi^{\dagger}_{\downarrow} (0) | \Psi_{BCS} \rangle
\end{equation}
is the pair-correlation-function for opposite spins and
$| \Psi_{BCS} \rangle$ is the usual BCS trial function.
For the 2D model under consideration the general
expression (\ref{size.general}) gives \cite{Randeria.2D,Quick}
\begin{equation}
\xi_{0}^{2} = \frac{\hbar^{2}}{4 m} \frac{1}{\Delta}
\left[ \frac{\mu}{\Delta} +
\frac{\mu^{2} + 2 \Delta^{2}}{\mu^{2} + \Delta^{2}}
\left(\frac{\pi}{2} + \tan^{-1} \frac{\mu}{\Delta} \right)^{-1} \right],
				   \label{size}
\end{equation}
where $\Delta$ and $\mu$ were given by (\ref{solution.2D}).
So, we are ready now to compare (\ref{coherence}) and (\ref{size}).

	   Again to understand the underlying physics it is worth to
look at two extremes of (\ref{coherence}) and (\ref{size}).
For high carrier densities $\epsilon_{F} \gg |\varepsilon_{b}|$
one finds that $\xi \sim \xi_{0} \sim \bar v_{F}/\Delta$, i.e.
the well-known Pippard's result is reproduced correctly.
Moreover, if, according to \cite{Miyake}, we introduce the pair size
$\xi_{b} = \hbar (m |\varepsilon_{b}|)^{-1/2}$ in vacuum,
then it is clear that $\xi \sim \xi_{0} \ltwid \xi_{b}$ when
$|\varepsilon_{b}| \ltwid \epsilon_{F}$. Therefore, both $\xi$ and
$\xi_{0}$ prove to be close to the pair size which is much larger
than the interparticle spacing. The latter statement one can see from the
value of the dimensionless parameter $\xi k_{F} \sim \xi_{0} k_{F} \sim
\epsilon_{F}/\Delta \gg 1$.

	     In the opposite limit $|\varepsilon_{b}| \gg \epsilon_{F}$
of very low density one can see that $\xi_{b} \ll \xi \sim k_{F}^{-1}$
while $\xi_{0} \sim \xi_{b}$. Consequently the correct interpretation of
$\xi_{0}$ is the pair size (in presence of the Fermi sea
(\ref{size.general})) rather than the coherence length.  The former in
the extreme Bose regime is much smaller than the mean interparticle
spacing, since $\xi_{0} k_{F} \ll 1$, while always $\xi k_{F} \sim 1$.
The meaning of $\xi$ is indeed the coherence length because it remains
finite and comparable with the mean interparticle spacing even when
$|\varepsilon_{b}|$ goes to infinity. This situation is consistent with
the case of $^{4}$He where the coherence length is nonzero and comparable
with the mean inter-atomic distance although $|\varepsilon_{b}|$
(or energy of nucleon-nucleon binding) is really extremely large.
Since $\xi_{b} k_{F} \sim \xi_{0} k_F$ and since $\xi_{b} k_{F}$ is
directly related to the dimensionless ratio
$\epsilon_{F}/|\varepsilon_{b}|$ (which was discussed in Section~3.1)
it can be inferred that $\xi_0 k_{F}$ is a physical parameter which can
correctly determine the type of pairing.

        Calculations of the $\lambda_{H}$ value require one to
introduce the external magnetic field $\mbox{\bf H}$ and also the
usual extension of the derivatives
$\hbar \nabla \to \hbar \nabla - (2 i e/c) \mbox{\bf A}$
($e$ is the electron charge, $c$ is the light velocity,
$\mbox{\bf A}$ is the vector potential). Adding then the energy of the
magnetic field
\begin{equation}
\int d \tau \int d \mbox{\bf r}
\frac{\mbox{\bf H}^{2}}{8 \pi z} \label{magnetic}
\end{equation}
to the effective action (\ref{kinetic+potential}) by direct calculation
one gets
\begin{equation}
\lambda_{H} = \left[ \frac{c^{2}}
{32 \pi e^{2} T_{1} (\Delta^{2}) z \Delta^{2}} \right]^{1/2}.
			      \label{penetration}
\end{equation}
Note that in (\ref{magnetic}) and (\ref{penetration})
$z$ is the number of superconducting layers per unit length
taking into account the 3D character of the external field action.
Hence, Ginzburg-Landau parameter (see (\ref{coherence}),
(\ref{penetration})) is equal to
\begin{equation}
\kappa \equiv \frac{\lambda_{H}}{\xi} = \frac{1}{\hbar}
\left[ \frac{c^{2} a}{32 \pi e^{2} T_{1}(\Delta^{2}) z \Delta^{2}}
\right]^{1/2}.          \label{Ginzburg}
\end{equation}
Substituting here the accurate expressions (\ref{solution.2D}),
(\ref{coefficients.energy}) and (\ref{a}) for $\Delta$,
$T_{1}(\Delta^{2})$ and $a$ it is easy to find that (\ref{Ginzburg})
takes a simple enough form
\begin{equation}
\kappa = \frac{c}{e} \sqrt{\frac{m}{z}
\frac{2 |\varepsilon_{b}|}{2 \epsilon_{F} + |\varepsilon_{b}|}},
			    \label{Ginzburg.final}
\end{equation}
from which the explicit dependence of this parameter versus the ratio
between $\epsilon_{F}$ and $|\varepsilon_{b}|$ follows immediately.
So, for $z \sim (10^{7} - 10^{8})$cm$^{-1}$
(as occurs in real cuprates) the formula (\ref{Ginzburg.final})
for a large range of values for $\epsilon_{F}$ gives
$\kappa \sim 10^{2}$. This means that 2D metals with low
$\epsilon_{F}$ (so called underdoped case) turn out to be very strongly
type-II superconductors. However, it follows from
the same formula that with increasing $\epsilon_{F}$ the parameter
$\kappa$ decreases and in the limit
$\epsilon_{F} \gg |\varepsilon_{b}|$ (strongly overdoped case)
it is possible (in principle) to obtain the value $\kappa \ltwid 1$
or in the other words, to change the type of the superconductor.

     We estimate finally the value of the second critical field
$H_{c2}$. Indeed, equating the energy gain (see (\ref{energy})) of
the state with $\Delta \neq 0$ and the field energy one can
show that
\begin{displaymath}
H_{c2} = \frac{2}{\hbar}
\epsilon_{F}\sqrt{m z}.
\end{displaymath}
For the same parameters we find
that $H_{c2} \sim (10^{5}-10^{6})$Oe $= 10-10^{2}$T which also
corresponds to the critical magnetic field values for the HTSC
metal-oxides \cite{Kopnin,Micnas}. Finally, note that the
concentration dependence of $\partial H_{c2}/\partial T$ was
studied also \cite{Shovkovy}.

\section{Crossover in the quasi-2D systems}
\setcounter{equation}{0}
    As we have already mentioned in the introduction there are many
ways which allow one to extend the zero-temperature analysis of the 2D
model from the previous Chapter to the case $T \neq 0$. Every
way has some advantages as well as disadvantages. So, different
generalizations of the model are to be investigated. Here we shall
study a quasi-2D extension, which might be relevant for HTSCs
with a relatively low anisotropy of the conductivity as, for
example, is present in the Y-based cuprates. On the other hand,
the quasi-2D model can be considered as an extension of the 3D
model from Chapter~2. Our treatment is based on the paper
\cite{GLSh.Cond} (see also its detailed versions
\cite{Sharapov,Sharapov.PhD}).

\subsection{Model}
The simplest model Hamiltonian density for the carriers
in the quasi-2D system reads
\begin{eqnarray}
\cal{H} = & - & \psi_{\sigma}^{\dagger}(x) \left[
\frac{\nabla _{\perp}^{2}}{2 m_{\perp}}
  - \frac{1}{m_{z} d_{z}^{2}} \cos{(\imath \nabla _{z} d_{z})} + \mu \right]
    \psi_{\sigma}(x)
\nonumber                              \\
& - & V \psi_{\uparrow}^{\dagger}(x) \psi_{\downarrow}^{\dagger}(x)
      \psi_{\downarrow}(x) \psi_{\uparrow}(x),
                       \label{Hamiltonian.quasi}
\end{eqnarray}
where $x \equiv \tau, \mbox{\bf r}_{\perp}, r_{z}$
(with $\mbox{\bf r}_{\perp}$ being a 2D vector);
$\psi_{\sigma}(x)$, $\mu$ and $V$ are already determined;
$m_{\perp}$ is an carrier effective mass in the planes
(for example, CuO$_{2}$ ones);
$m_{z}$ is an effective mass in the $z$-direction;
$d_{z}$ is the interlayer distance.

	The Hamiltonian proposed proves to be very convenient
for study of fluctuation stabilization by weak 3D one-particle
inter-plane tunnelling. The extension of the 2D system in the
third-direction could give rise to a possibility that it may undergo
the 3D superconducting transition with the establishment of homogeneous
long-range order. This transition does not contradict the
2D theorems \cite{Coleman}. We omitted in (\ref{Hamiltonian.quasi})
the two-particle (Josephson) tunnelling considering it here to be
smaller than the one-particle already taken into account.
There can however be situations when Josephson tunnelling becomes
more essential then the one-particle coherent one. In addition some
authors consider that the most important mechanism for HTSC is the
incoherent inter-plane hopping (through, for instance, the impurity
(localized) states or due to the assistance of phonons.)

   It is significant that the large anisotropy of the conductivity cannot
be identified with the similar anisotropy of the effective masses
$m_{z}$ and $m_{\perp}$. In particular,
HTSC with rather large anisotropy in $z$-direction do not display
the usual metal behaviour at low temperatures. It means that interplane
motion of the fermions is incoherent and the BKT transition could
take place before the 3D superconducting transition (see the next
Chapter). But, as will be seen, for justification of approximations used
below the modest value of the ratio
$m_{z} / m_{\perp} \geq 10^{2}$ is already sufficient.
Such a value is present for instance in the HTSC
YBa$_{2}$Cu$_{3}$O$_{6+\delta}$ (1-2-3).  Because of this we shall study
the simplest case of tunnelling which is described by the second term
in (\ref{Hamiltonian.quasi}).

         The method for the study of the Hamiltonian
(\ref{Hamiltonian.quasi}) mainly coincides with that used for the 3D
model (\ref{Hamiltonian.3D}). So, we shall often refer to the similar
formulae from Chapter~2.

\subsection{The mean field approximation}
       The effective potential (\ref{Omega.Potential.expr}) obtained
in Appendix~A is in fact model independent. Thus, one has to use it
together with the dispersion law
\begin{equation}
\xi (\mbox{\bf k}) = \varepsilon (\mbox{\bf k}) - \mu =
\frac{\mbox{\bf k}_{\perp}^{2}}{2 m_{\perp}} +
\frac{1}{m_{z} d_{z}^{2}} \cos{k_{z} d_{z}} - \mu
\label{dispersion.law}
\end{equation}
in accordance with the Hamiltonian (\ref{Hamiltonian.quasi}).

       Again the stationary condition (\ref{gap}) results in the gap
equation (\ref{gap.3D}) where now $\xi(\mbox{\bf k})$ is determined
by (\ref{dispersion.law}).

       It is necessary to emphasize that fot the temperatures of
       interest, the band width in the $z$-direction is
$m_{z}^{-1} d_{z}^{-2} \ll T_{c}^{MF}$, i.e. the system is really a quasi-2D
one.  As for the last inequality, it is easy to see, that at $m_{z} \approx
10^{2} m_{e}$ and $d_{z} = 10 \dot A$ the value $\hbar^{2} / (m_{z} d_{z}^{2}
k_{B}) \sim 10$K is really far less then the usual critical temperatures in
HTSC compounds.

       Now we need to describe how one can regulate the ultraviolet
divergence in the  gap equation (\ref{gap.3D}). As  has been
already noted in Section~2.1 for the case of the local-pair
superconductivity the cutoff of BCS type cannot be applied. Moreover,
the regularization by the 3D scattering length $a_{s}$ (see the
definition (\ref{scattering})) is unsuitable for the quasi-2D system.
It turns out that in order to eliminate the divergences in
(\ref{gap.3D}) it is convenient to introduce, as in the 2D
case, the energy  of the two-particle bound state
(compare with (\ref{bound.energy}))
\begin{equation}
\varepsilon _{b} = - 2 W \exp
\left(- \frac{4 \pi d_{z}}{m_{\perp} V} \right),
                                \label{bound.energy.quasi}
\end{equation}
where $W = \mbox{\bf k}_{\perp B}^{2} /2 m_{\perp}$
is the bandwidth in the plane
\footnote{Note that in the region of the parameters considered these
bound states are formed without any threshold.}.
The factor $d_{z}$ in the index of the exponent is necessary to
preserve the dimensionlessness.

         Again, contrary to the usual BCS approach, the value of the
chemical potential should be consistently defined from the
equation (\ref{number.mean}), which leads to the second, or number,
equation (\ref{number.mean.expr}).

         Taking into account the above-mentioned inequality
$m_{z}^{-1} d_{z}^{-2} \ll T_{c}^{MF}$ the number equation
(\ref{number.mean.expr}) in the limit $W \to \infty$
takes the following final form:
\begin{equation}
2 T_{c}^{MF} \ln \left[1 + \exp \left( \frac{\mu}{T_{c}^{MF}} \right)
\right] = 2 \epsilon_{F},
\label{number.mean.quasi}
\end{equation}
where (here) $\epsilon_{F} = \pi n_{f} d_{z}/m_{\perp}$ is the Fermi energy
of free quasi-2D fermions with
$\varepsilon (\mbox{\bf k}) \sim \mbox{\bf k}_{\perp}^{2}$.

         So, one needs to solve simultaneously the system of the
equations (\ref{gap.3D}) with the dispersion law
(\ref{dispersion.law}) and (\ref{number.mean.quasi}) with two unknown
variables, $T_{c}^{MF}$ and $\mu$, respectively.

         At high carrier concentrations $n_{f}$, such that
$\mu \gg T_{c}^{MF}$, the equality $\mu \simeq \epsilon_{F}$ is indeed
the solution of (\ref{number.mean.quasi}). Than taking into
account the regularization procedure, it follows from (\ref{gap.3D})
that (compare with (\ref{temperature.mean}))
\begin{equation}
T_{c}^{MF} \simeq \frac{\gamma}{\pi}
\sqrt{2 |\varepsilon _{b}| \epsilon _{F}} =
\frac{2 \gamma}{\pi} \sqrt{W \epsilon_{F}}
\exp \left( -\frac{2 \pi d_{z}}{m_{\perp} V} \right).
                             \label{temperature.mean.quasi}
\end{equation}
This is just the well-known BCS result for a 2D metal \cite{GGL.Preprint}.

            In the opposite case of small $n_{f}$, such as
$-\mu \gg T_{c}^{MF}$, the roles of the gap (\ref{gap.3D})
and number (\ref{number.mean.quasi}) equations are as above in some
sense reversed:  the equation  (\ref{gap.3D}) determines $\mu$, while
the (\ref{number.mean.quasi}) determines the value of $T_{c}^{MF}$.
Now, using the definition (\ref{bound.energy.quasi}) one obtains from
(\ref{gap.3D})
\begin{equation}
\ln \frac{2 W}{|\varepsilon_{b}|} = \frac{1}{2 \pi}
\int_{0}^{2 \pi} d t \int_{0}^{2 W} d x
\frac{1}{x - 2 \mu + (2/m_{z} d_{z}^{2}) \cos t}.
                                 \label{gap.quasi.low}
\end{equation}
Integrating the right side of (\ref{gap.quasi.low}) and in the limit $W \to
\infty$ we immediately arrive at the final expression
\begin{equation}
\mu = -\frac{|\varepsilon _{b}|}{2} -
\frac{1}{2 (m_{z} d_{z}^{2})^{2} |\varepsilon _{b}|}.
                            \label{chemical.low}
\end{equation}
This expression, except for the second term, is identical to the
result described in Section~2.3.  The second term is directly connected to
the quasi-2D character of the model and, despite its far smaller magnitude,
is very important  when  the fluctuations are taken into account.  At
$|\varepsilon _{b}| \gg \epsilon _{F}$ the equation
(\ref{number.mean.quasi}) transforms into the transcendental one:
\begin{equation}
\frac{|\varepsilon _{b}|}{2 T_{c}^{MF}} =
\ln \frac{T_{c}^{MF}}{\epsilon _{F}}.
\label{temperature.mean.low}
\end{equation}

Its follows from (\ref{temperature.mean.low}) that we obtain
a similar result to that in Section~2.3:  for fixed
$\epsilon_{F}$, the value of $T_{c}^{MF}$
grows rapidly as the coupling constant $V$ increases. Thus, for the
case of small carrier density, the temperature $T_{c}^{MF}$ is not
connected to the critical one $T_c$,  but in fact corresponds to
the temperature of composite boson dissociation (compare with
Section~2.3).

\subsection{The role of Gaussian fluctuations}
      In order to investigate the effects of composite bosons
formed in the system, one should again take into account the order
parameter fluctuations. As was discussed in the previous Section,
one can simply use the expressions from Chapter~2 with
$\xi(\mbox{\bf k})$ determined by (\ref{dispersion.law}). Thus, the
expressions (\ref{thermopotential.final}) and (\ref{Gamma.momentum})
may be used directly if one applies the ultraviolet regularization
by the energy (\ref{bound.energy.quasi}) as has been done in
Section~4.2. Once again, from (\ref{thermopotential.final}), we arrive
at the "conservation law" (\ref{number.fluctuation}).

          Let us consider now the influence of $n_{B} (\mu , T)$
determined by (\ref{boson.definition}) on the behaviour of the system
as the carrier density changes. We should solve self-consistently
the system of equations (\ref{gap.3D}) and
(\ref{number.fluctuation}).

          At high enough carrier density the contribution of
$n_{B} (\mu , T)$ to (\ref{number.fluctuation}) is negligible,
and one arrives at the equality $T_c \simeq T_{c}^{MF}$,
where $T_{c}^{MF}$ is given by (\ref{temperature.mean.quasi}).

         Recall (see Section~2.4) that in a  more consistent scheme one
         should take into consideration the correction $\partial (\Omega -
\Omega_{pot}) / \partial \Phi $ to the gap equation (\ref{gap}), and
consequently the fluctuations modify the equation. This correction may be
especially important in the quasi-2D case. However, because of the
conditions $\mu = \epsilon _{F} \gg T_c$ and $m_{z} d_{z}^ {2} T_c \gg
1$, one can convince oneself that this correction changes $T_c$ rather
weakly \cite{Kats} (see also \cite{Vonsovsky}).

        Note that the consideration of the
        correction to the gap equation (\ref{gap}) should be very
interesting because based on the general 2D theorems one knows that
$T_{c} \to 0$ when $m_{z} \to \infty$ (2D case) even though $\mu \simeq
\epsilon_{F}$.  To trace this limit the above-mentioned correction must
be taken into account.  From the other side, when $m_{z} \to \infty$
the BKT scenario will take place at any density $n_{f}$ and we have
to modify the approach from Chapter~2 used here to study the formation
of the inhomogeneous BKT phase.  The latter will be accurately done in
the next Chapter.

           Turning back to the equality $T_{c} \simeq T_{c}^{MF}$ one
can see, using (\ref{temperature.mean.quasi})
(the $T_{c}^{MF}$ coincides with the expression obtained in
\cite{GGL.Preprint} for a pure 2D superconductor due to the band
narrowness in the $k_z$-direction), that $T_c$ increases with the
growth of coupling constant $V$. This behaviour of $T_{c}$ is just as
in the 3D case (see (\ref{temperature.mean})).

           At small concentrations, such that $|\mu| / T_c \gg 1$
taking into account the definition (\ref{bound.energy.quasi})
in the limit $W \to \infty$ one obtains from
(\ref{Gamma.momentum}) that
\begin{eqnarray}
& & \! \! \! \! \! \! \! \! \! \! \! \! \! \! \! \! \! \!
\Gamma^{-1} (i \Omega_{n} , \mbox{\bf K} )  =
\frac{m_{\perp}}{4 \pi d_{z}}
\times             \nonumber \\
& & \! \! \! \! \! \! \! \! \! \! \! \! \! \! \! \!
\ln{\frac{ \frac{\ds \mbox{\bf K}_{\perp}^{2}}{\ds 4 m_{\perp}} -
2 \mu - i \Omega_{n}  +
\sqrt{ \left(\frac{\ds \mbox{\bf K}_{\perp}^{2}}{\ds 4 m_{\perp}} -
2 \mu - i \Omega_{n} \right)^{2} +
\left( \frac{\ds 2}{\ds m_{z} d_{z}^{2}} \cos{\frac{\ds K_z d_{z}}{\ds 2}}
\right)^{2} } } {2 |\varepsilon _{b}|}} .
                                  \label{Gamma.renormalized.quasi}
\end{eqnarray}
After introducing the phase
$\delta(\omega, \mbox{\bf K}) \equiv
- \arg \Gamma^{-1}(\omega+i0, \mbox{\bf K})$ it can again be written
as:
\begin{eqnarray}
\delta(\omega , \mbox{\bf K}) = \pi \theta \left( \omega -
\frac{\mbox{\bf K}_{\perp}^{2}}{4 m_{\perp}} \right. & + &
\frac{1}{2 |\varepsilon _{b}| (m_{z} d_{z}^{2})^{2}} \cos{K_z d_{z}}
+ 2 \mu + |\varepsilon_{b}|
\nonumber                           \\
& + & \left.\frac{1}{2 |\varepsilon _{b}| (m_{z} d_{z}^{2})^{2}}
\right).
                               \label{phase.quasi}
\end{eqnarray}
At last, after substitution of the expression (\ref{chemical.low})
for the chemical potential into (\ref{phase.quasi}) one
arrives at the final equation for $T_{c}$, namely:
\begin{equation}
n_{B} (\mu, T_c) \equiv  \int \frac{d \mbox{\bf K}}{(2 \pi)^{3}}
n_{B} \left( \frac{\mbox{\bf K}_{\perp}^{2}}{4 m_{\perp}} + \frac{1}{2
|\varepsilon _{b}| (m_{z} d_{z}^{2})^{2}} (1 - \cos{K_z d_{z}}) \right )
\simeq \frac{n_{f}}{2}.
                               \label{number.boson.quasi}
\end{equation}
It is easy to see that the boson effective mass  for its motion in
the plane retains the value $2 m_{\perp}$. As for the motion
between the planes, the effective boson mass increases considerably:
$2 |\varepsilon _{b}| m_{z}^{2} d_{z}^{2} ( \gg m_{z})$.  It can be
stressed that this increase has a dynamical character as is
testified simply by the presence of $|\varepsilon _{b}|$. Physically, it is
ensured by the one-particle character (see equation
(\ref{Hamiltonian.quasi})) of the tunnelling between planes which only takes
place through the virtual breakup of a pair for which the energy loss is of
the order $|\varepsilon _{b}|$.

Now, using the formula for the BEC of an
ideal quasi-2D Bose-gas \cite{Wen} (see also the review \cite{Micnas}),
it is easy to write down an equation for $T_c$, which here
takes the simple form
\begin{equation}
T_c \simeq \frac{\pi n_{f} d_{z}}{2 m_{\perp}
\ln{(2 T_c |\varepsilon _{b}| m_{z}^{2} d_{z}^{4})}} =
\frac{\epsilon_{F}}{2 \ln{(2 T_c |\varepsilon _{b}| m_{z}^{2} d_{z}^{4})}}.
                          \label{temperature.bose.quasi}
\end{equation}
The last equation describes the characteristic properties of a
quasi-2D superconductor with  small carrier density:

\noindent
{\bf i)} firstly, the critical temperature
$T_c \sim \epsilon _{F}$, or (see above) $T_c \sim n_{f}/2$,
or the number of the composite bosons,
as it should be  in 2D case (recall that in a 3D one $T_c \sim
n_{f}^{2/3}$ (see (\ref{temperature.bose.3D})), in contrast to the MF
approximation 2D $T_{c}^{MF} \sim \sqrt{n_{f}} \gg T_{c}$
\cite{GGL.Preprint} (see also (\ref{solution.2D}) because usually
$T_{c}^{MF} \sim \Delta$);

\noindent
{\bf ii)} secondly, contrary to the case of the 3D superconductor
where $T_c$ does not depend on $V$  at all
(see (\ref{temperature.bose.3D})), in a quasi-2D system $T_c$ does
depend on $V$, namely: $T_c$ decreases with the growth of $V$.  As
was stated above, the reason for this is the dynamical increase of
the composite boson mass along the third direction. Thus, the growth
of $|\varepsilon _{b}|$ (or of $V$, which is the same) "makes" the
system more and more two-dimensional even for the simplest case
of a quasi-2D metal with a local four-fermion interaction.

         It is interesting to note that a decreasing $T_c$ can also
take place in the case when the local pairs are bipolarons
\cite{Alexandrov}; then, the increasing of coupling with phonons,
which makes the pairs more massive, also leads to $T_c$ decreasing,
(rather than increasing) as the electron-phonon coupling grows.

           The main results here are not only the expressions
(\ref{temperature.mean.quasi}) and (\ref{temperature.bose.quasi})
for $T_c$ in the different limiting cases, namely the cases of large
and small $V$. No less important is the comparison of these expressions
which show that for a given density of fermions (i.e. a given $\epsilon_{F}$)
there are two essential regions.  If $|\varepsilon_{b}| \ll \epsilon_{F}$,
then even in the case of small (by absolute value $n_{f}$) densities the BCS
formula is valid and $T_c$ grows with increasing $|\varepsilon_{b}|$
(see (\ref{temperature.mean.quasi})). In the opposite case,
$|\varepsilon_{b}| \gg \epsilon_{F}$, $T_c$ decreases with the growth
of $|\varepsilon_{b}|$. It shows that in the case of quasi-2D systems
(it seems that the HTSC belong to this case) $T_{c}(|\varepsilon_{b}|)$
has a maximum. Consequently, there is a region
(for fixed $|\varepsilon_{b}|$) of values of
$\epsilon_{F}$ for which $T_c$ increases.  If we
also take into account the two-particle tunnelling in
(\ref{Hamiltonian.quasi}), then the previous result will only be a
lower bound for $T_c$ for large $|\varepsilon_{b}|$. The region of
$\epsilon_{F} \approx |\varepsilon_{b}|$ needs special study
(see e.g. \cite{Emery,LSh.FNT.1997}) because of the presence of both
strongly developed fluctuations and the possible distinction of properties
of such a Fermi-liquid from the Fermi-liquid of Landau type.

\section{2D crossover: finite temperature}
\setcounter{equation}{0}
        We have already studied the finite temperature (or more
exactly $T_{c}$) crossover for the 3D and quasi-2D models. The 2D
crossover at $T = 0$ crossover was addressed also. So, we are ready
to discuss the 2D crossover at $T \neq 0$. As  was stressed in the
Introduction (see also \cite{Randeria.book}) an analytical treatment
of the finite $T$ crossover problem in 2D is quite a  difficult task.
This is primarily related to the necessity to treat $T_{c}$ as the
BKT temperature $T_{BKT}$ below which there is an algebraic
(power decay) order and a finite superfluid density \cite{Berezinskii}
(see also the review \cite{Minnhagen} and the book \cite{Izyumov}).
Nevertheless, some insight into the peculiarities of the formation
of the BKT has been gained \cite{JETPLetters}.

\subsection{Model and Formalism}
          Most of previous analyses \cite{Varma,Serene,Tokumitu}
of the behaviour of the 2D systems at $T \neq 0$ have been based on
a Nozieres-Schmitt-Rink approach \cite{Nozieres}. As shown above, this
is simply the Gaussian approximation to the functional integral which
perhaps explains the difficulties faced in these calculations. On the
one hand, Gaussian fluctuations destroy long-range order in 2D and if
one searches for the $T_{c}^{2D}$ at which the order sets in one
should get zero in accordance with the above-mentioned theorems
\cite{Coleman}. On the other hand, taking into account Gaussian
fluctuations is completely inadequate to describe the BKT transition.

          Nonetheless, several steps have been made even in this direction.
For example, the BKT transition in the relativistic $2+1$-theory was
studied in \cite{MacKenzie}, and even the crossover from
superconductivity to superfluidity was considered in \cite{Drechsler}
(see also \cite{Zwerger})
according to the value of the carrier density $n_{f}$
(recall that $n_{f} = m \epsilon_{F}/\pi$). However, the method
employed in \cite{Drechsler} to the temperature $T_{BKT}$ has a
number of drawbacks.

           Specifically, the equation for $T_{BKT}$ was
obtained neglecting the existence of a neutral (real) order
parameter $\rho$, whose appearance at finite $T$, being due to the
breaking of only a discrete symmetry, is consistent with
Coleman-Mermin-Wagher-Hohenberg theorem \cite{Coleman}. As we shall
see below, $\rho$ gives the modulus of a multivalued complex order
parameter of a 2D system as a whole, and only the modulus determines
the possibility of the formation of nonuniform (including vortex)
configurations in the system.

            However, as a result of allowing for
a neutral order parameter, a region where $\rho$ decays gradually to
zero appears in the phase diagram of the system. This region separates
the standard normal phase with $\rho = 0$ from the BKT phase,
where there is the power decay of correlations. Despite of the
exponential decay of the correlations in it, this new region of
states very likely possesses unusual properties, since $\rho$ appears
in all expressions in the same manner as does the energy gap $\Delta$
in the theory of ordinary superconductors, though to calculate the
observed single-particle spectrum, of course, the carrier losses due
to scattering of carriers by fluctuations of the phase of the order
parameter and, in case of real systems, by dopants must be taken into
account \cite{Loktev.Pogorelov}. The possible existence of such a phase,
which is also in some sense normal, might shed light of the frequently
anomalous (see, for example, the reviews
\cite{Loktev.review,Pines.review}) behaviour of the normal state of
HTSC, specifically, the temperature dependencies of the spin
susceptibility, resistivity, specific heat, photoemission spectra,
and so on \cite{Levi,Randeria.Nature,Uemura}, for the explanation of
which the idea of a pseudogap (and also spin gap) in the region
$T > T_{c}$ is now widely employed.

             Thus, our objective in this Chapter is to calculate
$T_{BKT}$ and $T_{\rho}$ ($T_{\rho}$ is the temperature at which
$\rho \to 0$) as functions of $n_{f}$ and to establish a form of
this new region, which lies in the temperature interval
$T_{BKT} < T < T_{\rho}$. Besides, we will try to demonstrate using
the example of the static spin susceptibility that
this phase may really be used to explain the above-mentioned
anomalous properties of the HTSC. That is the reason why the phase
was called the "anomalous normal" phase.

            There is no need to write down the model Hamiltonian
which is studied here, because it is identical with that of described
in Chapter~3.1.

            The desired phase diagram consisting from normal,
anomalous and superconducting
phases was calculated firstly in \cite{JETPLetters} employing the
Hubbard-Stratonovich method (see Section 2.2, the equations
(\ref{partition.Hubbard}) -- (\ref{Effective.Action})).
In the 2D case, however, instead of using the accepted method for
calculation the partition function $Z(v, \mu, T,)$
(see (\ref{partition.Hubbard})), one must to perform the calculation
in  modulus-phase variables. This prevents us from the subsequent
treatment of the phase fluctuations at Gaussian level only.
Thus, we will be able to take into account the phase degree of freedom
with needed accuracy.

           The modulus-phase variables were introduced in accordance
with \cite{Thouless}, where the parametrization
\begin{equation}
\Phi(x) = \rho(x) e^{-2 i \theta(x)}, \qquad
\Phi^{\ast}(x) = \rho(x) e^{2 i \theta(x)},
                      \label{parametrization.Hubbard}
\end{equation}
was used. One can easily see that (\ref{parametrization.Hubbard})
corresponds to the obvious transformation of the initial Fermi-fields,
namely
\begin{equation}
\psi_{\sigma}(x) = \chi_{\sigma}(x) e^{i \theta(x)}, \qquad
\psi_{\sigma}^{\dagger}(x) = \chi_{\sigma}^{\dagger}(x) e^{-i \theta(x)},
                        \label{parametrization.fermi}
\end{equation}
where the field operator $\chi_{\sigma}(x)$ describes neutral fermions
and $\exp[i \theta(x)]$ corresponds to the charge degree of freedom.
In Nambu variables (\ref{Nambu.variables}) the transformation
(\ref{parametrization.fermi}) takes the following form
\begin{equation}
\Psi(x) = e^{i \tau_{3} \theta(x)} \Upsilon(x), \qquad
\Psi^{\dagger}(x) = \Upsilon^{\dagger}(x) e^{-i \tau_{3} \theta(x)}.
                          \label{parametrization.Nambu}
\end{equation}
Making corresponding substitutions (\ref{parametrization.Nambu})
in the representation (\ref{partition.Hubbard}) and integrating over
the fermi-fields $\Upsilon$ and $\Upsilon^{\dagger}$ we arrive at the
expression (compare with (\ref{partition.effective}) and
(\ref{Effective.Action})
\begin{equation}
Z(v, \mu, T) = \int \rho {\cal D} \rho {\cal D} \theta
\exp{[-\beta \Omega (v, \mu, T, \rho(x), \partial \theta (x))]},
                                \label{partition.2D.effective}
\end{equation}
where
\begin{equation}
\beta \Omega (v, \mu, T, \rho (x), \partial \theta (x)) =
\frac{1}{V} \int_{0}^{\beta}
d \tau \int d \mbox{\bf r} \rho^{2}(x) -
\mbox{Tr Ln} G^{-1} + \mbox{Tr Ln} G_{0}^{-1}
                        \label{Effective.Action.2D}
\end{equation}
is as (\ref{Effective.Action}) the one-loop effective action, which,
however, depends on the modulus-phase variables. The action
(\ref{Effective.Action.2D}) is expressed through the Green function of
the initial (charged) fermions that has in the new variables the
following operator form
\footnote{It may be obtained as a solution of some differential
equation with the antiperiodic boundary conditions
(see (\ref{Green.fermion}) and (\ref{boundary})).}
\begin{eqnarray}
G^{-1} & = & -\hat{I} \partial_{\tau} +
\tau_{3} \left(\frac{\nabla^{2}}{2 m} + \mu \right) +
\tau_{1} \rho(\tau, \mbox{\bf r}) -
\nonumber                    \\
&& \! \! \! \! \! \! \! \! \! \! \! \! \! \! \! \! \! \! \!
\tau_{3} \left(i \partial_{\tau} \theta(\tau, \mbox{\bf r}) +
\frac{(\nabla \theta(\tau, \mbox{\bf r}))^{2}}{2 m} \right) +
\hat{I} \left(\frac{i \nabla^{2} \theta(\tau, \mbox{\bf r})}{2 m} +
\frac{i \nabla \theta(\tau, \mbox{\bf r}) \nabla}{m} \right).
                          \label{Green.fermion.phase}
\end{eqnarray}
The free fermion Green function $G_{0}$ that provides the convenient
regularization in the process of calculation was defined in
(\ref{Green.free}). It is important that neither the smallness nor
slowness of the variation of the phase of the order parameter was
assumed in obtaining expression (\ref{Green.fermion.phase}).

            Since the low-energy dynamics in the phases in which
$\rho \neq 0$ is determined mainly by the long-wavelength fluctuations
of $\theta(x)$, only the lowest order derivatives of the phase need be
retained in the expansion of
$\Omega(v, \mu, T, \rho(x), \partial \theta(x))$:
\begin{equation}
\Omega (v, \mu, \rho(x), \partial \theta(x))  \simeq
\Omega _{kin} (v, \mu, T, \rho, \partial \theta(x)) +
\Omega _{pot} (v, \mu, T, \rho)
                  \label{kinetic.phase+potential}
\end{equation}
where
\begin{equation}
\Omega _{kin} (v, \mu, T, \rho, \partial \theta(x))
 =  T \mbox{Tr} \sum_{n=1}^{\infty}
\left. \frac{1}{n} ({\cal G} \Sigma)^{n} \right|_{\rho = const}
               \label{Omega.Kinetic.phase}
\end{equation}
and (see (\ref{Omega.Potential}))
\begin{equation}
\Omega _{pot} (v, \mu, T, \rho)  =
\left. \left(\frac{1}{V} \int d \mbox{\bf r} \rho^{2} -
T \mbox{Tr Ln} {\cal G}^{-1} +
T \mbox{Tr Ln} G_{0}^{-1} \right) \right|_{\rho = const}.
                 \label{Omega.Potential.modulus}
\end{equation}
The kinetic $\Omega_{kin}$ and potential $\Omega_{pot}$ parts
are expressed through the Green function of the neutral fermions
which obeys the equation (compare with (\ref{Green.fermion}))
\begin{equation}
\left[-\hat I \partial_{\tau} +
\tau_{3} \left(\frac{\nabla^{2}}{2 m} + \mu \right)
+ \tau_{1} \rho \right]
{\cal G}(\tau, \mbox{\bf r}) = \delta(\tau) \delta(\mbox{\bf r})
          \label{Green.fermion.modulus}
\end{equation}
and the operator
\begin{equation}
\Sigma(\partial \theta) \equiv
\tau_{3} \left(i \partial_{\tau} \theta +
\frac{(\nabla \theta)^{2}}{2 m} \right) -
\hat{I} \left(\frac{i \nabla^{2} \theta}{2 m} +
\frac{i \nabla \theta(\tau, \mbox{\bf r}) \nabla}{m} \right).
                          \label{Sigma}
\end{equation}

          The representation (\ref{kinetic.phase+potential}) allows
one to get a full set of the equations which are necessary to find out
$T_{BKT}$, $\rho(T_{BKT})$ and $\mu(T_{BKT})$ at given $\epsilon_{F}$
(or, for example, $\rho(T)$ and $\mu(T)$ at given $T$ and
$\epsilon_{F}$).
While the equation for $T_{BKT}$ will be written using the kinetic
part (\ref{Omega.Kinetic.phase}) of the effective action, the equations
for $\rho(T_{BKT})$ and $\mu(T_{BKT})$ (or $\rho(T)$ and $\mu(T)$)
could be obtained using the mean field potential
(\ref{Omega.Potential.modulus}). It turns out that in the phase where
$\rho \neq 0$ the mean-field approximation for the modulus variable
describes the system quite well. This is mainly related with a
nonperturbative character of the Hubbard-Stratonovich method, i.e. most
of effects are taken over by a nonzero value of $\rho$.

\subsection{The equation for $T_{BKT}$}
If our model were reduced to a some known model describing the BKT
phase transition, we would be easily write the equation for $T_{BKT}$.
Indeed, in the lowest orders the kinetic term (\ref{Omega.Kinetic.phase})
coincides with so called classical XY model \cite{Minnhagen,Izyumov}
which has the following continuum Hamiltonian
\begin{equation}
H = \frac{J}{2}
\int d \mbox{\bf r} (\nabla \theta (\mbox{\bf r}))^{2} .
                                  \label{XY.Hamiltonian}
\end{equation}
Here $J$ is the some constant (in the original classical discrete
XY model it is the value of spin) and $\theta$ is the angle (phase)
of the two-component vector in plane.

            The temperature of the BKT transition is, in fact, known
for this model, namely
\begin{equation}
T_{BKT} = \frac{\pi}{2} J.
                        \label{BKT.temperature}
\end{equation}
Despite the very simple form
\footnote{The exponentially small correction is omitted here.}
of the equation (\ref{BKT.temperature}), it was derived
\cite{Berezinskii} (see also \cite{Minnhagen,Izyumov}) using
the renormalization group technique, which takes into account the
non-single-valuedness of the phase $\theta$. Thus, the fluctuations
of the phase are taken into account at a higher approximation than
the Gaussian one.

            To expand $\Omega _{kin}$ up to
$\sim (\nabla \theta)^{2}$, it would be sufficient
to restrict  ourselves to terms with $n=1,2$ in the expansion
(\ref{Omega.Kinetic.phase}). The procedure of calculation (see Appendix~B)
is similar to that employed in \cite{Schakel}, where the case of large
densities $n_f$ at $T=0$ was considered, and gives
\footnote{A total derivative with respect to  $\tau$ is omitted.}
\begin{equation}
\Omega _{kin} =
\frac{T}{2}
\int_{0}^{\beta} d \tau \int d \mbox{\bf r}
\left[
J(\mu, T, \rho(\mu, T)) (\nabla \theta)^{2} +
K(\mu, T, \rho(\mu, T)) (\partial_{\tau} \theta)^{2}
\right],
                 \label{Omega.Kinetic.phase.final}
\end{equation}
where
\begin{equation}
J(\mu, T, \rho) = \frac{1}{m} n_{F}(\mu, T, \rho) -
\frac{T}{\pi} \int_{-\mu/2T}^{\infty} dx
\frac{x + \mu/2T}{\cosh^{2} \sqrt{x^{2} +
\frac{\ds \rho^{2}}{\ds 4 T^{2}}}}
                              \label{J}
\end{equation}
characterizes the stiffness of neutral condensate,
\begin{equation}
K(\mu, T, \rho) = \frac{m}{2 \pi} \left(
1 + \frac{\mu}{\sqrt{\mu^2 + \rho^2}}
\tanh{\frac{\sqrt{\mu^2 + \rho^2}}{2T}} \right),
                               \label{K}
\end{equation}
and a value
\begin{equation}
n_{F}(\mu, T, \rho) = \frac{m}{2 \pi} \left\{
\sqrt{\mu^{2} + \rho^{2}} + \mu +
2 T \ln{\left[
1 + \exp{\left(-\frac{\sqrt{\mu^{2} + \rho^{2}}}{T}\right)}
\right]}\right\}
                 \label{fermion.density.2D}
\end{equation}
has a sense of fermi-quasiparticles density
(for $\rho = 0$ the expression (\ref{fermion.density.2D}) is
simply the density of the free fermions). Note that
$J(\mu,T, \rho=0) = 0$.

            A direct comparison of the expression
(\ref{Omega.Kinetic.phase.final}) with the Hamiltonian of XY model
(\ref{XY.Hamiltonian}) makes it possible to write a equation for
$T_{BKT}$:
\begin{equation}
\frac{\pi}{2} J(\mu, T_{BKT}, \rho(\mu, T_{BKT})) = T_{BKT}.
                              \label{BKT.equation}
\end{equation}
Although mathematically the problem reduces to a well-known problem,
the analogy is incomplete. Indeed, in the standard XY model
(as well as the nonlinear $\sigma$ model) the vector (spin)
subject to ordering is assumed to be a unit vector with no dependence
\footnote{There is no doubts that in certain situations (for example,
very high $T$) it also can become a thermodynamical variable,
i.e. dependent on $T$, as happens in problems of phase transitions
between ordered (magnetic) and disordered (paramagnetic) phases when
the spin itself vanishes. Specifically, for quasi-2D spin systems it is
virtually obvious that as one proceeds from high-$T$ regions, at first
a spin modulus forms in 2D clasters of finite size and only then does
global 3D ordering occur.}
on $T$. In our case this is fundamentally not the case, and a
self-consistent calculation of $T_{BKT}$ as a function of $n_{f}$
requires additional equations for $\rho$ and $\mu$, which together
with (\ref{BKT.equation}) form a complete system.

\subsection{The effective potential and the equations for $\rho$ and $\mu$}
            There is no need to repeat the calculation of the effective
potential. The point is that the effective potential
(\ref{Omega.Potential.expr}) calculated in Appendix~A depends on
(see Section~3.2) the invariant product $\Phi \Phi^{\ast} = \rho^{2}$
only. Thus, one may immediately write
\begin{equation}
\Omega _{pot}(v, \mu, T, \rho) = {v} \left[
\frac{\rho^2}{V} -
\int \frac{d \mbox{\bf k}}{(2 \pi)^{2}}
\left(2T
\ln\cosh{\frac{\sqrt{\xi^{2}(\mbox{\bf k})     + \rho^{2}}}{2 T}} -
\xi(\mbox{\bf k}) \right) \right],
                     \label{Omega.Potential.modulus.final}
\end{equation}
where $\xi(\mbox{\bf k}) = \mbox{\bf k}^{2}/2m - \mu$.
Then the desired missing equations are
the condition
$\partial \Omega_{pot}(\rho)/ \partial \rho = 0$
that the potential (\ref{Omega.Potential.modulus.final})
be minimum and the equality
$v^{-1} \partial \Omega_{pot} /\partial \mu = -n_f$, which
fixes $n_f$.
For them we have, respectively:
\begin{equation}
\frac{1}{V} = \int \frac{d \mbox{\bf k}}{(2 \pi)^{2}}
\frac{1}{2\sqrt{\xi^{2}(\mbox{\bf k}) + \rho^{2}}}
\tanh{\frac{\sqrt{\xi^{2}(\mbox{\bf k}) +
\rho^{2}}}{2 T}},              \label{rho}
\end{equation}
\begin{equation}
n_{F}(\mu, T, \rho) = n_f.     \label{number.rho}
\end{equation}
The equations (\ref{rho}) and (\ref{number.rho}) obtained above
comprise a self-consistent system for determining the modulus $\rho$
of the order parameter and the chemical potential $\mu$ in the
mean-field approximation for fixed $T$ and $n_{f}$.

           As we have already discussed in Section~3.3, the energy
of two-particle bound states $\varepsilon_{b}$ (see its definition
(\ref{bound.energy})), is more convenient to use than the
four-fermi constant $V$. For example, using the identity
\begin{displaymath}
\tanh \frac{x}{2} = 1 - \frac{2}{\exp x + 1}
\end{displaymath}
one may easily go to the limits $W \to \infty$ and $V \to 0$ in the
equation (\ref{rho}), which after this renormalization becomes
\begin{equation}
\ln{\frac{|\varepsilon_{b}|}{\sqrt{\mu^{2} + \rho^{2}} - \mu }} =
2 \int_{-\mu/T}^{\infty} d u
\frac{1}{\sqrt{u^2 + (\frac{\ds \rho}{\ds T})^2}
\left[\exp{\sqrt{u^2 + (\frac{\ds \rho}{\ds T})^2}} + 1 \right]},
                 \label{rho.bound}
\end{equation}
Thus, in practice, we will solve numerically the system of the
equations (\ref{BKT.equation}), (\ref{rho.bound}) and (\ref{number.rho})
to study $T_{BKT}$ as function of $n_{f}$.

            It is easy to show that at $T = 0$ the system
(\ref{rho.bound}), (\ref{number.rho}) transforms into the system
(\ref{gap+number.2D}) which was already addressed. Recall that its
solution is $\rho = \sqrt{2 |\varepsilon_{b}| \epsilon_{F}}$ and
$\mu = - |\varepsilon_{b}|/2 + \epsilon_{F}$. This will be useful for
studying the concentration dependencies of $2 \Delta/T_{BKT}$ and
$2 \Delta/T_{\rho}$, where $\Delta$ is the zero-temperature gap in
the quasiparticle excitation spectrum
\cite{Leggett,Randeria.2D,Randeria.book}
\begin{equation}
\Delta =  \left\{
\begin{array}{cc}
\rho,                      & \mu > 0; \\
\sqrt{\mu^{2} + \rho^{2}}, & \mu < 0.
\end{array}
\right.
                             \label{energy.gap}
\end{equation}

           Setting $\rho = 0$ in the equations (\ref{rho}) and
(\ref{number.rho}), we arrive (in the same approximation) at the
equations for the critical temperature $T_{\rho}$ and the corresponding
value of $\mu$:
\begin{equation}
\ln{\frac{|\varepsilon_{b}|}{T_{\rho}}\frac{\gamma}{\pi}} =
- \int_{0}^{\mu/2T_{\rho}} d u \frac{\tanh{u}}{u}
\qquad (\gamma = 1.781),
                 \label{temperature.rho}
\end{equation}
\begin{equation}
T_{\rho} \ln{\left[1 +
\exp{ \left( \frac{\mu}{T_{\rho}} \right)}\right]} = \epsilon_{F}.
                   \label{number.temperature.rho}
\end{equation}
Note that these equations coincide with the system which determines
$T_{c}^{(2D) MF}$ and $\mu(T_{c}^{(2D) MF})$  (see \cite{GGL.Preprint}).
This is evidently related with mean-field approximation used here.
There is, however, a crucial difference between these values.
Namely, if one takes into account the fluctuations, the value of
$T_{c}^{2D}$ should go to zero, while the value of $T_{\rho}$ should
stay finite. That is the reason why the temperature $T_{\rho}$ has
its own physical meaning: the incoherent (local or Cooper) pairs
begin to be formed formed below $T_{\rho}$. At higher temperatures
there are these pair fluctuations only (see e.g. \cite{LSh.FNT.1997}).

\subsection{The phase diagram}
          The numerical investigation of the systems (\ref{BKT.equation}),
(\ref{rho.bound}), (\ref{number.rho})  and (\ref{temperature.rho}),
(\ref{number.temperature.rho}) gives the following very interesting
results, which are displayed graphically.
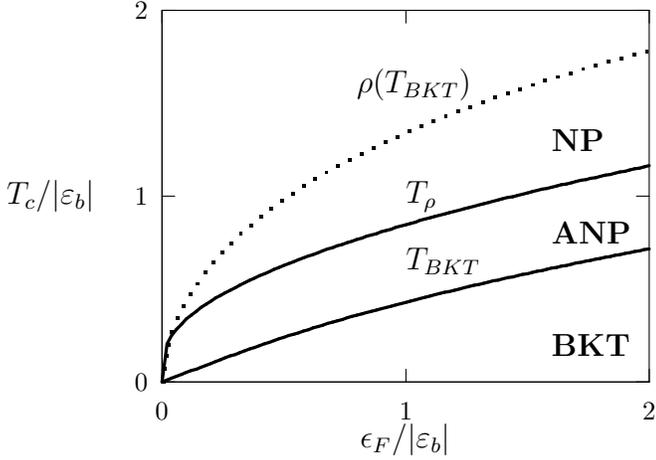
\begin{figure}[ht]
\setlength{\unitlength}{0.240900pt}
\ifx\plotpoint\undefined\newsavebox{\plotpoint}\fi
\sbox{\plotpoint}{\rule[-0.200pt]{0.400pt}{0.400pt}}%
\special{em:linewidth 0.4pt}%
\begin{picture}(1049,720)(0,0)
\font\gnuplot=cmr10 at 10pt
\gnuplot
\put(220,113){\special{em:moveto}}
\put(985,113){\special{em:lineto}}
\put(220,113){\special{em:moveto}}
\put(220,697){\special{em:lineto}}
\put(220,113){\special{em:moveto}}
\put(240,113){\special{em:lineto}}
\put(985,113){\special{em:moveto}}
\put(965,113){\special{em:lineto}}
\put(198,113){\makebox(0,0)[r]{0}}
\put(220,405){\special{em:moveto}}
\put(240,405){\special{em:lineto}}
\put(985,405){\special{em:moveto}}
\put(965,405){\special{em:lineto}}
\put(198,405){\makebox(0,0)[r]{1}}
\put(220,697){\special{em:moveto}}
\put(240,697){\special{em:lineto}}
\put(985,697){\special{em:moveto}}
\put(965,697){\special{em:lineto}}
\put(198,697){\makebox(0,0)[r]{2}}
\put(220,113){\special{em:moveto}}
\put(220,133){\special{em:lineto}}
\put(220,697){\special{em:moveto}}
\put(220,677){\special{em:lineto}}
\put(220,68){\makebox(0,0){0}}
\put(603,113){\special{em:moveto}}
\put(603,133){\special{em:lineto}}
\put(603,697){\special{em:moveto}}
\put(603,677){\special{em:lineto}}
\put(603,68){\makebox(0,0){1}}
\put(985,113){\special{em:moveto}}
\put(985,133){\special{em:lineto}}
\put(985,697){\special{em:moveto}}
\put(985,677){\special{em:lineto}}
\put(985,68){\makebox(0,0){2}}
\put(220,113){\special{em:moveto}}
\put(985,113){\special{em:lineto}}
\put(985,697){\special{em:lineto}}
\put(220,697){\special{em:lineto}}
\put(220,113){\special{em:lineto}}
\put(45,405){\makebox(0,0){$T_c/|\varepsilon_{b}|$}}
\put(602,23){\makebox(0,0){$\epsilon_{F}/|\varepsilon_{b}|$}}
\put(526,580){\makebox(0,0)[l]{$\rho (T_{BKT})$}}
\put(603,405){\makebox(0,0)[l]{$T_{\rho}$}}
\put(603,303){\makebox(0,0)[l]{$T_{BKT}$}}
\put(832,171){\makebox(0,0)[l]{${\bf BKT}$}}
\put(832,347){\makebox(0,0)[l]{${\bf ANP}$}}
\put(832,493){\makebox(0,0)[l]{${\bf NP}$}}
\sbox{\plotpoint}{\rule[-0.600pt]{1.200pt}{1.200pt}}%
\special{em:linewidth 1.2pt}%
\put(220,113){\special{em:moveto}}
\put(228,173){\special{em:lineto}}
\put(235,186){\special{em:lineto}}
\put(243,196){\special{em:lineto}}
\put(251,204){\special{em:lineto}}
\put(258,212){\special{em:lineto}}
\put(266,218){\special{em:lineto}}
\put(274,224){\special{em:lineto}}
\put(281,230){\special{em:lineto}}
\put(289,235){\special{em:lineto}}
\put(297,240){\special{em:lineto}}
\put(304,245){\special{em:lineto}}
\put(312,249){\special{em:lineto}}
\put(319,253){\special{em:lineto}}
\put(327,257){\special{em:lineto}}
\put(335,261){\special{em:lineto}}
\put(342,265){\special{em:lineto}}
\put(350,269){\special{em:lineto}}
\put(358,273){\special{em:lineto}}
\put(365,276){\special{em:lineto}}
\put(373,280){\special{em:lineto}}
\put(381,283){\special{em:lineto}}
\put(388,286){\special{em:lineto}}
\put(396,290){\special{em:lineto}}
\put(404,293){\special{em:lineto}}
\put(411,296){\special{em:lineto}}
\put(419,299){\special{em:lineto}}
\put(427,302){\special{em:lineto}}
\put(434,305){\special{em:lineto}}
\put(442,308){\special{em:lineto}}
\put(450,311){\special{em:lineto}}
\put(457,313){\special{em:lineto}}
\put(465,316){\special{em:lineto}}
\put(472,319){\special{em:lineto}}
\put(480,322){\special{em:lineto}}
\put(488,324){\special{em:lineto}}
\put(495,327){\special{em:lineto}}
\put(503,329){\special{em:lineto}}
\put(511,332){\special{em:lineto}}
\put(518,334){\special{em:lineto}}
\put(526,337){\special{em:lineto}}
\put(534,339){\special{em:lineto}}
\put(541,342){\special{em:lineto}}
\put(549,344){\special{em:lineto}}
\put(557,346){\special{em:lineto}}
\put(564,349){\special{em:lineto}}
\put(572,351){\special{em:lineto}}
\put(580,353){\special{em:lineto}}
\put(587,356){\special{em:lineto}}
\put(595,358){\special{em:lineto}}
\put(603,360){\special{em:lineto}}
\put(610,362){\special{em:lineto}}
\put(618,364){\special{em:lineto}}
\put(625,367){\special{em:lineto}}
\put(633,369){\special{em:lineto}}
\put(641,371){\special{em:lineto}}
\put(648,373){\special{em:lineto}}
\put(656,375){\special{em:lineto}}
\put(664,377){\special{em:lineto}}
\put(671,379){\special{em:lineto}}
\put(679,381){\special{em:lineto}}
\put(687,383){\special{em:lineto}}
\put(694,385){\special{em:lineto}}
\put(702,387){\special{em:lineto}}
\put(710,389){\special{em:lineto}}
\put(717,391){\special{em:lineto}}
\put(725,393){\special{em:lineto}}
\put(733,395){\special{em:lineto}}
\put(740,397){\special{em:lineto}}
\put(748,399){\special{em:lineto}}
\put(755,401){\special{em:lineto}}
\put(763,403){\special{em:lineto}}
\put(771,404){\special{em:lineto}}
\put(778,406){\special{em:lineto}}
\put(786,408){\special{em:lineto}}
\put(794,410){\special{em:lineto}}
\put(801,412){\special{em:lineto}}
\put(809,414){\special{em:lineto}}
\put(817,416){\special{em:lineto}}
\put(824,417){\special{em:lineto}}
\put(832,419){\special{em:lineto}}
\put(840,421){\special{em:lineto}}
\put(847,423){\special{em:lineto}}
\put(855,424){\special{em:lineto}}
\put(863,426){\special{em:lineto}}
\put(870,428){\special{em:lineto}}
\put(878,430){\special{em:lineto}}
\put(886,431){\special{em:lineto}}
\put(893,433){\special{em:lineto}}
\put(901,434){\special{em:lineto}}
\put(908,436){\special{em:lineto}}
\put(916,438){\special{em:lineto}}
\put(924,439){\special{em:lineto}}
\put(931,441){\special{em:lineto}}
\put(939,443){\special{em:lineto}}
\put(947,444){\special{em:lineto}}
\put(954,446){\special{em:lineto}}
\put(962,448){\special{em:lineto}}
\put(970,449){\special{em:lineto}}
\put(977,451){\special{em:lineto}}
\put(985,453){\special{em:lineto}}
\put(220,113){\special{em:moveto}}
\put(228,116){\special{em:lineto}}
\put(235,119){\special{em:lineto}}
\put(243,122){\special{em:lineto}}
\put(251,125){\special{em:lineto}}
\put(258,128){\special{em:lineto}}
\put(266,131){\special{em:lineto}}
\put(274,133){\special{em:lineto}}
\put(281,136){\special{em:lineto}}
\put(289,139){\special{em:lineto}}
\put(297,142){\special{em:lineto}}
\put(304,145){\special{em:lineto}}
\put(312,148){\special{em:lineto}}
\put(319,151){\special{em:lineto}}
\put(327,154){\special{em:lineto}}
\put(335,156){\special{em:lineto}}
\put(342,159){\special{em:lineto}}
\put(350,162){\special{em:lineto}}
\put(358,165){\special{em:lineto}}
\put(365,167){\special{em:lineto}}
\put(373,170){\special{em:lineto}}
\put(381,173){\special{em:lineto}}
\put(388,175){\special{em:lineto}}
\put(396,178){\special{em:lineto}}
\put(404,181){\special{em:lineto}}
\put(411,183){\special{em:lineto}}
\put(419,186){\special{em:lineto}}
\put(427,188){\special{em:lineto}}
\put(434,190){\special{em:lineto}}
\put(442,193){\special{em:lineto}}
\put(450,195){\special{em:lineto}}
\put(457,198){\special{em:lineto}}
\put(465,200){\special{em:lineto}}
\put(472,202){\special{em:lineto}}
\put(480,205){\special{em:lineto}}
\put(488,207){\special{em:lineto}}
\put(495,209){\special{em:lineto}}
\put(503,211){\special{em:lineto}}
\put(511,214){\special{em:lineto}}
\put(518,216){\special{em:lineto}}
\put(526,218){\special{em:lineto}}
\put(534,220){\special{em:lineto}}
\put(541,222){\special{em:lineto}}
\put(549,224){\special{em:lineto}}
\put(557,226){\special{em:lineto}}
\put(564,228){\special{em:lineto}}
\put(572,230){\special{em:lineto}}
\put(580,232){\special{em:lineto}}
\put(587,234){\special{em:lineto}}
\put(595,236){\special{em:lineto}}
\put(603,238){\special{em:lineto}}
\put(610,240){\special{em:lineto}}
\put(618,242){\special{em:lineto}}
\put(625,244){\special{em:lineto}}
\put(633,246){\special{em:lineto}}
\put(641,248){\special{em:lineto}}
\put(648,250){\special{em:lineto}}
\put(656,252){\special{em:lineto}}
\put(664,254){\special{em:lineto}}
\put(671,255){\special{em:lineto}}
\put(679,257){\special{em:lineto}}
\put(687,259){\special{em:lineto}}
\put(694,261){\special{em:lineto}}
\put(702,263){\special{em:lineto}}
\put(710,265){\special{em:lineto}}
\put(717,266){\special{em:lineto}}
\put(725,268){\special{em:lineto}}
\put(733,270){\special{em:lineto}}
\put(740,272){\special{em:lineto}}
\put(748,273){\special{em:lineto}}
\put(755,275){\special{em:lineto}}
\put(763,277){\special{em:lineto}}
\put(771,278){\special{em:lineto}}
\put(778,280){\special{em:lineto}}
\put(786,282){\special{em:lineto}}
\put(794,283){\special{em:lineto}}
\put(801,285){\special{em:lineto}}
\put(809,287){\special{em:lineto}}
\put(817,288){\special{em:lineto}}
\put(824,290){\special{em:lineto}}
\put(832,292){\special{em:lineto}}
\put(840,293){\special{em:lineto}}
\put(847,295){\special{em:lineto}}
\put(855,296){\special{em:lineto}}
\put(863,298){\special{em:lineto}}
\put(870,300){\special{em:lineto}}
\put(878,301){\special{em:lineto}}
\put(886,303){\special{em:lineto}}
\put(893,304){\special{em:lineto}}
\put(901,306){\special{em:lineto}}
\put(908,307){\special{em:lineto}}
\put(916,309){\special{em:lineto}}
\put(924,310){\special{em:lineto}}
\put(931,312){\special{em:lineto}}
\put(939,313){\special{em:lineto}}
\put(947,315){\special{em:lineto}}
\put(954,316){\special{em:lineto}}
\put(962,318){\special{em:lineto}}
\put(970,319){\special{em:lineto}}
\put(977,321){\special{em:lineto}}
\put(985,322){\special{em:lineto}}
\sbox{\plotpoint}{\rule[-0.500pt]{1.000pt}{1.000pt}}%
\special{em:linewidth 1.0pt}%
\put(220,113){\usebox{\plotpoint}}
\multiput(220,113)(2.836,20.561){3}{\usebox{\plotpoint}}
\multiput(228,171)(5.596,19.987){2}{\usebox{\plotpoint}}
\put(242.83,213.61){\usebox{\plotpoint}}
\multiput(243,214)(9.282,18.564){0}{\usebox{\plotpoint}}
\put(252.09,232.18){\usebox{\plotpoint}}
\put(262.18,250.27){\usebox{\plotpoint}}
\multiput(266,256)(12.208,16.786){0}{\usebox{\plotpoint}}
\put(274.15,267.23){\usebox{\plotpoint}}
\put(285.99,284.24){\usebox{\plotpoint}}
\multiput(289,288)(12.966,16.207){0}{\usebox{\plotpoint}}
\put(298.92,300.47){\usebox{\plotpoint}}
\multiput(304,307)(14.676,14.676){0}{\usebox{\plotpoint}}
\put(312.77,315.88){\usebox{\plotpoint}}
\put(326.99,330.99){\usebox{\plotpoint}}
\multiput(327,331)(14.676,14.676){0}{\usebox{\plotpoint}}
\put(341.66,345.66){\usebox{\plotpoint}}
\multiput(342,346)(15.620,13.668){0}{\usebox{\plotpoint}}
\put(357.72,358.79){\usebox{\plotpoint}}
\multiput(358,359)(14.676,14.676){0}{\usebox{\plotpoint}}
\multiput(365,366)(16.604,12.453){0}{\usebox{\plotpoint}}
\put(373.41,372.30){\usebox{\plotpoint}}
\multiput(381,378)(15.759,13.508){0}{\usebox{\plotpoint}}
\put(389.63,385.23){\usebox{\plotpoint}}
\multiput(396,390)(16.604,12.453){0}{\usebox{\plotpoint}}
\put(406.28,397.63){\usebox{\plotpoint}}
\multiput(411,401)(17.601,11.000){0}{\usebox{\plotpoint}}
\put(423.68,408.92){\usebox{\plotpoint}}
\multiput(427,411)(16.889,12.064){0}{\usebox{\plotpoint}}
\put(440.98,420.37){\usebox{\plotpoint}}
\multiput(442,421)(17.601,11.000){0}{\usebox{\plotpoint}}
\multiput(450,426)(16.889,12.064){0}{\usebox{\plotpoint}}
\put(458.29,431.81){\usebox{\plotpoint}}
\multiput(465,436)(18.021,10.298){0}{\usebox{\plotpoint}}
\put(476.28,442.14){\usebox{\plotpoint}}
\multiput(480,444)(17.601,11.000){0}{\usebox{\plotpoint}}
\put(494.21,452.55){\usebox{\plotpoint}}
\multiput(495,453)(18.564,9.282){0}{\usebox{\plotpoint}}
\multiput(503,457)(18.564,9.282){0}{\usebox{\plotpoint}}
\put(512.70,461.97){\usebox{\plotpoint}}
\multiput(518,465)(18.564,9.282){0}{\usebox{\plotpoint}}
\put(531.11,471.55){\usebox{\plotpoint}}
\multiput(534,473)(18.021,10.298){0}{\usebox{\plotpoint}}
\multiput(541,477)(18.564,9.282){0}{\usebox{\plotpoint}}
\put(549.48,481.18){\usebox{\plotpoint}}
\multiput(557,484)(18.021,10.298){0}{\usebox{\plotpoint}}
\put(568.37,489.64){\usebox{\plotpoint}}
\multiput(572,491)(18.564,9.282){0}{\usebox{\plotpoint}}
\multiput(580,495)(19.077,8.176){0}{\usebox{\plotpoint}}
\put(587.28,498.14){\usebox{\plotpoint}}
\multiput(595,502)(19.434,7.288){0}{\usebox{\plotpoint}}
\put(606.11,506.78){\usebox{\plotpoint}}
\multiput(610,509)(19.434,7.288){0}{\usebox{\plotpoint}}
\multiput(618,512)(19.077,8.176){0}{\usebox{\plotpoint}}
\put(625.11,515.04){\usebox{\plotpoint}}
\multiput(633,518)(18.564,9.282){0}{\usebox{\plotpoint}}
\put(644.25,522.93){\usebox{\plotpoint}}
\multiput(648,524)(18.564,9.282){0}{\usebox{\plotpoint}}
\put(663.41,530.78){\usebox{\plotpoint}}
\multiput(664,531)(19.957,5.702){0}{\usebox{\plotpoint}}
\multiput(671,533)(18.564,9.282){0}{\usebox{\plotpoint}}
\put(682.65,538.37){\usebox{\plotpoint}}
\multiput(687,540)(19.957,5.702){0}{\usebox{\plotpoint}}
\multiput(694,542)(19.434,7.288){0}{\usebox{\plotpoint}}
\put(702.27,545.10){\usebox{\plotpoint}}
\multiput(710,548)(19.077,8.176){0}{\usebox{\plotpoint}}
\put(721.58,552.72){\usebox{\plotpoint}}
\multiput(725,554)(20.136,5.034){0}{\usebox{\plotpoint}}
\multiput(733,556)(19.077,8.176){0}{\usebox{\plotpoint}}
\put(741.16,559.43){\usebox{\plotpoint}}
\multiput(748,562)(19.957,5.702){0}{\usebox{\plotpoint}}
\put(760.77,566.17){\usebox{\plotpoint}}
\multiput(763,567)(19.434,7.288){0}{\usebox{\plotpoint}}
\multiput(771,570)(19.957,5.702){0}{\usebox{\plotpoint}}
\put(780.39,572.90){\usebox{\plotpoint}}
\multiput(786,575)(20.136,5.034){0}{\usebox{\plotpoint}}
\put(799.99,579.57){\usebox{\plotpoint}}
\multiput(801,580)(20.136,5.034){0}{\usebox{\plotpoint}}
\multiput(809,582)(19.434,7.288){0}{\usebox{\plotpoint}}
\put(819.76,585.79){\usebox{\plotpoint}}
\multiput(824,587)(19.434,7.288){0}{\usebox{\plotpoint}}
\put(839.57,591.89){\usebox{\plotpoint}}
\multiput(840,592)(19.957,5.702){0}{\usebox{\plotpoint}}
\multiput(847,594)(19.434,7.288){0}{\usebox{\plotpoint}}
\put(859.35,598.09){\usebox{\plotpoint}}
\multiput(863,599)(19.957,5.702){0}{\usebox{\plotpoint}}
\multiput(870,601)(19.434,7.288){0}{\usebox{\plotpoint}}
\put(879.14,604.28){\usebox{\plotpoint}}
\multiput(886,606)(19.957,5.702){0}{\usebox{\plotpoint}}
\put(899.21,609.55){\usebox{\plotpoint}}
\multiput(901,610)(19.077,8.176){0}{\usebox{\plotpoint}}
\multiput(908,613)(20.136,5.034){0}{\usebox{\plotpoint}}
\put(918.96,615.74){\usebox{\plotpoint}}
\multiput(924,617)(19.957,5.702){0}{\usebox{\plotpoint}}
\multiput(931,619)(20.136,5.034){0}{\usebox{\plotpoint}}
\put(939.03,621.01){\usebox{\plotpoint}}
\multiput(947,623)(19.957,5.702){0}{\usebox{\plotpoint}}
\put(958.93,626.85){\usebox{\plotpoint}}
\multiput(962,628)(20.136,5.034){0}{\usebox{\plotpoint}}
\multiput(970,630)(19.957,5.702){0}{\usebox{\plotpoint}}
\put(978.89,632.47){\usebox{\plotpoint}}
\end{picture}

\caption{$T_{BKT}$ and $T_{\rho}$ versus the free fermion density.
The dots represent the function $\rho(\epsilon_{F})$ at $T = T_{BKT}$.
The regions of the normal phase (NP), anomalous normal phase (ANP) and BKT
phase are indicated.}
\end{figure}

\noindent
{\bf a)} The anomalous phase region (see Fig.1) in the present model is
commensurate with the BKT region. But it has not been ruled out that in the
case of the quasi-2D model this region will disappear as $n_{f}$ increases.
For example, in the case of an indirect interaction it was shown
\cite{Turkowskii} that the anomalous phase region really exists at the low
carrier density only, i.e. it shrinks when the doping increases.

\noindent
{\bf b)} For low $\epsilon_{F} (\ll |\varepsilon_{b}|)$ the function
$T_{BKT}(\epsilon_{F})$ is linear, as it also confirmed by the
analytical solution of the system (\ref{BKT.equation}), (\ref{rho.bound})
and (\ref{number.rho}), which gives $T_{BKT} = \epsilon_{F}/2$.
Surprisingly such behaviour (although for $T_{c}$) is copied
in different families of non-conventional ("exotic") superconductors
(HTSC including) with comparatively small Fermi energy \cite{Uemura}.

        We note that in this limit the temperature $T_{c}$ of formation of
a homogeneous order parameter for the quasi-2D model (see Chapter~4,
the equation (\ref{temperature.bose.quasi})) can be written in the
following form \cite{Turkowskii} (see also \cite{Uemura})
\begin{equation}
T_{c} \approx \frac{T_{BKT}}
{\ln (\epsilon_{F} |\varepsilon_{b}|/4 t_{\mid
\mid}^{2})},
\label{temperature.quasi}
\end{equation}
where
$t_{\mid \mid} = 1/(m_{z} d_{z}^{2})$ is the inter-plane hopping
(coherent tunnelling) constant. This shows that when $T_{c} < T_{BKT}$ the
weak three-dimensionalization can preserve (in any case, for low $n_{f}$)
the regions of the anomalous and BKT phases, which, for example, happens
in the relativistic quasi-2D model \cite{Ichinose}. At the same time,
as the three-dimensionalization parameter $t_{\mid \mid}$ increases,
when $T_{c} > T_{BKT}$ the BKT phase can vanish, provide, however,
that the anomalous phase region and both temperatures $T_{\rho}$ and $T_{c}$
are preserved. It follows from (\ref{temperature.quasi}) that
the BKT phase vanishes when
$t_{\mid \mid} > \sqrt{2 |\varepsilon_{b}| \epsilon_{F}} (= \Delta)$.
\begin{figure}[ht]
\setlength{\unitlength}{0.240900pt}
\ifx\plotpoint\undefined\newsavebox{\plotpoint}\fi
\sbox{\plotpoint}{\rule[-0.200pt]{0.400pt}{0.400pt}}%
\special{em:linewidth 0.4pt}%
\begin{picture}(1500,1350)(0,0)
\font\gnuplot=cmr10 at 10pt
\gnuplot
\put(176,488){\special{em:moveto}}
\put(1436,488){\special{em:lineto}}
\put(176,68){\special{em:moveto}}
\put(176,1327){\special{em:lineto}}
\put(176,68){\special{em:moveto}}
\put(196,68){\special{em:lineto}}
\put(1436,68){\special{em:moveto}}
\put(1416,68){\special{em:lineto}}
\put(154,68){\makebox(0,0)[r]{-0.5}}
\put(176,488){\special{em:moveto}}
\put(196,488){\special{em:lineto}}
\put(1436,488){\special{em:moveto}}
\put(1416,488){\special{em:lineto}}
\put(154,488){\makebox(0,0)[r]{0}}
\put(176,907){\special{em:moveto}}
\put(196,907){\special{em:lineto}}
\put(1436,907){\special{em:moveto}}
\put(1416,907){\special{em:lineto}}
\put(154,907){\makebox(0,0)[r]{0.5}}
\put(176,1327){\special{em:moveto}}
\put(196,1327){\special{em:lineto}}
\put(1436,1327){\special{em:moveto}}
\put(1416,1327){\special{em:lineto}}
\put(154,1327){\makebox(0,0)[r]{1}}
\put(176,68){\special{em:moveto}}
\put(176,88){\special{em:lineto}}
\put(176,1327){\special{em:moveto}}
\put(176,1307){\special{em:lineto}}
\put(176,23){\makebox(0,0){0}}
\put(806,68){\special{em:moveto}}
\put(806,88){\special{em:lineto}}
\put(806,1327){\special{em:moveto}}
\put(806,1307){\special{em:lineto}}
\put(806,23){\makebox(0,0){1}}
\put(1436,68){\special{em:moveto}}
\put(1436,88){\special{em:lineto}}
\put(1436,1327){\special{em:moveto}}
\put(1436,1307){\special{em:lineto}}
\put(1436,23){\makebox(0,0){2}}
\put(176,68){\special{em:moveto}}
\put(1436,68){\special{em:lineto}}
\put(1436,1327){\special{em:lineto}}
\put(176,1327){\special{em:lineto}}
\put(176,68){\special{em:lineto}}
\put(680,152){\makebox(0,0)[l]{$1$}}
\put(680,320){\makebox(0,0)[l]{$2$}}
\put(680,572){\makebox(0,0)[l]{$3$}}
\put(649,781){\makebox(0,0)[l]{$4$}}
\put(995,907){\makebox(0,0)[l]{$5$}}
\put(1121,1075){\makebox(0,0)[l]{$6$}}
\put(1184,1260){\makebox(0,0)[l]{$7$}}
\put(270,739){\makebox(0,0)[l]{${\bf BKT}$}}
\put(995,656){\makebox(0,0)[l]{${\bf NP}$}}
\put(491,572){\makebox(0,0)[l]{${\bf ANP}$}}
\put(1184,-15){\makebox(0,0)[l]{$T/T_{\rho}$}}
\put(-12,236){\makebox(0,0)[l]{$\mu/|\varepsilon_{b}|$}}
\put(-12,1159){\makebox(0,0)[l]{$\mu/\epsilon_{F}$}}
\put(201,110){\special{em:moveto}}
\put(226,110){\special{em:lineto}}
\put(252,110){\special{em:lineto}}
\put(277,110){\special{em:lineto}}
\put(302,110){\special{em:lineto}}
\put(327,110){\special{em:lineto}}
\put(352,110){\special{em:lineto}}
\put(378,110){\special{em:lineto}}
\put(403,110){\special{em:lineto}}
\put(428,110){\special{em:lineto}}
\put(453,111){\special{em:lineto}}
\put(478,111){\special{em:lineto}}
\put(504,112){\special{em:lineto}}
\put(529,112){\special{em:lineto}}
\put(554,113){\special{em:lineto}}
\put(579,114){\special{em:lineto}}
\put(604,115){\special{em:lineto}}
\put(630,116){\special{em:lineto}}
\put(655,118){\special{em:lineto}}
\put(680,119){\special{em:lineto}}
\put(705,121){\special{em:lineto}}
\put(730,122){\special{em:lineto}}
\put(756,124){\special{em:lineto}}
\put(781,126){\special{em:lineto}}
\put(806,128){\special{em:lineto}}
\put(831,114){\special{em:lineto}}
\put(856,90){\special{em:lineto}}
\put(879,68){\special{em:lineto}}
\put(201,236){\special{em:moveto}}
\put(226,236){\special{em:lineto}}
\put(252,236){\special{em:lineto}}
\put(277,236){\special{em:lineto}}
\put(302,236){\special{em:lineto}}
\put(327,236){\special{em:lineto}}
\put(352,236){\special{em:lineto}}
\put(378,237){\special{em:lineto}}
\put(403,237){\special{em:lineto}}
\put(428,238){\special{em:lineto}}
\put(453,240){\special{em:lineto}}
\put(478,241){\special{em:lineto}}
\put(504,243){\special{em:lineto}}
\put(529,246){\special{em:lineto}}
\put(554,248){\special{em:lineto}}
\put(579,251){\special{em:lineto}}
\put(604,255){\special{em:lineto}}
\put(630,258){\special{em:lineto}}
\put(655,262){\special{em:lineto}}
\put(680,266){\special{em:lineto}}
\put(705,271){\special{em:lineto}}
\put(730,275){\special{em:lineto}}
\put(756,280){\special{em:lineto}}
\put(781,285){\special{em:lineto}}
\put(806,291){\special{em:lineto}}
\put(831,273){\special{em:lineto}}
\put(856,246){\special{em:lineto}}
\put(882,219){\special{em:lineto}}
\put(907,192){\special{em:lineto}}
\put(932,164){\special{em:lineto}}
\put(957,135){\special{em:lineto}}
\put(982,106){\special{em:lineto}}
\put(1008,77){\special{em:lineto}}
\put(1015,68){\special{em:lineto}}
\put(201,446){\special{em:moveto}}
\put(226,446){\special{em:lineto}}
\put(252,446){\special{em:lineto}}
\put(277,446){\special{em:lineto}}
\put(302,446){\special{em:lineto}}
\put(327,446){\special{em:lineto}}
\put(352,447){\special{em:lineto}}
\put(378,448){\special{em:lineto}}
\put(403,449){\special{em:lineto}}
\put(428,451){\special{em:lineto}}
\put(453,454){\special{em:lineto}}
\put(478,457){\special{em:lineto}}
\put(504,461){\special{em:lineto}}
\put(529,465){\special{em:lineto}}
\put(554,470){\special{em:lineto}}
\put(579,475){\special{em:lineto}}
\put(604,481){\special{em:lineto}}
\put(630,487){\special{em:lineto}}
\put(655,502){\special{em:lineto}}
\put(680,518){\special{em:lineto}}
\put(705,534){\special{em:lineto}}
\put(730,552){\special{em:lineto}}
\put(756,570){\special{em:lineto}}
\put(781,589){\special{em:lineto}}
\put(806,609){\special{em:lineto}}
\put(831,565){\special{em:lineto}}
\put(856,505){\special{em:lineto}}
\put(882,467){\special{em:lineto}}
\put(907,439){\special{em:lineto}}
\put(932,410){\special{em:lineto}}
\put(957,380){\special{em:lineto}}
\put(982,349){\special{em:lineto}}
\put(1008,318){\special{em:lineto}}
\put(1033,287){\special{em:lineto}}
\put(1058,254){\special{em:lineto}}
\put(1083,222){\special{em:lineto}}
\put(1108,188){\special{em:lineto}}
\put(1134,154){\special{em:lineto}}
\put(1159,120){\special{em:lineto}}
\put(1184,85){\special{em:lineto}}
\put(1196,68){\special{em:lineto}}
\put(201,628){\special{em:moveto}}
\put(226,628){\special{em:lineto}}
\put(252,628){\special{em:lineto}}
\put(277,628){\special{em:lineto}}
\put(302,628){\special{em:lineto}}
\put(327,628){\special{em:lineto}}
\put(352,629){\special{em:lineto}}
\put(378,631){\special{em:lineto}}
\put(403,634){\special{em:lineto}}
\put(428,638){\special{em:lineto}}
\put(453,644){\special{em:lineto}}
\put(478,650){\special{em:lineto}}
\put(504,658){\special{em:lineto}}
\put(529,666){\special{em:lineto}}
\put(554,676){\special{em:lineto}}
\put(579,687){\special{em:lineto}}
\put(604,698){\special{em:lineto}}
\put(630,711){\special{em:lineto}}
\put(655,724){\special{em:lineto}}
\put(680,738){\special{em:lineto}}
\put(705,753){\special{em:lineto}}
\put(730,769){\special{em:lineto}}
\put(756,785){\special{em:lineto}}
\put(781,802){\special{em:lineto}}
\put(806,820){\special{em:lineto}}
\put(831,792){\special{em:lineto}}
\put(856,747){\special{em:lineto}}
\put(882,700){\special{em:lineto}}
\put(907,653){\special{em:lineto}}
\put(932,604){\special{em:lineto}}
\put(957,554){\special{em:lineto}}
\put(982,503){\special{em:lineto}}
\put(1008,466){\special{em:lineto}}
\put(1033,434){\special{em:lineto}}
\put(1058,401){\special{em:lineto}}
\put(1083,368){\special{em:lineto}}
\put(1108,334){\special{em:lineto}}
\put(1134,300){\special{em:lineto}}
\put(1159,265){\special{em:lineto}}
\put(1184,229){\special{em:lineto}}
\put(1209,193){\special{em:lineto}}
\put(1234,156){\special{em:lineto}}
\put(1260,119){\special{em:lineto}}
\put(1285,81){\special{em:lineto}}
\put(1294,68){\special{em:lineto}}
\put(201,907){\special{em:moveto}}
\put(226,907){\special{em:lineto}}
\put(252,907){\special{em:lineto}}
\put(277,907){\special{em:lineto}}
\put(302,907){\special{em:lineto}}
\put(327,908){\special{em:lineto}}
\put(352,909){\special{em:lineto}}
\put(378,910){\special{em:lineto}}
\put(403,913){\special{em:lineto}}
\put(428,916){\special{em:lineto}}
\put(453,920){\special{em:lineto}}
\put(478,926){\special{em:lineto}}
\put(504,932){\special{em:lineto}}
\put(529,939){\special{em:lineto}}
\put(554,947){\special{em:lineto}}
\put(579,956){\special{em:lineto}}
\put(604,965){\special{em:lineto}}
\put(630,976){\special{em:lineto}}
\put(655,986){\special{em:lineto}}
\put(680,998){\special{em:lineto}}
\put(705,1010){\special{em:lineto}}
\put(730,1023){\special{em:lineto}}
\put(756,1036){\special{em:lineto}}
\put(781,1050){\special{em:lineto}}
\put(806,1064){\special{em:lineto}}
\put(831,1046){\special{em:lineto}}
\put(856,1020){\special{em:lineto}}
\put(882,992){\special{em:lineto}}
\put(907,964){\special{em:lineto}}
\put(932,935){\special{em:lineto}}
\put(957,905){\special{em:lineto}}
\put(982,874){\special{em:lineto}}
\put(1008,843){\special{em:lineto}}
\put(1033,810){\special{em:lineto}}
\put(1058,777){\special{em:lineto}}
\put(1083,743){\special{em:lineto}}
\put(1108,708){\special{em:lineto}}
\put(1134,673){\special{em:lineto}}
\put(1159,637){\special{em:lineto}}
\put(1184,600){\special{em:lineto}}
\put(1209,562){\special{em:lineto}}
\put(1234,524){\special{em:lineto}}
\put(1260,485){\special{em:lineto}}
\put(1285,446){\special{em:lineto}}
\put(1310,406){\special{em:lineto}}
\put(1335,365){\special{em:lineto}}
\put(1360,324){\special{em:lineto}}
\put(1386,282){\special{em:lineto}}
\put(1411,240){\special{em:lineto}}
\put(1436,197){\special{em:lineto}}
\put(201,1117){\special{em:moveto}}
\put(226,1117){\special{em:lineto}}
\put(252,1117){\special{em:lineto}}
\put(277,1117){\special{em:lineto}}
\put(302,1117){\special{em:lineto}}
\put(327,1117){\special{em:lineto}}
\put(352,1118){\special{em:lineto}}
\put(378,1119){\special{em:lineto}}
\put(403,1121){\special{em:lineto}}
\put(428,1123){\special{em:lineto}}
\put(453,1126){\special{em:lineto}}
\put(478,1129){\special{em:lineto}}
\put(504,1134){\special{em:lineto}}
\put(529,1139){\special{em:lineto}}
\put(554,1145){\special{em:lineto}}
\put(579,1151){\special{em:lineto}}
\put(604,1158){\special{em:lineto}}
\put(630,1165){\special{em:lineto}}
\put(655,1173){\special{em:lineto}}
\put(680,1182){\special{em:lineto}}
\put(705,1191){\special{em:lineto}}
\put(730,1200){\special{em:lineto}}
\put(756,1210){\special{em:lineto}}
\put(781,1220){\special{em:lineto}}
\put(806,1230){\special{em:lineto}}
\put(831,1220){\special{em:lineto}}
\put(856,1208){\special{em:lineto}}
\put(882,1195){\special{em:lineto}}
\put(907,1182){\special{em:lineto}}
\put(932,1168){\special{em:lineto}}
\put(957,1154){\special{em:lineto}}
\put(982,1139){\special{em:lineto}}
\put(1008,1124){\special{em:lineto}}
\put(1033,1108){\special{em:lineto}}
\put(1058,1092){\special{em:lineto}}
\put(1083,1075){\special{em:lineto}}
\put(1108,1058){\special{em:lineto}}
\put(1134,1040){\special{em:lineto}}
\put(1159,1022){\special{em:lineto}}
\put(1184,1003){\special{em:lineto}}
\put(1209,984){\special{em:lineto}}
\put(1234,964){\special{em:lineto}}
\put(1260,944){\special{em:lineto}}
\put(1285,924){\special{em:lineto}}
\put(1310,903){\special{em:lineto}}
\put(1335,882){\special{em:lineto}}
\put(1360,860){\special{em:lineto}}
\put(1386,838){\special{em:lineto}}
\put(1411,816){\special{em:lineto}}
\put(1436,793){\special{em:lineto}}
\put(201,1243){\special{em:moveto}}
\put(226,1243){\special{em:lineto}}
\put(252,1243){\special{em:lineto}}
\put(277,1243){\special{em:lineto}}
\put(302,1243){\special{em:lineto}}
\put(327,1243){\special{em:lineto}}
\put(352,1243){\special{em:lineto}}
\put(378,1244){\special{em:lineto}}
\put(403,1244){\special{em:lineto}}
\put(428,1246){\special{em:lineto}}
\put(453,1247){\special{em:lineto}}
\put(478,1249){\special{em:lineto}}
\put(504,1251){\special{em:lineto}}
\put(529,1254){\special{em:lineto}}
\put(554,1257){\special{em:lineto}}
\put(579,1260){\special{em:lineto}}
\put(604,1264){\special{em:lineto}}
\put(630,1268){\special{em:lineto}}
\put(655,1273){\special{em:lineto}}
\put(680,1278){\special{em:lineto}}
\put(705,1283){\special{em:lineto}}
\put(730,1289){\special{em:lineto}}
\put(756,1295){\special{em:lineto}}
\put(781,1301){\special{em:lineto}}
\put(806,1307){\special{em:lineto}}
\put(831,1305){\special{em:lineto}}
\put(856,1301){\special{em:lineto}}
\put(882,1297){\special{em:lineto}}
\put(907,1294){\special{em:lineto}}
\put(932,1289){\special{em:lineto}}
\put(957,1285){\special{em:lineto}}
\put(982,1280){\special{em:lineto}}
\put(1008,1275){\special{em:lineto}}
\put(1033,1270){\special{em:lineto}}
\put(1058,1264){\special{em:lineto}}
\put(1083,1259){\special{em:lineto}}
\put(1108,1253){\special{em:lineto}}
\put(1134,1247){\special{em:lineto}}
\put(1159,1240){\special{em:lineto}}
\put(1184,1233){\special{em:lineto}}
\put(1209,1226){\special{em:lineto}}
\put(1234,1219){\special{em:lineto}}
\put(1260,1212){\special{em:lineto}}
\put(1285,1204){\special{em:lineto}}
\put(1310,1196){\special{em:lineto}}
\put(1335,1188){\special{em:lineto}}
\put(1360,1180){\special{em:lineto}}
\put(1386,1171){\special{em:lineto}}
\put(1411,1163){\special{em:lineto}}
\put(1436,1154){\special{em:lineto}}
\sbox{\plotpoint}{\rule[-0.400pt]{0.800pt}{0.800pt}}%
\special{em:linewidth 0.8pt}%
\put(176,68){\special{em:moveto}}
\put(242,118){\special{em:lineto}}
\put(281,169){\special{em:lineto}}
\put(312,219){\special{em:lineto}}
\put(338,270){\special{em:lineto}}
\put(360,321){\special{em:lineto}}
\put(380,372){\special{em:lineto}}
\put(397,423){\special{em:lineto}}
\put(413,475){\special{em:lineto}}
\put(426,561){\special{em:lineto}}
\put(438,641){\special{em:lineto}}
\put(449,706){\special{em:lineto}}
\put(459,761){\special{em:lineto}}
\put(468,808){\special{em:lineto}}
\put(477,847){\special{em:lineto}}
\put(484,882){\special{em:lineto}}
\put(491,912){\special{em:lineto}}
\put(498,939){\special{em:lineto}}
\put(504,963){\special{em:lineto}}
\put(510,984){\special{em:lineto}}
\put(515,1003){\special{em:lineto}}
\put(520,1020){\special{em:lineto}}
\put(525,1036){\special{em:lineto}}
\put(529,1050){\special{em:lineto}}
\put(533,1064){\special{em:lineto}}
\put(537,1076){\special{em:lineto}}
\put(541,1087){\special{em:lineto}}
\put(545,1097){\special{em:lineto}}
\put(548,1106){\special{em:lineto}}
\put(552,1115){\special{em:lineto}}
\put(555,1124){\special{em:lineto}}
\put(558,1131){\special{em:lineto}}
\put(561,1138){\special{em:lineto}}
\put(564,1145){\special{em:lineto}}
\put(566,1151){\special{em:lineto}}
\put(569,1157){\special{em:lineto}}
\put(571,1163){\special{em:lineto}}
\put(574,1168){\special{em:lineto}}
\put(576,1173){\special{em:lineto}}
\put(578,1178){\special{em:lineto}}
\put(580,1183){\special{em:lineto}}
\put(582,1187){\special{em:lineto}}
\put(584,1191){\special{em:lineto}}
\put(586,1195){\special{em:lineto}}
\put(588,1198){\special{em:lineto}}
\put(590,1202){\special{em:lineto}}
\put(592,1205){\special{em:lineto}}
\put(594,1208){\special{em:lineto}}
\put(595,1211){\special{em:lineto}}
\put(597,1214){\special{em:lineto}}
\put(599,1217){\special{em:lineto}}
\put(600,1220){\special{em:lineto}}
\put(602,1222){\special{em:lineto}}
\put(603,1225){\special{em:lineto}}
\put(605,1227){\special{em:lineto}}
\put(606,1229){\special{em:lineto}}
\put(607,1232){\special{em:lineto}}
\put(609,1234){\special{em:lineto}}
\put(610,1236){\special{em:lineto}}
\put(611,1238){\special{em:lineto}}
\put(613,1240){\special{em:lineto}}
\put(614,1241){\special{em:lineto}}
\put(615,1243){\special{em:lineto}}
\put(616,1245){\special{em:lineto}}
\put(617,1246){\special{em:lineto}}
\put(619,1248){\special{em:lineto}}
\put(620,1250){\special{em:lineto}}
\put(621,1251){\special{em:lineto}}
\put(622,1252){\special{em:lineto}}
\put(623,1254){\special{em:lineto}}
\put(624,1255){\special{em:lineto}}
\put(625,1256){\special{em:lineto}}
\put(626,1258){\special{em:lineto}}
\put(627,1259){\special{em:lineto}}
\put(628,1260){\special{em:lineto}}
\put(629,1261){\special{em:lineto}}
\put(630,1262){\special{em:lineto}}
\put(630,1263){\special{em:lineto}}
\put(631,1264){\special{em:lineto}}
\put(632,1266){\special{em:lineto}}
\put(633,1266){\special{em:lineto}}
\put(634,1267){\special{em:lineto}}
\put(635,1268){\special{em:lineto}}
\put(636,1269){\special{em:lineto}}
\put(636,1270){\special{em:lineto}}
\put(637,1271){\special{em:lineto}}
\put(638,1272){\special{em:lineto}}
\put(639,1273){\special{em:lineto}}
\put(639,1274){\special{em:lineto}}
\put(640,1274){\special{em:lineto}}
\put(641,1275){\special{em:lineto}}
\put(642,1276){\special{em:lineto}}
\put(642,1277){\special{em:lineto}}
\put(643,1277){\special{em:lineto}}
\put(644,1278){\special{em:lineto}}
\put(644,1279){\special{em:lineto}}
\put(645,1279){\special{em:lineto}}
\put(646,1280){\special{em:lineto}}
\put(646,1281){\special{em:lineto}}
\put(647,1281){\special{em:lineto}}
\put(648,1282){\special{em:lineto}}
\put(806,68){\special{em:moveto}}
\put(806,1327){\special{em:lineto}}
\end{picture}

\caption{$\mu(T)$ for different values of
$\epsilon_{F}/|\varepsilon_{b}|$:
1 --- 0.05; 2 --- 0.2; 3 --- 0.45; 4 --- 0.6; 5 --- 1; 6 --- 2; 7 --- 5.
(For $\mu > 0$ and $\mu < 0$ the chemical potential was scaled to
$\epsilon_{F}$ and $|\varepsilon_{b}|$, respectively.) The thick lines
bound regions of the BKT, anomalous normal and normal phases.}
\end{figure}
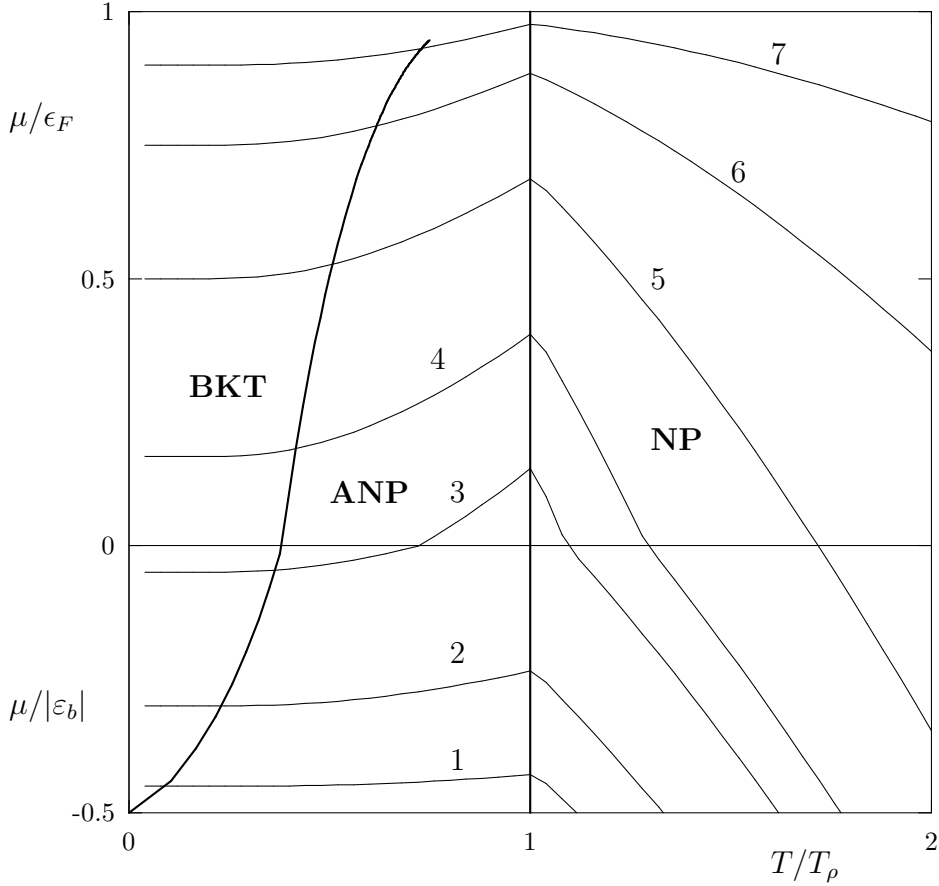

\noindent
{\bf c)} Figure~2 shows the values $n_{f}$ for which $\mu$ differs
substantially from $\epsilon_{F}$ and, in other words, the Landau
Fermi-liquid theory becomes inapplicable for metals with low or
intermediate carrier density. As expected, the kink $\mu$ at
$T = T_{\rho}$, experiments on observation which were discussed in
\cite{Marel} and have been interpreted for the 1-2-3 cuprate
\cite{Dotsenko}, becomes increasingly less pronounced as $\epsilon_{F}$
increases. But in the present case it is interesting that in the
approximation employed it happens at the normal-anomalous phases
boundary or before superconductivity really appears. Therefore it
would be of great interest to perform experiments which would reveal
the temperature dependence $\mu(T)$ especially for strongly anisotropic
and relatively weakly doped cuprates.

\noindent
{\bf d)} It follows from curve 3 in Fig.~2 that the transition
(change in sign of $\mu$) from local to Cooper pairs is possible not
only as $\epsilon_{F}$ increases, which is more or less obvious, but
also (for some $n_{f}$) as $T$ increases.

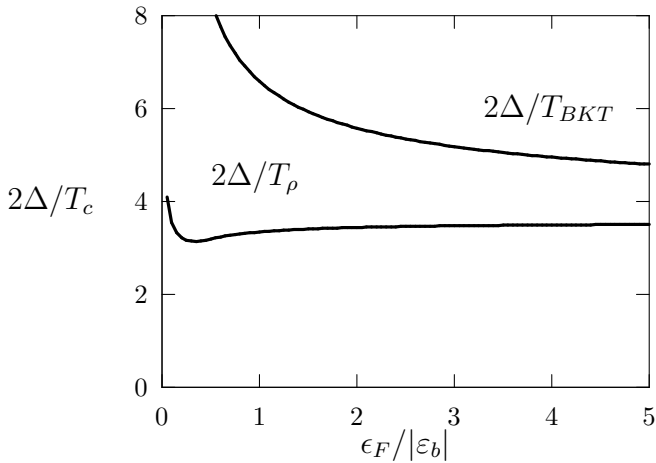
\begin{figure}[ht]
\setlength{\unitlength}{0.240900pt}
\ifx\plotpoint\undefined\newsavebox{\plotpoint}\fi
\sbox{\plotpoint}{\rule[-0.200pt]{0.400pt}{0.400pt}}%
\special{em:linewidth 0.4pt}%
\begin{picture}(1049,720)(0,0)
\font\gnuplot=cmr10 at 10pt
\gnuplot
\put(220,113){\special{em:moveto}}
\put(985,113){\special{em:lineto}}
\put(220,113){\special{em:moveto}}
\put(220,697){\special{em:lineto}}
\put(220,113){\special{em:moveto}}
\put(240,113){\special{em:lineto}}
\put(985,113){\special{em:moveto}}
\put(965,113){\special{em:lineto}}
\put(198,113){\makebox(0,0)[r]{0}}
\put(220,259){\special{em:moveto}}
\put(240,259){\special{em:lineto}}
\put(985,259){\special{em:moveto}}
\put(965,259){\special{em:lineto}}
\put(198,259){\makebox(0,0)[r]{2}}
\put(220,405){\special{em:moveto}}
\put(240,405){\special{em:lineto}}
\put(985,405){\special{em:moveto}}
\put(965,405){\special{em:lineto}}
\put(198,405){\makebox(0,0)[r]{4}}
\put(220,551){\special{em:moveto}}
\put(240,551){\special{em:lineto}}
\put(985,551){\special{em:moveto}}
\put(965,551){\special{em:lineto}}
\put(198,551){\makebox(0,0)[r]{6}}
\put(220,697){\special{em:moveto}}
\put(240,697){\special{em:lineto}}
\put(985,697){\special{em:moveto}}
\put(965,697){\special{em:lineto}}
\put(198,697){\makebox(0,0)[r]{8}}
\put(220,113){\special{em:moveto}}
\put(220,133){\special{em:lineto}}
\put(220,697){\special{em:moveto}}
\put(220,677){\special{em:lineto}}
\put(220,68){\makebox(0,0){0}}
\put(373,113){\special{em:moveto}}
\put(373,133){\special{em:lineto}}
\put(373,697){\special{em:moveto}}
\put(373,677){\special{em:lineto}}
\put(373,68){\makebox(0,0){1}}
\put(526,113){\special{em:moveto}}
\put(526,133){\special{em:lineto}}
\put(526,697){\special{em:moveto}}
\put(526,677){\special{em:lineto}}
\put(526,68){\makebox(0,0){2}}
\put(679,113){\special{em:moveto}}
\put(679,133){\special{em:lineto}}
\put(679,697){\special{em:moveto}}
\put(679,677){\special{em:lineto}}
\put(679,68){\makebox(0,0){3}}
\put(832,113){\special{em:moveto}}
\put(832,133){\special{em:lineto}}
\put(832,697){\special{em:moveto}}
\put(832,677){\special{em:lineto}}
\put(832,68){\makebox(0,0){4}}
\put(985,113){\special{em:moveto}}
\put(985,133){\special{em:lineto}}
\put(985,697){\special{em:moveto}}
\put(985,677){\special{em:lineto}}
\put(985,68){\makebox(0,0){5}}
\put(220,113){\special{em:moveto}}
\put(985,113){\special{em:lineto}}
\put(985,697){\special{em:lineto}}
\put(220,697){\special{em:lineto}}
\put(220,113){\special{em:lineto}}
\put(45,405){\makebox(0,0){$2\Delta / T_c$}}
\put(602,23){\makebox(0,0){$\epsilon_{F}/|\varepsilon_{b}|$}}
\put(297,442){\makebox(0,0)[l]{$2 \Delta/T_{\rho}$}}
\put(725,551){\makebox(0,0)[l]{$2 \Delta/T_{BKT}$}}
\sbox{\plotpoint}{\rule[-0.600pt]{1.200pt}{1.200pt}}%
\special{em:linewidth 1.2pt}%
\put(228,412){\special{em:moveto}}
\put(235,372){\special{em:lineto}}
\put(243,356){\special{em:lineto}}
\put(251,348){\special{em:lineto}}
\put(258,344){\special{em:lineto}}
\put(266,343){\special{em:lineto}}
\put(274,342){\special{em:lineto}}
\put(281,343){\special{em:lineto}}
\put(289,344){\special{em:lineto}}
\put(297,346){\special{em:lineto}}
\put(304,348){\special{em:lineto}}
\put(312,349){\special{em:lineto}}
\put(319,351){\special{em:lineto}}
\put(327,352){\special{em:lineto}}
\put(335,353){\special{em:lineto}}
\put(342,354){\special{em:lineto}}
\put(350,355){\special{em:lineto}}
\put(358,356){\special{em:lineto}}
\put(365,356){\special{em:lineto}}
\put(373,357){\special{em:lineto}}
\put(381,358){\special{em:lineto}}
\put(388,358){\special{em:lineto}}
\put(396,359){\special{em:lineto}}
\put(404,359){\special{em:lineto}}
\put(411,360){\special{em:lineto}}
\put(419,360){\special{em:lineto}}
\put(427,361){\special{em:lineto}}
\put(434,361){\special{em:lineto}}
\put(442,361){\special{em:lineto}}
\put(450,362){\special{em:lineto}}
\put(457,362){\special{em:lineto}}
\put(465,362){\special{em:lineto}}
\put(472,363){\special{em:lineto}}
\put(480,363){\special{em:lineto}}
\put(488,363){\special{em:lineto}}
\put(495,363){\special{em:lineto}}
\put(503,364){\special{em:lineto}}
\put(511,364){\special{em:lineto}}
\put(518,364){\special{em:lineto}}
\put(526,364){\special{em:lineto}}
\put(534,364){\special{em:lineto}}
\put(541,365){\special{em:lineto}}
\put(549,365){\special{em:lineto}}
\put(557,365){\special{em:lineto}}
\put(564,365){\special{em:lineto}}
\put(572,365){\special{em:lineto}}
\put(580,365){\special{em:lineto}}
\put(587,365){\special{em:lineto}}
\put(595,366){\special{em:lineto}}
\put(603,366){\special{em:lineto}}
\put(610,366){\special{em:lineto}}
\put(618,366){\special{em:lineto}}
\put(625,366){\special{em:lineto}}
\put(633,366){\special{em:lineto}}
\put(641,366){\special{em:lineto}}
\put(648,366){\special{em:lineto}}
\put(656,367){\special{em:lineto}}
\put(664,367){\special{em:lineto}}
\put(671,367){\special{em:lineto}}
\put(679,367){\special{em:lineto}}
\put(687,367){\special{em:lineto}}
\put(694,367){\special{em:lineto}}
\put(702,367){\special{em:lineto}}
\put(710,367){\special{em:lineto}}
\put(717,367){\special{em:lineto}}
\put(725,367){\special{em:lineto}}
\put(733,367){\special{em:lineto}}
\put(740,367){\special{em:lineto}}
\put(748,367){\special{em:lineto}}
\put(756,368){\special{em:lineto}}
\put(763,368){\special{em:lineto}}
\put(771,368){\special{em:lineto}}
\put(778,368){\special{em:lineto}}
\put(786,368){\special{em:lineto}}
\put(794,368){\special{em:lineto}}
\put(801,368){\special{em:lineto}}
\put(809,368){\special{em:lineto}}
\put(817,368){\special{em:lineto}}
\put(824,368){\special{em:lineto}}
\put(832,368){\special{em:lineto}}
\put(840,368){\special{em:lineto}}
\put(847,368){\special{em:lineto}}
\put(855,368){\special{em:lineto}}
\put(863,368){\special{em:lineto}}
\put(870,368){\special{em:lineto}}
\put(878,368){\special{em:lineto}}
\put(886,368){\special{em:lineto}}
\put(893,368){\special{em:lineto}}
\put(901,369){\special{em:lineto}}
\put(909,369){\special{em:lineto}}
\put(916,369){\special{em:lineto}}
\put(924,369){\special{em:lineto}}
\put(931,369){\special{em:lineto}}
\put(939,369){\special{em:lineto}}
\put(947,369){\special{em:lineto}}
\put(954,369){\special{em:lineto}}
\put(962,369){\special{em:lineto}}
\put(970,369){\special{em:lineto}}
\put(977,369){\special{em:lineto}}
\put(985,369){\special{em:lineto}}
\put(305,697){\special{em:moveto}}
\put(312,680){\special{em:lineto}}
\put(319,664){\special{em:lineto}}
\put(327,650){\special{em:lineto}}
\put(335,638){\special{em:lineto}}
\put(342,627){\special{em:lineto}}
\put(350,618){\special{em:lineto}}
\put(358,609){\special{em:lineto}}
\put(365,601){\special{em:lineto}}
\put(373,594){\special{em:lineto}}
\put(381,587){\special{em:lineto}}
\put(388,581){\special{em:lineto}}
\put(396,576){\special{em:lineto}}
\put(404,571){\special{em:lineto}}
\put(411,566){\special{em:lineto}}
\put(419,561){\special{em:lineto}}
\put(427,557){\special{em:lineto}}
\put(434,553){\special{em:lineto}}
\put(442,550){\special{em:lineto}}
\put(450,546){\special{em:lineto}}
\put(457,543){\special{em:lineto}}
\put(465,540){\special{em:lineto}}
\put(472,537){\special{em:lineto}}
\put(480,534){\special{em:lineto}}
\put(488,532){\special{em:lineto}}
\put(495,529){\special{em:lineto}}
\put(503,527){\special{em:lineto}}
\put(511,525){\special{em:lineto}}
\put(518,522){\special{em:lineto}}
\put(526,520){\special{em:lineto}}
\put(534,518){\special{em:lineto}}
\put(541,516){\special{em:lineto}}
\put(549,515){\special{em:lineto}}
\put(557,513){\special{em:lineto}}
\put(564,511){\special{em:lineto}}
\put(572,509){\special{em:lineto}}
\put(580,508){\special{em:lineto}}
\put(587,506){\special{em:lineto}}
\put(595,505){\special{em:lineto}}
\put(603,503){\special{em:lineto}}
\put(610,502){\special{em:lineto}}
\put(618,501){\special{em:lineto}}
\put(625,499){\special{em:lineto}}
\put(633,498){\special{em:lineto}}
\put(641,497){\special{em:lineto}}
\put(648,496){\special{em:lineto}}
\put(656,495){\special{em:lineto}}
\put(664,493){\special{em:lineto}}
\put(671,492){\special{em:lineto}}
\put(679,491){\special{em:lineto}}
\put(687,490){\special{em:lineto}}
\put(694,489){\special{em:lineto}}
\put(702,488){\special{em:lineto}}
\put(710,487){\special{em:lineto}}
\put(717,486){\special{em:lineto}}
\put(725,485){\special{em:lineto}}
\put(733,485){\special{em:lineto}}
\put(740,484){\special{em:lineto}}
\put(748,483){\special{em:lineto}}
\put(756,482){\special{em:lineto}}
\put(763,481){\special{em:lineto}}
\put(771,480){\special{em:lineto}}
\put(778,480){\special{em:lineto}}
\put(786,479){\special{em:lineto}}
\put(794,478){\special{em:lineto}}
\put(801,477){\special{em:lineto}}
\put(809,477){\special{em:lineto}}
\put(817,476){\special{em:lineto}}
\put(824,475){\special{em:lineto}}
\put(832,475){\special{em:lineto}}
\put(840,474){\special{em:lineto}}
\put(847,473){\special{em:lineto}}
\put(855,473){\special{em:lineto}}
\put(863,472){\special{em:lineto}}
\put(870,472){\special{em:lineto}}
\put(878,471){\special{em:lineto}}
\put(886,470){\special{em:lineto}}
\put(893,470){\special{em:lineto}}
\put(901,469){\special{em:lineto}}
\put(909,469){\special{em:lineto}}
\put(916,468){\special{em:lineto}}
\put(924,467){\special{em:lineto}}
\put(931,467){\special{em:lineto}}
\put(939,466){\special{em:lineto}}
\put(947,466){\special{em:lineto}}
\put(954,465){\special{em:lineto}}
\put(962,465){\special{em:lineto}}
\put(970,464){\special{em:lineto}}
\put(977,464){\special{em:lineto}}
\put(985,464){\special{em:lineto}}
\end{picture}

\caption{$2 \Delta/T_{BKT}$ and $2 \Delta/T_{\rho}$ versus the
free fermion density.}
\end{figure}

\noindent
{\bf e)} Finally the calculations showed (see Fig.~3) that the ratio
$2 \Delta/T_{BKT}$ is always greater than 4.4.
The value $2 \Delta/T_{\rho} (= 2\Delta/T_{c}^{MF})$ is, however,
somewhat lower and reaches the BCS theory limit of 3.52 only for
$\epsilon_{F} \gg |\varepsilon_{b}|$.
It is interesting that this concentration behaviour is consistent
with numerous measurements of this ratio in HTSC
\cite{Devereaux,Kendziora}. Note that the divergencies
of $2 \Delta/T_{BKT}$ and $2 \Delta/T_{\rho}$ at $\epsilon_{F} \to 0$
are directly related with the definition (\ref{energy.gap}).

Discussing the phase diagram obtained it can be
emphasized that the qualitative ideas about the crucial importance
of phase fluctuations for underdoped HTSC was, to our
knowledge, first discussed by Emery and Kivelson \cite{Emery}.
These authors, starting from the experimental data of Uemura et al.
\cite{Uemura} and some others (see the Refs. cited in \cite{Emery}),
and from the well-known general observation
the superconducting state is characterized
by a complex order parameter, introduced the temperature $T_{\theta}$
at which phase order can occur. They have also argued that at
low superconducting carrier density and poor screening (i.e. in bad metals)
phase fluctuations become more significant than all other fluctuations, so
that the classical XY-model is suitable for underdoped cuprate oxides. When
$T_c \ll T_{\theta}^{max}$ ($T_{\theta}^{max}$ is the temperature at which
the phase order disappears if the disordering effects of all the other
degrees of freedom are ignored), phase fluctuations must be
relatively unimportant and the observable $T_c$ will be close to the
$T_c^{MF}$ predicted by BCS-Bogolyubov theory.  Otherwise, $T_c \simeq
T_{\theta}^{max} < T_c^{MF}$, and $T_c^{MF}$ is simply
the temperature at which local pairing occurs.

Undoubtedly, the approach developed in this Chapter is
self-consistent and a more-or-less complete (in the hydrodynamical
approximation) extension of the semiqualitative
results presented in \cite{Emery,Uemura}.
At the same time there is only one difference (but in our opinion an
essential one) between the behaviour of the function we obtain for
$T_c^{MF}(n_f) \equiv T_{\rho}(n_f)$ in the limit $n_f \to 0$ and
the function sketched for $T_c^{MF}$ in \cite{Emery}. It is
clear from the figure (see Fig.1) (and can be shown analytically)
that in our case the zero density limit is $T_{\rho}(0)=0$, in contrast
to Refs.  \cite{Emery,Uemura} where $T_c^{MF}$ grows when $n_f$ decreases.

On the other hand, the limit $T_{\rho}(n_f \to 0) \to 0$
(the same argument also applies to $T_{dissoc}$ in Section 2.3)
cannot be considered sufficiently regular due to the
strong increase, for small $n_f$, of the neutral order
parameter fluctuations which were not taken into account
in the approximation used. From the
physical point of view perhaps the most consistent limit for this,
extremely low (when $\mu<0$) fermion density, region is:
$T_{\rho}(n_f\to 0) \to T_P \approx 2 \sqrt{\rho^2+\mu^2}
\sim |\varepsilon_b|$ (see (\ref{energy.gap}) and Fig.2).

\subsection{"Spin-gap" behaviour in the anomalous normal phase}

          It would be very interesting to study how a nonzero value of
the neutral order parameter affects the observable properties of the
2D system. Does this really resemble the gap opening in the traditional
superconductors, except that it happens in the normal phase? Or,
in other words, does the pseudogap open?

           We shall demonstrate this phenomenon taking, as a case in
point, the paramagnetic susceptibility of the system
\footnote{That was done with V.P.~Gusynin.}.

           To study the system in the magnetic field $\mbox{\bf H}$
one has to add the paramagnetic term
\begin{equation}
{\cal H}_{PM} = - \mu_{B} H \left[
\psi^{\dagger}_{\uparrow}(\mbox{\bf r}) \psi_{\uparrow}(\mbox{\bf r}) -
\psi^{\dagger}_{\downarrow}(\mbox{\bf r}) \psi_{\downarrow}(\mbox{\bf r})
\right],
                        \label{Hamiltonian.magnetic}
\end{equation}
to the Hamiltonian (\ref{Hamiltonian.3D}). Here
$\mu_{B} = e \hbar/2mc$ is the Bohr magneton. Note that, using the isotropy
in the problem,  we chose the direction of field $\mbox{\bf H}$ to be
perpendicular to the plane containing the vectors $\mbox{\bf r}$. (Recall
that in this Chapter $d = 2$ and $\mbox{\bf r}$ is the 2D vector.)

             It is a very simple matter to rewrite ${\cal H}_{PM}$ in
the Nambu variables (\ref{Nambu.variables}), namely
\begin{equation}
{\cal H}_{PM} = - \mu_{B} H
\Psi^{\dagger}(\mbox{\bf r}) \Psi(\mbox{\bf r}) .
                       \label{Hamiltonian.magnetic.Nambu}
\end{equation}
Then, adding the corresponding term to the equation
(\ref{Green.fermion.modulus}) for the neutral fermion Green function,
it is easy to obtain that of in the momentum representation
(compare with(\ref{B2}))
\begin{eqnarray}
{\cal G}(i \omega_{n}, \mbox{\bf k}, H) & = &
\frac{1}
{(i \omega_{n} + \mu_{B} H) \hat I -
\tau_{3} \xi(\mbox{\bf k}) + \tau_{1} \rho}
\nonumber                             \\
& = & \frac{(i \omega_{n} + \mu_{B} H) \hat I +
\tau_{3} \xi(\mbox{\bf k}) - \tau_{1} \rho}
{(i \omega_{n} + \mu_{B} H)^{2} - \xi^{2} (\mbox{\bf k}) - \rho^{2}}.
                        \label{Green.momentum.magnetic}
\end{eqnarray}

              The static paramagnetic susceptibility is expressed
through the magnetization
\begin{equation}
\chi(\mu, T, \rho) =  \left. \frac{\partial M(\mu, T, \rho, H)}
{\partial H} \right|_{H = 0},
                \label{susceptibility.definition}
\end{equation}
which in the mean-field approximation may be derived from the
effective potential
\begin{equation}
M(\mu, T, \rho, H) = - \frac{1}{v}
\frac{\partial \Omega_{pot}(v, \mu, T, \rho, H)}{\partial H}.
                         \label{magnetization.definition}
\end{equation}
Thus from (\ref{magnetization.definition}) one obtains
\footnote{Note that one should use the first part of
(\ref{Green.momentum.magnetic}) to get this formula.}
\begin{equation}
M(\mu, T, \rho, H) = \mu_{B} T \sum_{n = - \infty}^{\infty}
\int \frac{d \mbox{\bf k}}{(2 \pi)^{2}} \mbox{tr}
[{\cal G} (i \omega_{n}, \mbox{\bf k}, H) \hat I].
                              \label{magnetization}
\end{equation}
Then using the definition (\ref{susceptibility.definition}) one
arrives at
\begin{equation}
\chi(\mu, T, \rho) = \mu_{B}^{2}
\int \frac{d \mbox{\bf k}}{(2\pi)^{2}}
2 T \sum _{n = - \infty}^{\infty}
\frac{\xi^{2}(\mbox{\bf k}) + \rho^{2} - \omega_{n}^{2}}
{[\omega_{n}^{2} + \xi^{2}(\mbox{\bf k}) + \rho^{2}]^{2}}.
                      \label{susceptibility.sum}
\end{equation}
The sum in (\ref{susceptibility.sum}) is calculated in Appendix~C
(see the equation (\ref{C5})) and, thus, we obtain the final result
\begin{equation}
\chi(\mu, T, \rho) = \chi_{Pauli} \frac{1}{2}
\int_{-\mu/2T}^{\infty}
\frac{dx}{\cosh^{2} \sqrt{x^{2} + \frac{\ds \rho^{2}}{4T^{2}}}},
                            \label{susceptibility.final}
\end{equation}
where $\chi_{Pauli} \equiv \mu_{B}^{2} m/\pi$ is the Pauli
paramagnetic susceptibility for the 2D system.

           To study $\chi$ as a function of $T$ and $n_{f}$
(or $\epsilon_{F}$) the formula (\ref{susceptibility.final}) should
be used together with the equations (\ref{rho.bound}) and
(\ref{number.rho}).

           For the case of the normal phase ($\rho = 0$) one can
investigate the system analytically. Thus (\ref{susceptibility.final})
takes form
\begin{equation}
\chi(\mu, T, \rho =0) = \chi_{Pauli} \frac{1}{1 + \exp(-\mu/T)},
                   \label{susceptibility.NP.mu}
\end{equation}
while $\mu$ is already determined by (\ref{number.temperature.rho}).
This system has the solution
\begin{equation}
\chi(\epsilon_{F}, T, \rho = 0) = \chi_{Pauli}
[1 - \exp(-\epsilon_{F}/T)],
                        \label{susceptibility.NP}
\end{equation}
which coincides with that known from the literature \cite{Kvasnikov}.
\begin{figure}[ht]
\setlength{\unitlength}{0.240900pt}
\ifx\plotpoint\undefined\newsavebox{\plotpoint}\fi
\sbox{\plotpoint}{\rule[-0.200pt]{0.400pt}{0.400pt}}%
\special{em:linewidth 0.4pt}%
\begin{picture}(1049,720)(0,0)
\font\gnuplot=cmr10 at 10pt
\gnuplot
\put(176,68){\special{em:moveto}}
\put(985,68){\special{em:lineto}}
\put(176,68){\special{em:moveto}}
\put(176,697){\special{em:lineto}}
\put(176,68){\special{em:moveto}}
\put(196,68){\special{em:lineto}}
\put(985,68){\special{em:moveto}}
\put(965,68){\special{em:lineto}}
\put(154,68){\makebox(0,0)[r]{0}}
\put(176,225){\special{em:moveto}}
\put(196,225){\special{em:lineto}}
\put(985,225){\special{em:moveto}}
\put(965,225){\special{em:lineto}}
\put(154,225){\makebox(0,0)[r]{0.25}}
\put(176,383){\special{em:moveto}}
\put(196,383){\special{em:lineto}}
\put(985,383){\special{em:moveto}}
\put(965,383){\special{em:lineto}}
\put(154,383){\makebox(0,0)[r]{0.5}}
\put(176,540){\special{em:moveto}}
\put(196,540){\special{em:lineto}}
\put(985,540){\special{em:moveto}}
\put(965,540){\special{em:lineto}}
\put(154,540){\makebox(0,0)[r]{0.75}}
\put(176,697){\special{em:moveto}}
\put(196,697){\special{em:lineto}}
\put(985,697){\special{em:moveto}}
\put(965,697){\special{em:lineto}}
\put(154,697){\makebox(0,0)[r]{1}}
\put(176,68){\special{em:moveto}}
\put(176,88){\special{em:lineto}}
\put(176,697){\special{em:moveto}}
\put(176,677){\special{em:lineto}}
\put(176,23){\makebox(0,0){0}}
\put(266,68){\special{em:moveto}}
\put(266,88){\special{em:lineto}}
\put(266,697){\special{em:moveto}}
\put(266,677){\special{em:lineto}}
\put(266,23){\makebox(0,0){0.5}}
\put(356,68){\special{em:moveto}}
\put(356,88){\special{em:lineto}}
\put(356,697){\special{em:moveto}}
\put(356,677){\special{em:lineto}}
\put(356,23){\makebox(0,0){1}}
\put(446,68){\special{em:moveto}}
\put(446,88){\special{em:lineto}}
\put(446,697){\special{em:moveto}}
\put(446,677){\special{em:lineto}}
\put(446,23){\makebox(0,0){1.5}}
\put(536,68){\special{em:moveto}}
\put(536,88){\special{em:lineto}}
\put(536,697){\special{em:moveto}}
\put(536,677){\special{em:lineto}}
\put(536,23){\makebox(0,0){2}}
\put(625,68){\special{em:moveto}}
\put(625,88){\special{em:lineto}}
\put(625,697){\special{em:moveto}}
\put(625,677){\special{em:lineto}}
\put(625,23){\makebox(0,0){2.5}}
\put(715,68){\special{em:moveto}}
\put(715,88){\special{em:lineto}}
\put(715,697){\special{em:moveto}}
\put(715,677){\special{em:lineto}}
\put(715,23){\makebox(0,0){3}}
\put(805,68){\special{em:moveto}}
\put(805,88){\special{em:lineto}}
\put(805,697){\special{em:moveto}}
\put(805,677){\special{em:lineto}}
\put(805,23){\makebox(0,0){3.5}}
\put(895,68){\special{em:moveto}}
\put(895,88){\special{em:lineto}}
\put(895,697){\special{em:moveto}}
\put(895,677){\special{em:lineto}}
\put(895,23){\makebox(0,0){4}}
\put(985,68){\special{em:moveto}}
\put(985,88){\special{em:lineto}}
\put(985,697){\special{em:moveto}}
\put(985,677){\special{em:lineto}}
\put(985,23){\makebox(0,0){4.5}}
\put(176,68){\special{em:moveto}}
\put(985,68){\special{em:lineto}}
\put(985,697){\special{em:lineto}}
\put(176,697){\special{em:lineto}}
\put(176,68){\special{em:lineto}}
\put(715,150){\makebox(0,0)[l]{$1$}}
\put(464,206){\makebox(0,0)[l]{$2$}}
\put(446,301){\makebox(0,0)[l]{$3$}}
\put(679,540){\makebox(0,0)[l]{$4$}}
\put(643,603){\makebox(0,0)[l]{$5$}}
\put(607,659){\makebox(0,0)[l]{$6$}}
\put(214,445){\makebox(0,0)[l]{${\bf BKT}$}}
\put(805,118){\makebox(0,0)[l]{$T/T_{BKT}$}}
\put(183,634){\makebox(0,0)[l]{$\chi/ \chi_{Pauli}$}}
\sbox{\plotpoint}{\rule[-0.400pt]{0.800pt}{0.800pt}}%
\special{em:linewidth 0.8pt}%
\put(191,68){\special{em:moveto}}
\put(207,68){\special{em:lineto}}
\put(222,68){\special{em:lineto}}
\put(238,68){\special{em:lineto}}
\put(253,68){\special{em:lineto}}
\put(269,68){\special{em:lineto}}
\put(284,68){\special{em:lineto}}
\put(300,68){\special{em:lineto}}
\put(315,68){\special{em:lineto}}
\put(331,68){\special{em:lineto}}
\put(346,69){\special{em:lineto}}
\put(362,69){\special{em:lineto}}
\put(377,70){\special{em:lineto}}
\put(393,71){\special{em:lineto}}
\put(408,73){\special{em:lineto}}
\put(424,75){\special{em:lineto}}
\put(439,77){\special{em:lineto}}
\put(455,79){\special{em:lineto}}
\put(470,83){\special{em:lineto}}
\put(486,86){\special{em:lineto}}
\put(501,90){\special{em:lineto}}
\put(516,94){\special{em:lineto}}
\put(532,98){\special{em:lineto}}
\put(547,103){\special{em:lineto}}
\put(563,108){\special{em:lineto}}
\put(579,114){\special{em:lineto}}
\put(594,119){\special{em:lineto}}
\put(609,125){\special{em:lineto}}
\put(625,132){\special{em:lineto}}
\put(640,138){\special{em:lineto}}
\put(656,145){\special{em:lineto}}
\put(671,152){\special{em:lineto}}
\put(687,159){\special{em:lineto}}
\put(702,166){\special{em:lineto}}
\put(718,173){\special{em:lineto}}
\put(733,181){\special{em:lineto}}
\put(749,188){\special{em:lineto}}
\put(764,196){\special{em:lineto}}
\put(780,204){\special{em:lineto}}
\put(795,212){\special{em:lineto}}
\put(811,220){\special{em:lineto}}
\put(826,228){\special{em:lineto}}
\put(842,236){\special{em:lineto}}
\put(857,244){\special{em:lineto}}
\put(873,253){\special{em:lineto}}
\put(888,261){\special{em:lineto}}
\put(904,270){\special{em:lineto}}
\put(919,278){\special{em:lineto}}
\put(934,287){\special{em:lineto}}
\put(950,296){\special{em:lineto}}
\put(965,298){\special{em:lineto}}
\put(981,295){\special{em:lineto}}
\put(985,294){\special{em:lineto}}
\put(185,68){\special{em:moveto}}
\put(193,68){\special{em:lineto}}
\put(202,68){\special{em:lineto}}
\put(210,68){\special{em:lineto}}
\put(219,68){\special{em:lineto}}
\put(228,68){\special{em:lineto}}
\put(236,68){\special{em:lineto}}
\put(245,68){\special{em:lineto}}
\put(254,68){\special{em:lineto}}
\put(262,69){\special{em:lineto}}
\put(271,70){\special{em:lineto}}
\put(279,71){\special{em:lineto}}
\put(288,73){\special{em:lineto}}
\put(297,75){\special{em:lineto}}
\put(305,78){\special{em:lineto}}
\put(314,82){\special{em:lineto}}
\put(322,86){\special{em:lineto}}
\put(331,91){\special{em:lineto}}
\put(340,97){\special{em:lineto}}
\put(348,103){\special{em:lineto}}
\put(357,110){\special{em:lineto}}
\put(365,118){\special{em:lineto}}
\put(374,126){\special{em:lineto}}
\put(383,134){\special{em:lineto}}
\put(391,143){\special{em:lineto}}
\put(400,152){\special{em:lineto}}
\put(408,162){\special{em:lineto}}
\put(417,172){\special{em:lineto}}
\put(426,182){\special{em:lineto}}
\put(434,193){\special{em:lineto}}
\put(443,203){\special{em:lineto}}
\put(452,214){\special{em:lineto}}
\put(460,226){\special{em:lineto}}
\put(469,237){\special{em:lineto}}
\put(477,249){\special{em:lineto}}
\put(486,261){\special{em:lineto}}
\put(495,273){\special{em:lineto}}
\put(503,285){\special{em:lineto}}
\put(512,297){\special{em:lineto}}
\put(520,309){\special{em:lineto}}
\put(529,322){\special{em:lineto}}
\put(538,334){\special{em:lineto}}
\put(546,347){\special{em:lineto}}
\put(555,359){\special{em:lineto}}
\put(563,372){\special{em:lineto}}
\put(572,385){\special{em:lineto}}
\put(581,398){\special{em:lineto}}
\put(589,410){\special{em:lineto}}
\put(598,423){\special{em:lineto}}
\put(607,436){\special{em:lineto}}
\put(615,434){\special{em:lineto}}
\put(624,429){\special{em:lineto}}
\put(632,425){\special{em:lineto}}
\put(641,421){\special{em:lineto}}
\put(650,417){\special{em:lineto}}
\put(658,413){\special{em:lineto}}
\put(667,409){\special{em:lineto}}
\put(675,405){\special{em:lineto}}
\put(684,401){\special{em:lineto}}
\put(693,397){\special{em:lineto}}
\put(701,393){\special{em:lineto}}
\put(710,390){\special{em:lineto}}
\put(719,386){\special{em:lineto}}
\put(727,383){\special{em:lineto}}
\put(736,380){\special{em:lineto}}
\put(744,376){\special{em:lineto}}
\put(753,373){\special{em:lineto}}
\put(762,370){\special{em:lineto}}
\put(770,367){\special{em:lineto}}
\put(779,364){\special{em:lineto}}
\put(787,361){\special{em:lineto}}
\put(796,358){\special{em:lineto}}
\put(805,355){\special{em:lineto}}
\put(813,352){\special{em:lineto}}
\put(822,349){\special{em:lineto}}
\put(830,347){\special{em:lineto}}
\put(839,344){\special{em:lineto}}
\put(848,341){\special{em:lineto}}
\put(856,339){\special{em:lineto}}
\put(865,336){\special{em:lineto}}
\put(874,334){\special{em:lineto}}
\put(882,331){\special{em:lineto}}
\put(891,329){\special{em:lineto}}
\put(899,326){\special{em:lineto}}
\put(908,324){\special{em:lineto}}
\put(917,322){\special{em:lineto}}
\put(925,320){\special{em:lineto}}
\put(934,317){\special{em:lineto}}
\put(942,315){\special{em:lineto}}
\put(951,313){\special{em:lineto}}
\put(960,311){\special{em:lineto}}
\put(968,309){\special{em:lineto}}
\put(977,307){\special{em:lineto}}
\put(985,305){\special{em:lineto}}
\put(183,68){\special{em:moveto}}
\put(190,68){\special{em:lineto}}
\put(197,68){\special{em:lineto}}
\put(204,68){\special{em:lineto}}
\put(211,68){\special{em:lineto}}
\put(218,68){\special{em:lineto}}
\put(226,68){\special{em:lineto}}
\put(233,68){\special{em:lineto}}
\put(240,68){\special{em:lineto}}
\put(247,69){\special{em:lineto}}
\put(254,70){\special{em:lineto}}
\put(261,71){\special{em:lineto}}
\put(268,73){\special{em:lineto}}
\put(275,76){\special{em:lineto}}
\put(282,80){\special{em:lineto}}
\put(289,84){\special{em:lineto}}
\put(296,89){\special{em:lineto}}
\put(303,95){\special{em:lineto}}
\put(310,102){\special{em:lineto}}
\put(317,110){\special{em:lineto}}
\put(325,118){\special{em:lineto}}
\put(332,127){\special{em:lineto}}
\put(339,136){\special{em:lineto}}
\put(346,146){\special{em:lineto}}
\put(353,157){\special{em:lineto}}
\put(360,168){\special{em:lineto}}
\put(367,179){\special{em:lineto}}
\put(374,191){\special{em:lineto}}
\put(381,203){\special{em:lineto}}
\put(388,216){\special{em:lineto}}
\put(395,228){\special{em:lineto}}
\put(402,242){\special{em:lineto}}
\put(410,255){\special{em:lineto}}
\put(417,269){\special{em:lineto}}
\put(424,282){\special{em:lineto}}
\put(431,296){\special{em:lineto}}
\put(438,310){\special{em:lineto}}
\put(445,325){\special{em:lineto}}
\put(452,339){\special{em:lineto}}
\put(459,354){\special{em:lineto}}
\put(466,368){\special{em:lineto}}
\put(473,383){\special{em:lineto}}
\put(480,398){\special{em:lineto}}
\put(487,413){\special{em:lineto}}
\put(494,428){\special{em:lineto}}
\put(501,443){\special{em:lineto}}
\put(508,458){\special{em:lineto}}
\put(516,473){\special{em:lineto}}
\put(523,488){\special{em:lineto}}
\put(530,503){\special{em:lineto}}
\put(537,500){\special{em:lineto}}
\put(544,495){\special{em:lineto}}
\put(551,491){\special{em:lineto}}
\put(558,487){\special{em:lineto}}
\put(565,483){\special{em:lineto}}
\put(572,478){\special{em:lineto}}
\put(579,474){\special{em:lineto}}
\put(586,470){\special{em:lineto}}
\put(593,466){\special{em:lineto}}
\put(600,462){\special{em:lineto}}
\put(607,459){\special{em:lineto}}
\put(615,455){\special{em:lineto}}
\put(622,451){\special{em:lineto}}
\put(629,447){\special{em:lineto}}
\put(636,444){\special{em:lineto}}
\put(643,440){\special{em:lineto}}
\put(650,437){\special{em:lineto}}
\put(657,434){\special{em:lineto}}
\put(664,430){\special{em:lineto}}
\put(671,427){\special{em:lineto}}
\put(678,424){\special{em:lineto}}
\put(685,420){\special{em:lineto}}
\put(693,417){\special{em:lineto}}
\put(700,414){\special{em:lineto}}
\put(707,411){\special{em:lineto}}
\put(714,408){\special{em:lineto}}
\put(721,405){\special{em:lineto}}
\put(728,402){\special{em:lineto}}
\put(735,400){\special{em:lineto}}
\put(742,397){\special{em:lineto}}
\put(749,394){\special{em:lineto}}
\put(756,391){\special{em:lineto}}
\put(763,389){\special{em:lineto}}
\put(770,386){\special{em:lineto}}
\put(777,383){\special{em:lineto}}
\put(784,381){\special{em:lineto}}
\put(792,378){\special{em:lineto}}
\put(799,376){\special{em:lineto}}
\put(806,373){\special{em:lineto}}
\put(813,371){\special{em:lineto}}
\put(820,369){\special{em:lineto}}
\put(827,366){\special{em:lineto}}
\put(834,364){\special{em:lineto}}
\put(841,362){\special{em:lineto}}
\put(848,360){\special{em:lineto}}
\put(855,357){\special{em:lineto}}
\put(862,355){\special{em:lineto}}
\put(869,353){\special{em:lineto}}
\put(876,351){\special{em:lineto}}
\put(883,349){\special{em:lineto}}
\put(181,68){\special{em:moveto}}
\put(186,68){\special{em:lineto}}
\put(191,68){\special{em:lineto}}
\put(196,68){\special{em:lineto}}
\put(201,68){\special{em:lineto}}
\put(205,68){\special{em:lineto}}
\put(210,68){\special{em:lineto}}
\put(215,68){\special{em:lineto}}
\put(220,68){\special{em:lineto}}
\put(225,69){\special{em:lineto}}
\put(230,70){\special{em:lineto}}
\put(235,71){\special{em:lineto}}
\put(240,73){\special{em:lineto}}
\put(245,76){\special{em:lineto}}
\put(250,80){\special{em:lineto}}
\put(255,84){\special{em:lineto}}
\put(260,90){\special{em:lineto}}
\put(264,97){\special{em:lineto}}
\put(269,105){\special{em:lineto}}
\put(274,114){\special{em:lineto}}
\put(279,124){\special{em:lineto}}
\put(284,134){\special{em:lineto}}
\put(289,146){\special{em:lineto}}
\put(294,159){\special{em:lineto}}
\put(299,172){\special{em:lineto}}
\put(304,186){\special{em:lineto}}
\put(309,201){\special{em:lineto}}
\put(314,217){\special{em:lineto}}
\put(318,233){\special{em:lineto}}
\put(323,250){\special{em:lineto}}
\put(328,267){\special{em:lineto}}
\put(333,285){\special{em:lineto}}
\put(338,303){\special{em:lineto}}
\put(343,322){\special{em:lineto}}
\put(348,341){\special{em:lineto}}
\put(353,360){\special{em:lineto}}
\put(358,380){\special{em:lineto}}
\put(363,400){\special{em:lineto}}
\put(368,420){\special{em:lineto}}
\put(372,441){\special{em:lineto}}
\put(378,461){\special{em:lineto}}
\put(382,482){\special{em:lineto}}
\put(387,504){\special{em:lineto}}
\put(392,525){\special{em:lineto}}
\put(397,546){\special{em:lineto}}
\put(402,568){\special{em:lineto}}
\put(407,589){\special{em:lineto}}
\put(412,611){\special{em:lineto}}
\put(417,633){\special{em:lineto}}
\put(422,655){\special{em:lineto}}
\put(427,656){\special{em:lineto}}
\put(431,653){\special{em:lineto}}
\put(436,651){\special{em:lineto}}
\put(441,649){\special{em:lineto}}
\put(446,647){\special{em:lineto}}
\put(451,644){\special{em:lineto}}
\put(456,642){\special{em:lineto}}
\put(461,640){\special{em:lineto}}
\put(466,637){\special{em:lineto}}
\put(471,635){\special{em:lineto}}
\put(476,632){\special{em:lineto}}
\put(481,630){\special{em:lineto}}
\put(486,628){\special{em:lineto}}
\put(490,625){\special{em:lineto}}
\put(495,623){\special{em:lineto}}
\put(500,620){\special{em:lineto}}
\put(505,618){\special{em:lineto}}
\put(510,615){\special{em:lineto}}
\put(515,613){\special{em:lineto}}
\put(520,611){\special{em:lineto}}
\put(525,608){\special{em:lineto}}
\put(530,606){\special{em:lineto}}
\put(535,603){\special{em:lineto}}
\put(540,601){\special{em:lineto}}
\put(545,598){\special{em:lineto}}
\put(549,596){\special{em:lineto}}
\put(554,593){\special{em:lineto}}
\put(559,591){\special{em:lineto}}
\put(564,589){\special{em:lineto}}
\put(569,586){\special{em:lineto}}
\put(574,584){\special{em:lineto}}
\put(579,581){\special{em:lineto}}
\put(584,579){\special{em:lineto}}
\put(589,577){\special{em:lineto}}
\put(594,574){\special{em:lineto}}
\put(599,572){\special{em:lineto}}
\put(604,570){\special{em:lineto}}
\put(608,567){\special{em:lineto}}
\put(613,565){\special{em:lineto}}
\put(618,563){\special{em:lineto}}
\put(623,560){\special{em:lineto}}
\put(628,558){\special{em:lineto}}
\put(633,556){\special{em:lineto}}
\put(638,553){\special{em:lineto}}
\put(643,551){\special{em:lineto}}
\put(648,549){\special{em:lineto}}
\put(653,547){\special{em:lineto}}
\put(658,545){\special{em:lineto}}
\put(662,542){\special{em:lineto}}
\put(667,540){\special{em:lineto}}
\put(181,68){\special{em:moveto}}
\put(185,68){\special{em:lineto}}
\put(190,68){\special{em:lineto}}
\put(194,68){\special{em:lineto}}
\put(199,68){\special{em:lineto}}
\put(203,68){\special{em:lineto}}
\put(208,68){\special{em:lineto}}
\put(212,68){\special{em:lineto}}
\put(217,68){\special{em:lineto}}
\put(221,69){\special{em:lineto}}
\put(226,70){\special{em:lineto}}
\put(230,71){\special{em:lineto}}
\put(235,73){\special{em:lineto}}
\put(239,76){\special{em:lineto}}
\put(244,79){\special{em:lineto}}
\put(248,84){\special{em:lineto}}
\put(253,90){\special{em:lineto}}
\put(257,97){\special{em:lineto}}
\put(262,104){\special{em:lineto}}
\put(266,113){\special{em:lineto}}
\put(271,123){\special{em:lineto}}
\put(275,134){\special{em:lineto}}
\put(280,146){\special{em:lineto}}
\put(284,159){\special{em:lineto}}
\put(289,172){\special{em:lineto}}
\put(293,187){\special{em:lineto}}
\put(298,202){\special{em:lineto}}
\put(302,218){\special{em:lineto}}
\put(307,234){\special{em:lineto}}
\put(311,252){\special{em:lineto}}
\put(316,270){\special{em:lineto}}
\put(320,288){\special{em:lineto}}
\put(325,307){\special{em:lineto}}
\put(329,327){\special{em:lineto}}
\put(334,347){\special{em:lineto}}
\put(338,367){\special{em:lineto}}
\put(343,388){\special{em:lineto}}
\put(348,409){\special{em:lineto}}
\put(352,430){\special{em:lineto}}
\put(356,452){\special{em:lineto}}
\put(361,474){\special{em:lineto}}
\put(365,497){\special{em:lineto}}
\put(370,519){\special{em:lineto}}
\put(375,542){\special{em:lineto}}
\put(379,565){\special{em:lineto}}
\put(384,588){\special{em:lineto}}
\put(388,612){\special{em:lineto}}
\put(393,635){\special{em:lineto}}
\put(397,659){\special{em:lineto}}
\put(402,683){\special{em:lineto}}
\put(406,684){\special{em:lineto}}
\put(411,683){\special{em:lineto}}
\put(415,682){\special{em:lineto}}
\put(420,681){\special{em:lineto}}
\put(424,680){\special{em:lineto}}
\put(429,678){\special{em:lineto}}
\put(433,677){\special{em:lineto}}
\put(438,676){\special{em:lineto}}
\put(442,675){\special{em:lineto}}
\put(447,673){\special{em:lineto}}
\put(451,672){\special{em:lineto}}
\put(456,671){\special{em:lineto}}
\put(460,670){\special{em:lineto}}
\put(465,668){\special{em:lineto}}
\put(469,667){\special{em:lineto}}
\put(474,665){\special{em:lineto}}
\put(478,664){\special{em:lineto}}
\put(483,662){\special{em:lineto}}
\put(487,661){\special{em:lineto}}
\put(492,659){\special{em:lineto}}
\put(497,658){\special{em:lineto}}
\put(501,656){\special{em:lineto}}
\put(506,655){\special{em:lineto}}
\put(510,653){\special{em:lineto}}
\put(515,652){\special{em:lineto}}
\put(519,650){\special{em:lineto}}
\put(524,648){\special{em:lineto}}
\put(528,647){\special{em:lineto}}
\put(532,645){\special{em:lineto}}
\put(537,644){\special{em:lineto}}
\put(542,642){\special{em:lineto}}
\put(546,640){\special{em:lineto}}
\put(551,639){\special{em:lineto}}
\put(555,637){\special{em:lineto}}
\put(560,635){\special{em:lineto}}
\put(564,634){\special{em:lineto}}
\put(569,632){\special{em:lineto}}
\put(573,630){\special{em:lineto}}
\put(578,628){\special{em:lineto}}
\put(582,627){\special{em:lineto}}
\put(587,625){\special{em:lineto}}
\put(591,623){\special{em:lineto}}
\put(596,622){\special{em:lineto}}
\put(600,620){\special{em:lineto}}
\put(605,618){\special{em:lineto}}
\put(609,616){\special{em:lineto}}
\put(614,615){\special{em:lineto}}
\put(618,613){\special{em:lineto}}
\put(623,611){\special{em:lineto}}
\put(627,610){\special{em:lineto}}
\put(180,68){\special{em:moveto}}
\put(184,68){\special{em:lineto}}
\put(189,68){\special{em:lineto}}
\put(193,68){\special{em:lineto}}
\put(197,68){\special{em:lineto}}
\put(201,68){\special{em:lineto}}
\put(206,68){\special{em:lineto}}
\put(210,68){\special{em:lineto}}
\put(214,68){\special{em:lineto}}
\put(218,69){\special{em:lineto}}
\put(223,70){\special{em:lineto}}
\put(227,71){\special{em:lineto}}
\put(231,73){\special{em:lineto}}
\put(235,76){\special{em:lineto}}
\put(240,79){\special{em:lineto}}
\put(244,84){\special{em:lineto}}
\put(248,90){\special{em:lineto}}
\put(252,96){\special{em:lineto}}
\put(257,104){\special{em:lineto}}
\put(261,113){\special{em:lineto}}
\put(265,123){\special{em:lineto}}
\put(269,134){\special{em:lineto}}
\put(274,145){\special{em:lineto}}
\put(278,158){\special{em:lineto}}
\put(282,172){\special{em:lineto}}
\put(286,186){\special{em:lineto}}
\put(290,202){\special{em:lineto}}
\put(295,218){\special{em:lineto}}
\put(299,234){\special{em:lineto}}
\put(303,252){\special{em:lineto}}
\put(307,270){\special{em:lineto}}
\put(312,288){\special{em:lineto}}
\put(316,307){\special{em:lineto}}
\put(320,327){\special{em:lineto}}
\put(324,347){\special{em:lineto}}
\put(329,368){\special{em:lineto}}
\put(333,389){\special{em:lineto}}
\put(337,411){\special{em:lineto}}
\put(341,433){\special{em:lineto}}
\put(346,455){\special{em:lineto}}
\put(350,477){\special{em:lineto}}
\put(354,500){\special{em:lineto}}
\put(358,524){\special{em:lineto}}
\put(363,547){\special{em:lineto}}
\put(367,571){\special{em:lineto}}
\put(371,595){\special{em:lineto}}
\put(375,619){\special{em:lineto}}
\put(380,643){\special{em:lineto}}
\put(384,668){\special{em:lineto}}
\put(388,692){\special{em:lineto}}
\put(392,694){\special{em:lineto}}
\put(396,694){\special{em:lineto}}
\put(401,694){\special{em:lineto}}
\put(405,693){\special{em:lineto}}
\put(409,693){\special{em:lineto}}
\put(413,693){\special{em:lineto}}
\put(418,692){\special{em:lineto}}
\put(422,692){\special{em:lineto}}
\put(426,691){\special{em:lineto}}
\put(430,691){\special{em:lineto}}
\put(435,691){\special{em:lineto}}
\put(439,690){\special{em:lineto}}
\put(443,690){\special{em:lineto}}
\put(447,689){\special{em:lineto}}
\put(452,688){\special{em:lineto}}
\put(456,688){\special{em:lineto}}
\put(460,687){\special{em:lineto}}
\put(464,687){\special{em:lineto}}
\put(468,686){\special{em:lineto}}
\put(473,685){\special{em:lineto}}
\put(477,685){\special{em:lineto}}
\put(481,684){\special{em:lineto}}
\put(486,683){\special{em:lineto}}
\put(490,683){\special{em:lineto}}
\put(494,682){\special{em:lineto}}
\put(498,681){\special{em:lineto}}
\put(502,680){\special{em:lineto}}
\put(507,680){\special{em:lineto}}
\put(511,679){\special{em:lineto}}
\put(515,678){\special{em:lineto}}
\put(519,677){\special{em:lineto}}
\put(524,676){\special{em:lineto}}
\put(528,675){\special{em:lineto}}
\put(532,674){\special{em:lineto}}
\put(536,673){\special{em:lineto}}
\put(541,673){\special{em:lineto}}
\put(545,672){\special{em:lineto}}
\put(549,671){\special{em:lineto}}
\put(553,670){\special{em:lineto}}
\put(558,669){\special{em:lineto}}
\put(562,668){\special{em:lineto}}
\put(566,667){\special{em:lineto}}
\put(570,666){\special{em:lineto}}
\put(575,665){\special{em:lineto}}
\put(579,664){\special{em:lineto}}
\put(583,663){\special{em:lineto}}
\put(587,662){\special{em:lineto}}
\put(591,661){\special{em:lineto}}
\put(596,660){\special{em:lineto}}
\put(600,659){\special{em:lineto}}
\sbox{\plotpoint}{\rule[-0.200pt]{0.400pt}{0.400pt}}%
\special{em:linewidth 0.4pt}%
\put(356,68){\special{em:moveto}}
\put(356,697){\special{em:lineto}}
\end{picture}

\caption{$\chi(T)$ for different values of
$\epsilon_{F}/|\varepsilon_{b}|$: 1 --- 0.2; 2 --- 0.6; 3 --- 1;
4 --- 5; 5 --- 10; 6 --- 20.}
\end{figure}
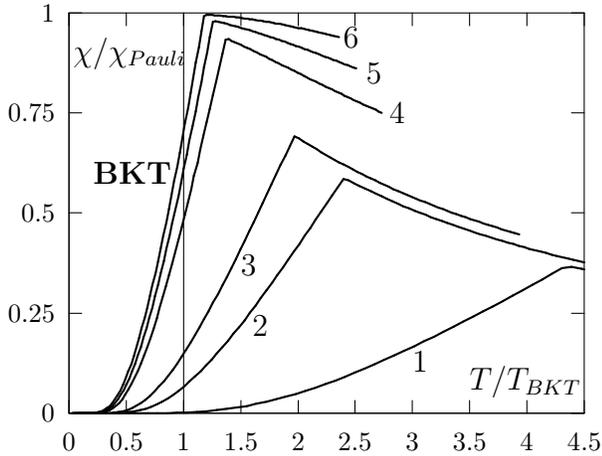

              The results of numerical study of the system
(\ref{susceptibility.final}), (\ref{rho.bound}) and (\ref{number.rho})
are presented in Fig.~4. One can see that the kink in $\chi$
happens at $T = T_{\rho}$ as in the dependence of $\mu$ on $T$.
Below $T_{\rho}$ the value of $\chi(T)$ decreases, although the system is
still normal. This means that the spin-gap (pseudogap) opens.
The size of the pseudogap region depends strongly on the doping
($\epsilon_{F}/ |\varepsilon_{b}|$), as it takes place for the real
HTSC \cite{Levi,Randeria.Nature,Pines.review}. For small values of
$\epsilon_{F}/ |\varepsilon_{b}|$ this region is large
($T_{\rho} > 2T_{BKT}$), while for the large ratio it is small.

\section{Concluding remarks}
To summarize we have discussed the crossover in the superconducting
transition between BCS- and Bose-like behaviour for the simplest
3D, quasi-2D and 2D models with s-wave direct nonretarded
attractive interaction.
It has been pointed out that optimally doped HTSC are still on the
BCS side of this crossover, although they are certainly far away
from the standard BCS description.

Above we emphasized the model description which in our opinion proves
to be the most suitable for the clarification of the diverse physical
properties peculiar to electronic systems with changeable
carrier density in any dimension. While there
is still no generally accepted microscopic theory of HTSC compounds and their
basic features (including the pairing mechanism), it seems to us that this
approach, although in a sense phenomenological, is of great interest since
it is able to cover the whole region of carrier concentrations
(and consequently the whole range of coupling constants),
temperatures and crystal anisotropies. It, as we tried to demonstrate, allows
one not only to propose a reasonable interpretation for the  observed
phenomena caused by doping but also to predict new phenomena (for example,
pseudogap phase formation as a new thermodynamically
equilibrium normal state of low dimensional conducting electronic systems).

           Evidently there are a number of important open questions.
They may be divided into two classes: the first one concerns the problem
of a better and more complete treatment of the models themselves.
The second class is related to the problem of to what extent these
model are applicable to HTSC compounds and what are the necessary
ingredients for a more realistic description.

           Regarding our treatment of the 3D and quasi-2D models, it is
obvious that one has to take into account the interaction between the
bosons also. In particular the Gaussian approximation is not sufficient
to give reliable results for $T_{c}$ at intermediate coupling when
the "size" of the bosons is comparable with the mean distance between
them. As for the 2D model, there is some unconfirmed numerical result
\cite{Deisz} based on a fully self-consistent determination of a phase
transition to a superconducting state in a conserving approximation,
which state that the superconducting transition is not the BKT
transition. Besides, it would be very interesting to obtain the
spectrum of the anomalous normal phase.

           Concerning the question to which the models considered
are really applicable to HTSC, it is obvious that most of the
complexity of these systems is neglected here. We did not take into
account an indirect nature of the interaction between the fermions
and d-wave pairing. Note, however, that some attempts to study the
crossover for these cases were made \cite{LSh.FNT.1996,LGN.1994,Zwerger}
(for general review see \cite{Loktev.review}).

           A lot of peculiarities of HTSC are now
connected with stripe structure of CuO$_2$ planes, which,
according to many experiments (see \cite{Bianconi} and references therein),
are divided onto bands of the normal and superconducting bands.
There is interesting and important problem how to investigate these
systems.

          The problem of the crossover from BEC to BCS
(especially for 2D systems) is such a rich one that without doubt will
bring us a lot of surprises in the near and far future.
One of them is perhaps the unified theory of
superconductivity and magnetism \cite{Zh} which has
already excited a lot of interest and criticism (e.g. \cite{And}).

\renewcommand{\theequation}{\Alph{section}.\arabic{equation}}
\appendix
\section{The effective potential}
\setcounter{equation}{0}
           Let us derive the effective potential
(\ref{Omega.Potential.expr}).
To obtain it one should write down the formal expression
(\ref{Omega.Potential}) in the momentum representation, so that
\begin{eqnarray}
&& \Omega_{pot}(v, \mu, T, \Phi,\Phi^{\ast}) = v \left\{
\frac{|\Phi|^{2}}{V} - T \sum_{n = -\infty}^{+\infty}
\int \frac{d {\mbox{\bf k}}}{(2 \pi)^{d}}
\mbox{tr} [\ln G^{-1} (i \omega_{n}, \mbox{\bf k})
e^{i \delta \omega_{n} \tau_{3}}]
\right.
\nonumber                                \\
&& \qquad
+ \left. T \sum_{n = -\infty}^{+\infty}
\int \frac{d {\mbox{\bf k}}}{(2 \pi)^{d}}
\mbox{tr} [\ln G_{0}^{-1} (i \omega_{n}, \mbox{\bf k})
e^{i \delta \omega_{n} \tau_{3}}] \right\}, \quad
\delta \to +0,                                 \label{A1}
\end{eqnarray}
where
\begin{equation}
G^{-1} (i \omega_{n}, \mbox{\bf k}) =
i \omega_{n} \hat I - \tau_{3} \xi(\mbox{\bf k}) +
\tau_{+} \Phi + \tau_{-} \Phi^{\ast} =
\left( \begin{array}{ccc}
i \omega_{n} - \xi(\mbox{\bf k}) & \Phi \\
\Phi^{\ast}                      & i \omega_{n} + \xi(\mbox{\bf k})
\end{array} \right)                             \label{A2}
\end{equation}
and
\begin{equation}
G_{0}^{-1} (i \omega_{n}, \mbox{\bf k}) =
\left. G^{-1} (i \omega_{n}, \mbox{\bf k})
\right|_{\Phi = \Phi^{\ast} = \mu =0} =
\left( \begin{array}{ccc}
i \omega_{n} - \varepsilon(\mbox{\bf k}) & 0 \\
0                      & i \omega_{n} + \varepsilon(\mbox{\bf k})
\end{array} \right)                             \label{A3}
\end{equation}
are the inverse Green functions.
The exponential factor  $e^{i \delta \omega_{n} \tau_{3}}$ is added
into (\ref{A1}) to provide a right regularization which is necessary
to perform the calculation with the Green functions (see
\cite{Abrikosov}). For instance, one obtains that
\begin{eqnarray}
&& \! \! \! \! \! \! \! \! \! \! \! \!
\lim_{\delta \to +0} \sum_{n = -\infty}^{+\infty}
\mbox{tr} [\ln G^{-1} (i \omega_{n}, \mbox{\bf k})
e^{i \delta \omega_{n} \tau_{3}}] =
\nonumber                   \\
&& \! \! \! \! \! \! \! \! \! \!
\lim_{\delta \to +0} \left\{ \sum_{n = -\infty}^{+\infty}
\mbox{tr} [\ln G^{-1} (i \omega_{n}, \mbox{\bf k})]
\cos \delta \omega_{n} +  \right.
\nonumber                    \\
&& \! \! \! \! \! \! \! \!
\left. i \sum_{\omega_{n} > 0} \sin \delta \omega_{n}
\mbox{tr} [(\ln G^{-1} (i \omega_{n}, \mbox{\bf k}) -
\ln G^{-1} (- i \omega_{n}, \mbox{\bf k})) \tau_{3}] \right\} =
\nonumber                      \\
&& \! \! \! \! \! \!
\sum_{n = -\infty}^{+\infty}
\mbox{tr} [\ln G^{-1} (i \omega_{n}, \mbox{\bf k})] -
\frac{\xi(\mbox{\bf k})}{T},                    \label{A4}
\end{eqnarray}
where the properties that
\begin{displaymath}
\ln G^{-1} (i \omega_{n}, \mbox{\bf k})  =
- \tau_{3} \frac{\xi(\mbox{\bf k})}{i \omega_{n}}, \quad
\omega_{n} \to \infty
\end{displaymath}
and
\begin{displaymath}
\sum_{\omega_{n} > 0} \frac{\sin \delta \omega_{n}}{\omega_{n}}
\simeq \frac{1}{2 \pi T} \int_{0}^{\infty} dx
\frac{\sin \delta x }{x} = \frac{1}{4 T} \mbox{sign} \delta
\end{displaymath}
were used.

	    To calculate the sum in (\ref{A4}) one have to firstly use
the identity $\mbox{tr} \ln \hat A = \ln \det \hat A$, so that
(\ref{A1}) takes the following form
\begin{eqnarray}
\Omega_{pot}(v, \mu, T, \Phi,\Phi^{\ast}) = v \left\{
\frac{|\Phi|^{2}}{V} \right. & - & T \sum_{n = -\infty}^{+\infty}
\int \frac{d {\mbox{\bf k}}}{(2 \pi)^{d}}
\ln \frac{\det G^{-1} (i \omega_{n}, \mbox{\bf k})}
{\det G_{0}^{-1} (i \omega_{n}, \mbox{\bf k})}
\nonumber                                \\
& - & \left. \int \frac{d {\mbox{\bf k}}}{(2 \pi)^{d}}
[-\xi (\mbox{\bf k}) + \varepsilon (\mbox{\bf k})] \right\}.
					  \label{A5}
\end{eqnarray}
Calculating the determinants of the Green functions (\ref{A2})
and (\ref{A3}) one gets
\begin{eqnarray}
\Omega_{pot}(v, \mu, T, \Phi,\Phi^{\ast}) = v \left\{
\frac{|\Phi|^{2}}{V} \right. & - & T \sum_{n = -\infty}^{+\infty}
\int \frac{d {\mbox{\bf k}}}{(2 \pi)^{d}}
\ln \frac{\omega_{n}^{2} + \xi^{2}(\mbox{\bf k}) + |\Phi|^{2}}
{\omega_{n}^{2} + \varepsilon^{2}(\mbox{\bf k})}
\nonumber                                \\
& - & \left. \int \frac{d {\mbox{\bf k}}}{(2 \pi)^{d}}
[-\xi (\mbox{\bf k}) + \varepsilon (\mbox{\bf k})] \right\},
					  \label{A6}
\end{eqnarray}
where the role of $G_{0} (i \omega_{n}, \mbox{\bf k})$ in the
regularization of $\Omega_{pot}$ is evident now. The summation
in (\ref{A6}) can be done if one uses the following representation
\begin{equation}
\ln \frac{\omega_{n}^{2} + a^{2}}{\omega_{n}^{2} + b^{2}} =
\int_{0}^{\infty} dx
\left(\frac{1}{\omega_{n}^{2} + a^{2} + x} -
      \frac{1}{\omega_{n}^{2} + b^{2} + x} \right). \label{A7}
\end{equation}
Then, the sum  \cite{Prudnikov}
\begin{equation}
\sum_{k = 0}^{\infty}\frac{1}{(2k+1)^{2} + c^{2}} =
\frac{\pi}{4c} \tanh \frac{\pi c}{2}       \label{A8}
\end{equation}
may be now applied and one obtains
\begin{eqnarray}
&& \! \! \! \! \! \! \! \! \! \! \! \!  \! \! \! \!  \! \! \! \!
\ln \frac{\omega_{n}^{2} + a^{2}}{\omega_{n}^{2} + b^{2}} =
\nonumber                   \\
\int_{0}^{\infty} dx  && \! \! \! \! \left(
\frac{1}{2 \sqrt{b^{2} + x}} \tanh \frac{\sqrt{b^{2} + x}}{2 T} -
\frac{1}{2 \sqrt{a^{2} + x}} \tanh \frac{\sqrt{a^{2} + x}}{2 T}
\right).                     \label{A9}
\end{eqnarray}
Integrating (\ref{A9}) over $x$  one thus arrives to the expression:
\begin{eqnarray}
&& \! \! \! \!  \! \! \! \!  \! \! \! \!
T \sum_{n = -\infty}^{+\infty}
\int \frac{d {\mbox{\bf k}}}{(2 \pi)^{d}}
\ln \frac{\omega_{n}^{2} + \xi(\mbox{\bf k})^{2} + |\Phi|^{2}}
{\omega_{n}^{2} + \varepsilon^{2}(\mbox{\bf k})} =
\nonumber                      \\
&& 2 T  \int \frac{d {\mbox{\bf k}}}{(2 \pi)^{d}}
\ln \frac{\cosh [\sqrt{\xi^{2}(\mbox{\bf k}) + |\Phi|^{2}}/2T ]}
{\cosh [\varepsilon(\mbox{\bf k})/2T ]}.       \label{A10}
\end{eqnarray}
Finally, substituting (\ref{A10}) into (\ref{A6})
we get (\ref{Omega.Potential.expr}).

\section{Low energy kinetic part of the effective action}
\setcounter{equation}{0}
         Here we derive the kinetic part (\ref{Omega.Kinetic.phase})
of the effective action (\ref{kinetic.phase+potential}).
To obtain it one should calculate
directly the first two terms of the series in  (\ref{Omega.Kinetic.phase})
which are formally written as
$\Omega_{kin}^{(1)} = T \mbox{Tr} ({\cal G} \Sigma)$ and
$\Omega_{kin}^{(2)} =
\frac{1}{2}T \mbox{Tr} ({\cal G} \Sigma {\cal G} \Sigma)$.
The straightforward calculation of $\Omega_{kin}^{(1)}$ gives
\begin{equation}
\Omega_{kin}^{(1)} = T \int_{0}^{\beta} d \tau \int d \mbox{\bf r}
\frac{T}{(2 \pi)^{2}} \sum_{n = - \infty}^{\infty}
\int d \mbox{\bf k}
\mbox{tr} [{\cal G}(i \omega_{n}, \mbox{\bf k}) \tau_{3}]
\left( i \partial_{\tau} \theta +
\frac{(\nabla \theta)^{2}}{2 m}\right),         \label{B1}
\end{equation}
where
\begin{equation}
{\cal G}(i \omega_{n}, \mbox{\bf k}) = - \frac{
i \omega_{n} \hat{I} + \tau_{3} \xi(\mbox{\bf k}) - \tau_{1} \rho}
{\omega_{n}^{2} + \xi^{2}(\mbox{\bf k}) + \rho^{2}}     \label{B2}
\end{equation}
is the Green function of the neutral fermions in the frequency-momentum
representation (compare with (\ref{Green.fermion.momentum})).
The summation over Matsubara
frequencies $\omega_{n} = \pi (2n + 1) T$ and integration over
$\mbox{\bf k}$ in (\ref{B1}) can be easily performed
using the sum (\ref{A8}) and thus one obtains
\begin{equation}
\Omega_{kin}^{(1)} =
T \int_{0}^{\beta} d \tau \int d \mbox{\bf r}
n_{F}(\mu, T, \rho) \left(
i \partial_{\tau} \theta + \frac{(\nabla \theta)^{2}}{2 m}
\right),                \label{B3}
\end{equation} where $n_{F}(\mu, T, \rho)$ is determined
by (\ref{fermion.density.2D}).  We note that $\Sigma$ has the following
structure $\Sigma = \tau_{3} O_{1} + \hat{I} O_{2}$ where $O_{1}$ and
$O_{2}$ are some differential operators (see (\ref{Sigma})).  One can
see, however, that the part of $\Sigma$, proportional to the unit
matrix $\hat{I}$, does not contribute in $\Omega_{kin}^{(1)}$.

For the case $T = 0$ \cite{Thouless,Schakel} when real time $t$
replaces imaginary time $\tau$, one can argue from the Galilean
invariance that the coefficient of $\partial_{t} \theta$
is rigidly related to the coefficient at $(\nabla \theta)^{2}$.
So it does not appear in $\Omega_{kin}^{(2)}$.
We wish, however, to stress that these arguments can not be used
to exclude the appearance of the term $(\nabla \theta)^{2}$
from $\Omega_{kin}^{(2)}$ when $T \neq 0$, thus we must calculate
it explicitly.

The $O_{1}$ term in $\Sigma$ yields
\begin{eqnarray}
\Omega_{kin}^{(2)} (O_{1}) & = & \frac{T}{2}
\int_{0}^{\beta} d \tau \int d \mbox{\bf r}
\frac{T}{(2 \pi)^{2}} \sum_{n = - \infty}^{\infty}
\int d \mbox{\bf k}
\mbox{tr} [{\cal G}(i \omega_{n}, \mbox{\bf k}) \tau_{3}
      {\cal G}(i \omega_{n}, \mbox{\bf k}) \tau_{3}] \times
\nonumber                                 \\
&& \left( i \partial_{\tau} \theta +
\frac{(\nabla \theta)^{2}}{2 m}\right)^{2},         \label{B4}
\end{eqnarray}
And from (\ref{B4}) we find that
\begin{equation}
\Omega_{kin}^{(2)} (O_{1}) = - \frac{T}{2}
\int_{0}^{\beta} d \tau \int d \mbox{\bf r}
K(\mu, T, \rho)
\left( i \partial_{\tau} \theta +
\frac{(\nabla \theta)^{2}}{2 m}\right)^{2},         \label{B5}
\end{equation}
where $K(\mu, T, \rho)$ was defined in (\ref{K}).
It is evident that $O_{1}$ term does not affect the coefficient
of $(\nabla \theta)^{2}$.
Further, it is easy to make sure that the cross term from
$O_{1}$ and $O_{2}$ in $\Omega_{kin}^{(2)}$ is absent.
Finally, the calculations of the $O_{2}$ term contribution
to $\Omega_{kin}^{2}$
\footnote{The higher than $(\nabla \theta)^{2}$ derivatives
were not found here.}
give
\begin{eqnarray}
\Omega_{kin}^{(2)} (O_{2}) = && \! \! \! \! \! \! T
\int_{0}^{\beta} d \tau \int d \mbox{\bf r}
\frac{T}{(2 \pi)^{2}} \sum_{n = - \infty}^{\infty}
\int d \mbox{\bf k} \mbox{\bf k}^{2}
\mbox{tr} [{\cal G}(i \omega_{n}, \mbox{\bf k}) \hat{I}
      {\cal G}(i \omega_{n}, \mbox{\bf k}) \hat{I}] \times
\nonumber               \\
&& \frac{(\nabla \theta)^{2}}{4 m^{2}}.                 \label{B6}
\end{eqnarray}
Thus, after summation over Matsubara frequencies (see Appendix~C)
\begin{equation}
\Omega_{kin}^{(2)} (O_{2}) = -
\int_{0}^{\beta} d \tau \int d \mbox{\bf r}
\frac{1}{32 \pi^{2} m^{2}}
\int d \mbox{\bf k}
\frac{\mbox{\bf k}^{2}}
{\cosh^{2} \frac{\ds \sqrt{\xi^{2}(\mbox{\bf k}) + \rho^{2}}}{\ds 2T}}
(\nabla \theta)^{2}.                                    \label{B7}
\end{equation}
As expected this term vanishes when $T \to 0$
but at finite $T$ it is comparable with (\ref{B3}).
Combining (\ref{B3}), (\ref{B5}) and (\ref{B7}) we obtain
(\ref{Omega.Kinetic.phase.final}).

\section{Summation over Matsubara frequencies}
\setcounter{equation}{0}
        Here we perform the summation over Matsubara frequencies in the
following expression
\begin{equation}
T \sum_{n = -\infty}^{\infty} \mbox{tr}
[{\cal G}(i \omega_{n}, \mbox{\bf k}) \hat I
{\cal G}(i \omega_{n}, \mbox{\bf k}) \hat I],
                               \label{C1}
\end{equation}
where the Green function ${\cal G}(i \omega_{n}, \mbox{\bf k})$ is given
by (\ref{B2}). At first, using the elementary properties of the Pauli
matrices one obtains
\begin{equation}
\mbox{tr} [{\cal G}(i \omega_{n}, \mbox{\bf k}) \hat I
{\cal G}(i \omega_{n}, \mbox{\bf k}) \hat I] =
\frac{2 [\xi^{2}(\mbox{\bf k}) + \rho^{2} - \omega_{n}^{2}]}
{[\omega_{n}^{2} + \xi^{2}(\mbox{\bf k}) + \rho^{2}]^{2}}.
                               \label{C2}
\end{equation}
Then, the summation can be easily carried out, if one uses the
following sums \cite{Prudnikov}
\begin{equation}
\sum_{k = 0}^{\infty} \frac{1}{[(2k+1)^{2} + a^{2}]^{2}} =
\frac{\pi}{8 a^{3}} \tanh \frac{\pi a}{2} -
\frac{\pi^{2}}{16 a^{2}} \frac{1}{\cosh^{2} \frac{\ds \pi a}{\ds 2}}
                                     \label{C3}
\end{equation}
and
\begin{equation}
\sum_{k = 0}^{\infty} \frac{(2k+1)^{2}}{[(2k+1)^{2} + a^{2}]^{2}} =
\frac{\pi}{8 a} \tanh \frac{\pi a}{2} +
\frac{\pi^{2}}{16} \frac{1}{\cosh^{2} \frac{\ds \pi a}{\ds 2}}.
                                     \label{C4}
\end{equation}
Assuming that
$a^{2} \equiv (\xi^{2} (\mbox{\bf k}) + \rho^{2})/ \pi^{2} T^{2}$
one directly arrives to the final result
\begin{equation}
2 T \sum _{n = - \infty}^{\infty}
\frac{\xi^{2}(\mbox{\bf k}) + \rho^{2} - \omega_{n}^{2}}
{[\omega_{n}^{2} + \xi^{2}(\mbox{\bf k}) + \rho^{2}]^{2}} =
- \frac{1}{2T} \frac{1}
{\cosh^{2} \frac{\ds \sqrt{\xi^{2}(\mbox{\bf k}) + \rho^{2}}}{\ds 2T}}.
                                           \label{C5}
\end{equation}

\end{document}